\documentclass[useAMS,usenatbib]{mn2e}
\usepackage{graphicx}
\usepackage{amsmath}
\usepackage{comment}
\usepackage{txfonts}
\usepackage{color}
\usepackage{float}
\usepackage[percent]{overpic} 

\title[Distances and Metallicities along the Giant Stellar Stream]{Major Substructure in the M31 Outer Halo: \\ Distances and Metallicities along the Giant Stellar Stream$^{\ddag}$}

\author[A. R. Conn]{A. R. Conn$^{1}$\thanks{E-mail:
anthony\_conn@hotmail.com}, B. McMonigal$^{1}$, N. F. Bate$^{2}$, G. F. Lewis$^{1}$, R. A. Ibata$^{3}$, N. F. Martin$^{3, 4}$, A. W. \newauthor McConnachie$^{5}$, A. M. N. Ferguson$^{6}$, M. J. Irwin$^{2}$, P. J. Elahi$^{1}$, K. A. Venn$^{7}$, A. D. Mackey$^{8}$\\
$^{1}$Sydney Institute for Astronomy, School of Physics, A28, The University of Sydney, Sydney, NSW 2006, Australia\\
$^{2}$Institute of Astronomy, University of Cambridge, Madingley Road, Cambridge CB3 0HA, UK\\
$^{3}$Observatoire astronomique de Strasbourg, Universit\'e de Strasbourg, CNRS, UMR 7550, 11 rue de l'Universit\'e, F-67000 Strasbourg\\
$^{4}$Max-Planck-Institut f\"ur Astronomie, K\"onigstuhl 17, D-69117 Heidelberg, Germany\\ 
$^{5}$ NRC Herzberg Institute of Astrophysics, 5071 West Saanich Road, Victoria, British Columbia, Canada V9E 2E7 \\
$^{6}$ Institute for Astronomy, University of Edinburgh, Royal Observatory, Blackford Hill, Edinburgh EH9 3HJ, UK \\
$^{7}$ Department of Physics and Astronomy, University of Victoria, 3800 Finnerty Road, Victoria, British Columbia, Canada V8P 5C2 \\ 
$^{8}$ RSAA, Australian National University, Mt. Stromlo Observatory, Cotter Road, Weston Creek, ACT 2611, Australia \\ \\
$^{\ddag}$Based on observations obtained with MegaPrime/MegaCam, a joint project of CFHT and CEA/DAPNIA, at the Canada-France-Hawaii Telescope (CFHT)\\
which is operated by the National Research Council (NRC) of Canada, the Institute National des Sciences de l'Univers of the Centre National de la Recherche\\
Scientifique of France, and the University of Hawaii.}

\begin{document}

\date{Accepted year month day; Received year month day; in original form year month day}

\pagerange{\pageref{firstpage}--\pageref{lastpage}} \pubyear{2015}

\maketitle

\label{firstpage}

\begin{abstract}
We present a renewed look at M31's Giant Stellar Stream along with the nearby structures Stream C and Stream D, exploiting a new algorithm capable of fitting to the red giant branch (RGB) of a structure in both colour and magnitude space. Using this algorithm, we are able to generate probability distributions in distance, metallicity and RGB width for a series of subfields spanning these structures. Specifically, we confirm a distance gradient of approximately 20 kpc per degree along a 6 degree extension of the Giant Stellar Stream, with the farthest subfields from M31 lying  $\sim$ 120 kpc more distant than the inner-most subfields. Further, we find a metallicity that steadily increases from $-0.7^{+0.1}_{-0.1}$ dex to $-0.2^{+0.2}_{-0.1}$ dex along the inner half of the stream before steadily dropping to a value of $-1.0^{+0.2}_{-0.2}$ dex at the farthest reaches of our coverage. The RGB width is found to increase rapidly from $0.4^{+0.1}_{-0.1}$ dex to $1.1^{+0.2}_{-0.1}$ dex in the inner portion of the stream before plateauing and decreasing marginally in the outer subfields of the stream. In addition, we estimate Stream C to lie at a distance between $794$ and $862$ kpc and Stream D between $758$ kpc and $868$ kpc. We estimate the median metallicity of Stream C to lie in the range $-0.7$ to $-1.6$ dex and a metallicity of $-1.1^{+0.3}_{-0.2}$ dex for Stream D. RGB widths for the two structures are estimated to lie in the range $0.4$ to $1.2$ dex and $0.3$ to $0.7$ dex respectively. In total, measurements are obtained for 19 subfields along the Giant Stellar Stream, 4 along Stream C, 5 along Stream D and 3 general M31 spheroid fields for comparison. We thus provide a higher resolution coverage of the structures in these parameters than has previously been available in the literature. 
\end{abstract}


\section{Introduction}

The Giant Stellar Stream (GSS - also known as the Giant Southern Stream) constitutes a major substructure in the halo of our neighbor galaxy M31. It was discovered in 2001 from a survey of the southeastern inner halo of M31 undertaken with the Wide Field Camera on the 2.5m Isaac Newton Telescope (\citealt{Ibata2001}; \citealt{Ferguson2002}). Followup observations with the 3.6m Canada-France-Hawaii Telescope (CFHT) further revealed the enormous extent of the stream, spanning at least $4^\circ$ of sky (\citealt{McConn2003}; \citealt{Ibata2007}). This corresponds to a projected size in excess of $50$ kpc at M31 halo distances. A high-density stellar stream of these proportions is a structure seldom seen in the Local Group and its importance for understanding the evolution of the M31 system cannot be overestimated.     

The GSS has proven to exhibit a complex morphology, with a wide spread in metallicities and evidence for more than one stellar population. Based on stellar isochrone fitting, \citet{Ibata2007} found evidence for a more metal rich core, surrounded by a sheath of bluer metal poor stars, which combine to produce a luminosity of $1.5 \times 10^8$ $L_{\odot}$ (a total absolute magnitude of $M_V \approx -15.6$). Similarly, studies such as \citet{Kalirai2006} and later \citet{Gilbert2009} find two kinematically separated populations in several inner fields of the stream, using data obtained with the DEIMOS spectrograph on the 10m Keck II telescope. \citet{Gilbert2009} again report a more metal poor envelope enclosing the core. \citet{Guhat2006} use data from the same source to deduce a mean metallicity of $[Fe/H] = -0.51$ toward the far end of the stream, suggesting the GSS is slightly more metal rich than the surrounding halo stars in this region. Using deep photometry obtained of an inner stream field via the Hubble Space Telescope's Advanced Camera for Surveys, \citet{Brown2006} compare their data with isochrone grids to ascertain a mean age of $\sim 8.8$ Gyr and a mean metallicity of $[Fe/H] = -0.7$ (slightly more metal poor than the spheroid population studied) but note a large spread in both parameters. Further to this, \citet{Bernard2015} have shown that star formation in the stream started early and quenched about 5 Gyr ago, by which time the metallicity of the stream progenitor had already reached Solar levels. On the basis of this, they propose an early type system as the stream progenitor, perhaps a dE or spiral bulge. Detailed age and metallicity distributions are also included in this contribution.  

By combining distance estimates for the stream, particularly those presented in \citet{McConn2003}, with kinematic data, it is possible to constrain the orbit of the stream progenitor, and also to measure the dark matter halo potential within the orbit. Numerous studies have been dedicated to these aims, such as that of \citet{Font2006} which uses the results of \citet{Guhat2006} to infer a highly elliptical orbit for the progenitor, viewed close to edge-on. Both \citet{Ibata2004} and, more recently, \citet{Fardal2013} have obtained mass estimates for M31 using the GSS, with the latter incorporating a mass estimate for the progenitor comparable to the mass of the Large Magellanic Cloud.   

Whilst the distance information presented in \citet{McConn2003} has been of great benefit to past studies, a more extensive data set, namely the Pan-Andromeda Archaeological Survey (PAndAS - \citealt{McConn2009}) is now available. This data set provides comprehensive coverage along the full extent of the GSS, as well as other structures in the vicinity, notably Stream C and Stream D \citep{Ibata2007}. Stream C is determined in that study to be a little brighter and substantially more metal rich than Stream D. Both streams exhibit distinct properties to the GSS and hence must be considered separate structures, despite their apparent intersection with the GSS on the sky. Given the reliance of the aforementioned orbital studies on high quality distance and metallicity information, and given the prominent role played by stellar streams as diagnostic tools within the paradigm of hierarchical galaxy formation, it is highly advantageous to further constrain the distance and metallicity as a function of position along the stream using these data. The following sections hence outline the results of a new tip of the red giant branch (TRGB) algorithm as applied to subfields lining the GSS and streams C and D. In section \S 2, we provide a description of this method, in \S 3 we present the results of this study and in \S 4 and \S 5 we conclude with a discussion and summary respectively. Note that this publication forms part of a series focusing on key substructure identified in the M31 outer halo. This series includes \citet{Bate2014}, \citet{Mackey2014} and \citet{McMonigal2016}.    


\begin{figure}
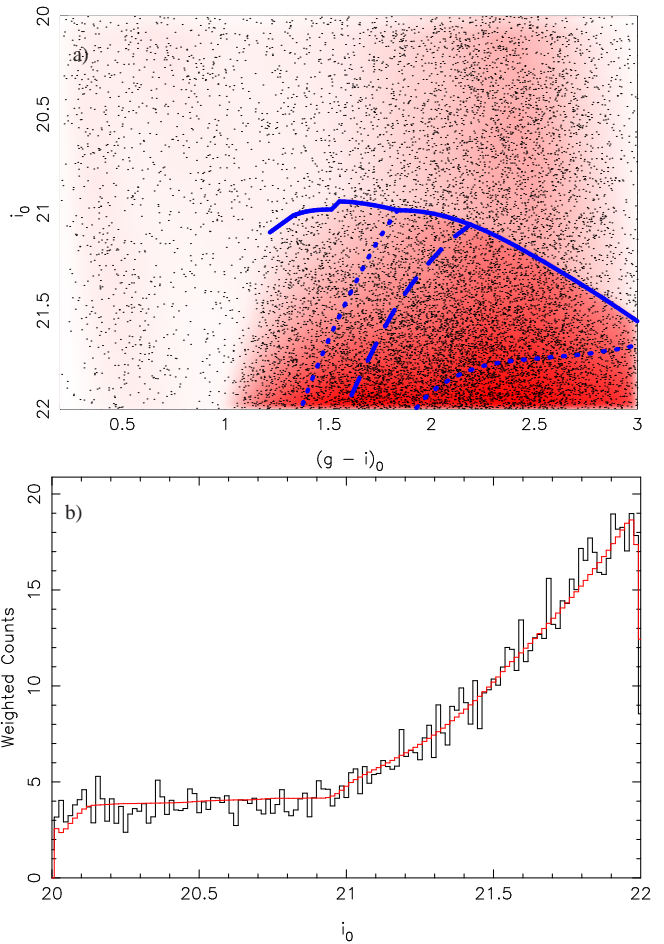

\begin{center} 
$ \begin{array}{c} 
\begin{overpic}[width = 0.35\textwidth,angle=-90]{GSS3_Best_Fit_CMD.ps}
\put(10,65){\small a)}
\end{overpic}
\\ \begin{overpic}[width = 0.35\textwidth,angle=-90]{GSS3_Best_Fit_LF_coarse_W.ps}
\put(10,65){\small b)}
\end{overpic}
\end{array}$
\end{center}
\caption{Model fits to the 2D Color-Magnitude Diagram (a) and 1D Luminosity Function (b) of a Giant Stellar Stream subfield (GSS3 - see Fig \ref{FieldMaps}). In panel (a), the model height (density) as a function of magnitude $i_0$ and color $(g-i)_0$ is indicated by the shade of red at that location. An isochrone representing the best-fit central metallicity of the data is shown as a blue dashed line. The blue dotted lines on either side are representative of the Gaussian $1\sigma$ spread in isochrone metallicities (the RGB width) - they \emph{do not} represent the uncertainty in the best-fit metallicity value. The solid blue line denotes the magnitude of the TRGB as a function of color, given the best-fit distance to that segment of the GSS returned by the algorithm. Panel (b) shows the one-dimensional model fit to the luminosity function, plotted by marginalizing over the color parameter (i.e. collapsing the \emph{x}-axis) in the CMD model fitted in panel (a). }
\label{Fit2GSS3}

\end{figure}


\begin{figure}
\begin{center}
\includegraphics[width = 0.35\textwidth,angle=-90]{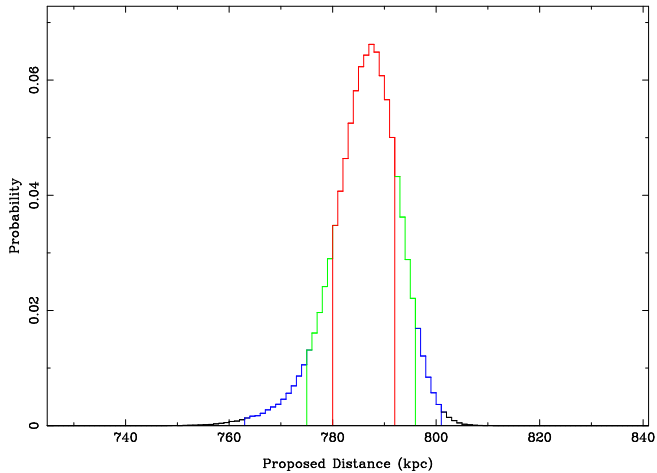}				  
\caption{Probability distribution function (PDF) in the distance to subfield GSS3 of the GSS. Red, green and blue segments of the distribution denote Gaussian $1\sigma$ (68.2 \%), 90\% and 99\% credibility intervals respectively.}
\label{GSS3_distance_pdf}
\end{center}

\end{figure}   


\section[]{A New Two-Dimensional TRGB algorithm}
\label{method}
Obtaining distances at closely spaced intervals along the Giant Stellar Stream has proven quite challenging, owing largely to the contrast of the stream with respect to the surrounding M31 halo stars, and also due to the wide spread in metallicities. Whilst the TRGB method presented in \citet{Conn2011} and \citet{Conn2012} provided the basis for the method we employ here, that method has its niche in application to metal poor populations with a low spread in metallicities. Hence for the GSS, a significant adaptation was necessary, as now discussed.  

In the earlier method, the luminosity function of the object in question was modeled using a truncated power law to represent the contribution from the object's red giant branch (RGB), as per Eq. \ref{eq1}:

\begin{equation}
	 \begin{split}
	\; L (m \ge m_{TRGB}) &= 10^{a(m - m_{TRGB})}  \\
	\; L (m < m_{TRGB}) &= 0 
	 \end{split}
	 \label{eq1}
\end{equation}
where $L$ represents the probability of finding a star at a given magnitude, $m$ is the (CFHT) $i$-band magnitude of the star in question, $m_{TRGB}$ is the TRGB magnitude and $a$ is the slope of the power law. To this power law was then added a polynomial fit to the luminosity function of a nearby field chosen to represent the contamination from non-object stars in the object field. This contamination component was then scaled relative to the object RGB component based on a comparison of the stellar density between the object and contamination fields. As this method is solely concerned with the $i$-band magnitude of a star, and does not take into account its color information, it is effectively a one-dimensional method in two-dimensional color-magnitude space. This means that the only metallicity information incorporated into the fit is that from the color-cut imposed on the stars beforehand. The dependence of the CFHT $i$-band TRGB magnitude on metallicity becomes an important consideration however for metallicities greater than $-1$ (see for example Fig. 6 of \citealt{Bellazzini2008} for the SDSS $i$-band which is comparable). For this reason, we have developed a two-dimensional approach to identifying the TRGB, one that incorporates a star's position in both color and magnitude space into the fitted model. 

For our two-dimensional model of the object RGB, we draw our basis from the isochrones provided in the Dartmouth Stellar Evolution Database \citep{Dotter2008}. Therein are provided the necessary theoretical isochrones for the CFHT $i$-band and $g$-band photometry provided by the PAndAS survey. Within this database, isochrones are provided for a range of ages ($1 \le age \le 15$ Gyr), metallicities ($-2.5 \le [Fe/H] \le 0.5$), helium abundances $y$ and alpha-enhancement $[\alpha/Fe]$ values. For use with our algorithm, we have generated a large set of $2257$ isochrones in CFHT $i$ vs $g-i$ space with $[Fe/H] = -2.50, -2.45, ..., 0.50$ for each of $age = 1.00, 1.25, ..., 5.00$ Gyr where $age \le 5$ Gyr and $age = 5.5, 6.0, ..., 15.0$ Gyr where $age > 5$ Gyr. All isochrones are generated with $y = 0.245 + 1.5 z$ and $[\alpha/Fe] = 0.00$. The model RGB can then be constructed via an interpolation of the isochrone grid corresponding to a given age.         

Using the set of Dartmouth isochrones as generated for any given age, we essentially have a field of points in 2D (i.e. those corresponding to the color and magnitude of a particular mass value within a given isochrone) which form the framework of our model. Each of these points can then be scaled relative to each other point, thus adding a third dimension which represents the model height or density at that location in the CMD. This model height can then be manipulated by a Markov Chain Monte Carlo (MCMC) algorithm by altering a number of parameters, as outlined below. The model surface in between the resulting points is then interpolated by taking adjacent sets of 3 points and fitting a triangular plane segment between them.  

In order to manipulate the model height at each point in color-magnitude space, 3 parameters are implemented. The first is the slope of a power law \emph{a} applied as a function of \emph{i}-band magnitude, as per Eq \ref{eq1}. The second and third denote the centre and width of a Gaussian weighting distribution applied as a function of metallicity (i.e. a function of both colour and magnitude). The slope parameter \emph{a} is a convenient, if crude measure for accounting for the increase in the stellar population as you move faint-ward from the TRGB. Significant time was invested in an effort to devise a more sophisticated approach taking into account the specific tracks of the isochrones, but the simplest approach of applying the slope directly as a function of \emph{i}-band magnitude remained the most effective and hence was used for all fits presented in this contribution. The Gaussian distribution applied as a function of metallicity is used to weight each isochrone based on the number of object stars lying along that isochrone. Each isochrone is hence given some constant height along all its constituent masses, with the slope parameter being used to discriminate between model heights within a single isochrone. The isochrones are weighted as follows:    
\begin{equation}
	W_{iso} = exp\left(- \frac{([Fe/H]_{iso} - [Fe/H]_0)^2}{2 \times w_{RGB}^2} \right)
	 \label{eq2}
\end{equation}
where $W_{iso}$ is the weight applied to isochrone $iso$, $[Fe/H]_0$ is the central metallicity of the population, $[Fe/H]_{iso}$ is the metallicity of the isochrone being weighted, and $w_{RGB}$ is the one sigma spread in the metallicity of the isochrones, which we shall refer to as the RGB width. We note that the metallicity distribution function can be far from Gaussian, but nevertheless hold that this simplified model is both efficient and adequate in its simplicity. In particular, the distribution for the general M31 spheroid is far from Gaussian and hence this component is essentially folded into the normalization of the field contamination. Our fitted streams are in contrast represented by far more Gaussian distribution functions, and hence are fitted as the signal component by our algorithm.  

With the model CMD for the object constructed in the aforementioned fashion, we now require the addition of a contamination model component. Here we use the PAndAS contamination models as provided in \citet{Martin2013}. Essentially they provide a measure of the intensity of the integrated Milky Way contamination in any given pixel in the PAndAS survey. Likewise, they allow the user to generate a model contamination CMD for any pixel in the survey. Whilst it is possible to derive a measure of the object-to-contamination ratio directly from these models, we find that given the low contrast in many of the GSS subfields, it is preferable to fit this ratio as a free parameter determined by the MCMC process. 

To generate our MCMC chains, we employ the Metropolis-Hastings algorithm. In summary, we determine the likelihood ${\cal L}_{proposed}$ of the model for a given set of parameters and compare with the likelihood of the most recent set of parameters in the chain ${\cal L}_{current}$. We then calculate the Metropolis Ratio $r$:

\begin{equation}
	r = \frac{{\cal L}_{proposed}}{{\cal L}_{current}}
	\label{e_MetroRat}
\end{equation}
and accept the proposed parameter set as the next in the chain if a new, uniform random deviate drawn from the interval $[0,1]$ is less than or equal to $r$. In order to step through the parameter space, we choose a fixed step size for each parameter that is large enough to traverse the whole probability space yet small enough to sample small features at a suitably high resolution. The new parameters are drawn from Gaussian distributions centered on the most recent accepted values in the chain, and with their width set equal to the step size. Upon the completion of the MCMC run, the chains are then inspected to insure that they are well mixed.  

Thus, we now have everything we need for our model CMD. At each iteration of the MCMC, we generate a model of the GSS red giant branch by using a grid of isochrones and manipulating their relative strengths using free parameters representing the central metallicity and RGB width of the stellar population combined with a parameter representing the slope in density as a function of \emph{i}-band magnitude. We then slide this model component over the top of the contamination model component, with their respective ratio set via a fourth free parameter. We restrict the fitted magnitude range to $20 \le i \le 22$ to provide adequate coverage of the range of distances we expect to encounter whilst retaining a relatively narrow, more easily simulated band across the CMD.  

The final fitted parameter then is the TRGB magnitude itself, which determines how far along the \emph{i}-band axis to slide the isochrone grid from it's default position at 10 pc (i.e. the isochrones are initially set to their absolute \emph{i}-band magnitudes). Thus it is actually the distance modulus of the population that we measure directly, since there is no fixed TRGB magnitude, but rather it is variable in color as exemplified in Fig. \ref{Fit2GSS3}. For the sake of presenting a specific TRGB magnitude (as all TRGB investigations traditionally have done), we define a reference TRGB apparent magnitude ($m_{TRGB}$), derived from the distance modulus assuming a fixed absolute magnitude of the TRGB ($M_{TRGB}$) of $i = -3.44$. This is a good approximation to the roughly constant value of $M_{TRGB}$ for intermediate to old, metal poor populations for which the TRGB standard candle has traditionally been used ($[Fe/H] \le -1$, see Fig. 6 of \citealt{Bellazzini2008}) and allows for direct comparison with other publications in this series. Clearly for the present study we are fitting populations that are often more metal rich than this, but it must be stressed that this adopted value is purely cosmetic with no bearing on the derived distance or any other determined parameter. 

The age of the isochrone grid is fixed at an appropriate value determined from the literature (9 Gyr in the case of the GSS, 9.5 Gyr for streams C and D and general spheroid fields and 7.5 Gyr for the M31 disk - all rounded from the values given in \citealt{Brown2006}). Initial tests of the algorithm with the population age added as a sixth free MCMC parameter revealed that the choice of age had no effect on the location of the parameter probability peaks returned by the MCMC, but only on their relative strengths. It was hence decided more efficient to fix the age at a suitable value for the target population, as determined from the literature. 

As an additional consideration, the model RGB is further convolved with a 2D Gaussian kernel to simulate the blurring effects of the photometric uncertainties. We assume a photometric uncertainty of 0.015 magnitudes for both $i$ and $g$ bands and set the dimensions of the Gaussian kernel accordingly. We note that whilst in the fitted range the photometric uncertainty lies in the range 0.005 to 0.025, the tip will generally be located in the range $20.5 \le i \le 21.5$ for the structures studied in this contribution, making the assumed uncertainty value the most suitable. Any issues of photometric blending must be resolved by excising any regions above some suitable density threshold, although such issues have only been observed at the centers of the densest structures in the PAndAS survey and were not an issue for this study. Similarly, care must be taken to insure that data incompleteness does not effect the fitted sample of stars, which was achieved in the present study by restricting the magnitude range of selected stars. 

Finally, at the conclusion of the MCMC run, a probability distribution function (PDF) in each free parameter is obtained by marginalizing over the other parameters. As an example, the distance PDF for the GSS3 subfield, which was obtained via sampling from the PDF in the reference TRGB magnitude, is presented in Fig. \ref{GSS3_distance_pdf}. The distance probability distribution is derived from that in the reference TRGB magnitude using the following equation:
\begin{equation}
	D = 10^ \frac{5 + m_{TRGB} - m_{ext} - M_{TRGB}}{5} 
	 \label{eq3}
\end{equation}
where $D$ is the distance in parsecs; $m_{TRGB}$ is the reference TRGB apparent magnitude, sampled from the PDF in this parameter produced by the MCMC; $m_{ext}$ is the extinction in magnitudes for the center of the field, as sampled from a Gaussian with a central value determined from the Schlegel extinction maps \citep{Schlegel1998} and a width equal to 10 \% of the central value; and $M_{TRGB}$ is the absolute magnitude of the TRGB. The uncertainty in $M_{TRGB}$ is a systematic quantity and we thus omit it from our calculations since we are primarily concerned in relative distances between subfields as opposed to absolute distances from Earth. We hence ignore any uncertainty in the absolute magnitude of the tip and note that all distances will have a systematic offset of not more than $50$ kpc (assuming an uncertainty of approximately $0.1$ magnitudes). All MCMC runs used for the results presented in this contribution were of $200,000$ iterations whilst the distance distributions are generated using $500,000$ samples of the $m_{TRGB}$ and $m_{ext}$ distributions.    

In conjunction with the results we present in the following section, we also provide an appendix to inform the interested reader as to any degeneracy between the key parameters of tip magnitude, metallicity and the RGB width. In Appendix A, we present contour plots illustrating the covariance between the tip magnitude and the metallicity for the GSS and Streams C and D. In Appendix B we present similar plots for the covariance between metallicity and RGB width for the same structures. In Appendix C we present both types of plot for our halo comparison fields which shall be referred to in the next section. It can be seen from these plots that any covariance between parameters is only minor. These plots are also extremely useful for visualizing the true probability space of the key parameters for each field, and provide an informative compliment to the results plotted in Figs. \ref{GSSDist} through \ref{Stream_D_parameters}. 


\begin{figure}
\begin{center} 
\begin{overpic}[width = 0.45\textwidth,angle=-0]{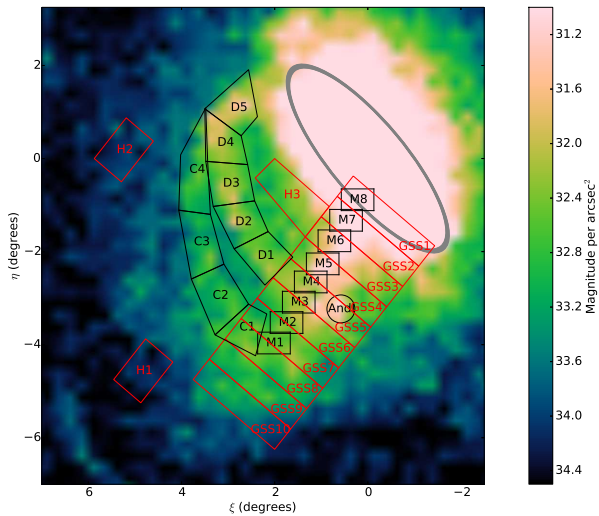}
\end{overpic}
\end{center}
\caption{Field placements for the Giant Stellar Stream, Streams C and D and all other fields pertinent to this study. Subfields GSS1-GSS10 (in red) cover the extent of the GSS, with fields M1 - M8 representing the fields from the \citet{McConn2003} study. The circular And I field is the exclusion zone omitted from subfields GSS4 and GSS5 due to the presence of the dwarf spheroidal galaxy Andromeda I. Fields H1-H3 are halo fields for comparison with the stream populations. Stream C (subfields C1 - C4) and Stream D (subfields D1 - D5) are delineated in black. The annulus used for our new M31 distance measurement is shown in grey. Field locations were chosen using enhanced brightness maps of the structures generated using models presented in an upcoming contribution (Martin et al., in prep).  Essentially, a wide range of stellar populations can be isolated on a pixel-by-pixel basis using these models. To best reveal our target structures, metallicity slices centered on [Fe/H] = -1.1, [Fe/H] = -1.2 and [Fe/H] = -1.3 have been used for this image.}
\label{FieldMaps}

\end{figure}


\section{Results}
\label{results}
The results we present in this section pertain to a number of separate structures. A field map illustrating the GSS subfields and Andromeda I exclusion zone as well as the fields utilized by \citet{McConn2003}, is presented in Fig. \ref{FieldMaps}. The subfield placements along Stream C and Stream D are also indicated in this figure. Our principal focus is the Giant Stellar Stream, which is contained within our field labeled `GSS'. Fields C and D enclose Streams C and D respectively; and Fields H1 through H3 are separate halo fields adjacent to our target fields which sample the general M31 spheroid for comparison purposes. 

\begin{figure*} 
\begin{center}
\includegraphics[width = 0.75\textwidth]{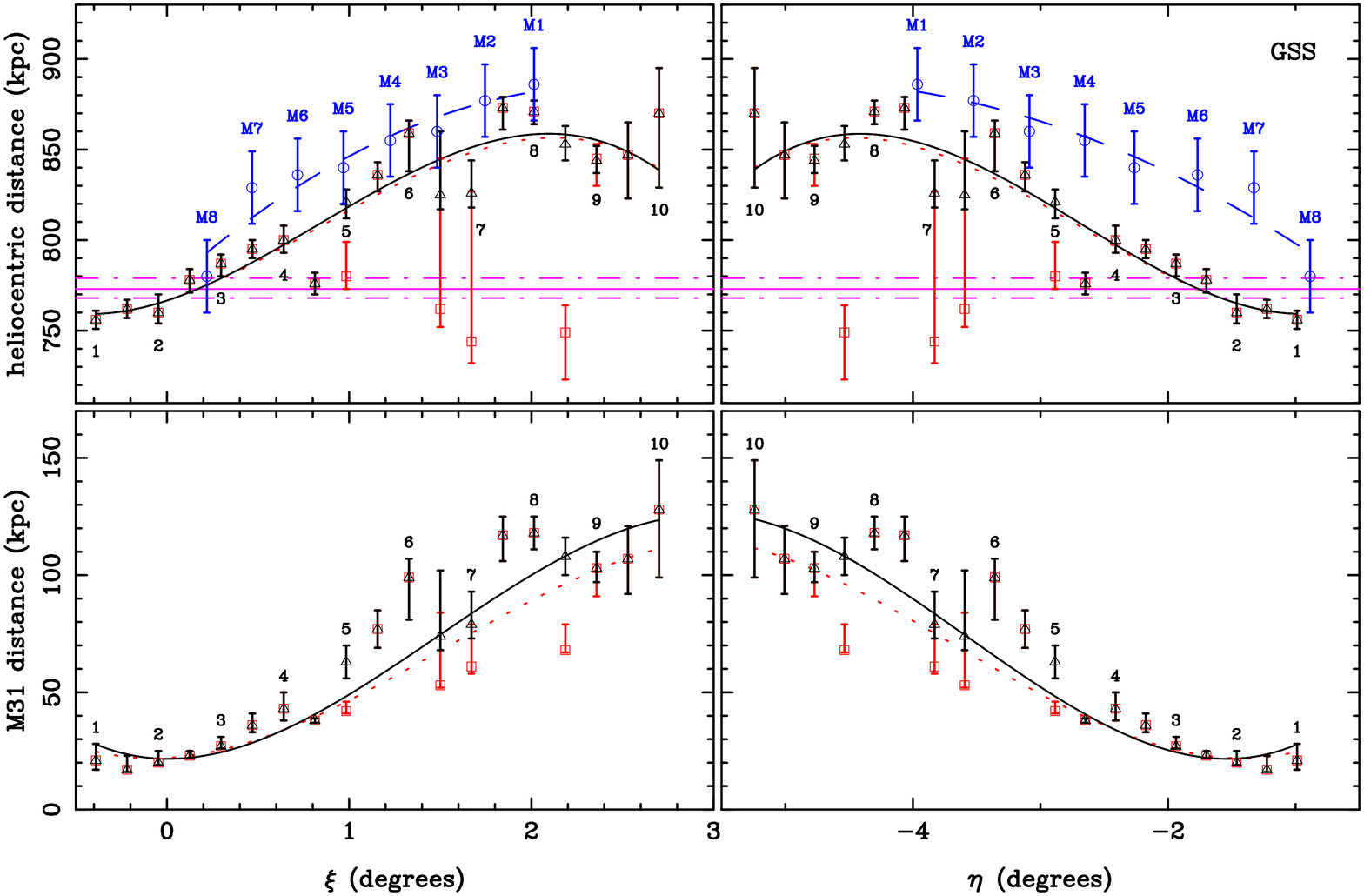}
\caption{Heliocentric distances of the GSS subfields and their distances from M31. Distances are plotted as a function of both $\xi$ and $\eta$. The heliocentric distance of M31 and its associated uncertainties are represented by \emph{solid} and \emph{dashed} horizontal purple lines respectively. Black triangle symbols and error bars denote our best parameter estimates derived via our new method. Square symbols indicate the most likely parameter values as determined from our unrestricted probability distributions. These measurements are shown in red in order to distinguish them from our preferred alternative measurements, derived by restricting the probability distribution function, where appropriate, to the most likely of the multiple peaks present. The results from these restricted-range distributions (triangle symbols) correspond directly to the square symbols where no restriction of the distribution was imposed. Blue circles and error bars represent the heliocentric distance measurements presented in \citet{McConn2003}.}
\label{GSSDist}
\end{center}

\end{figure*}

\begin{figure*} 
\begin{center}
\includegraphics[width = 0.75\textwidth]{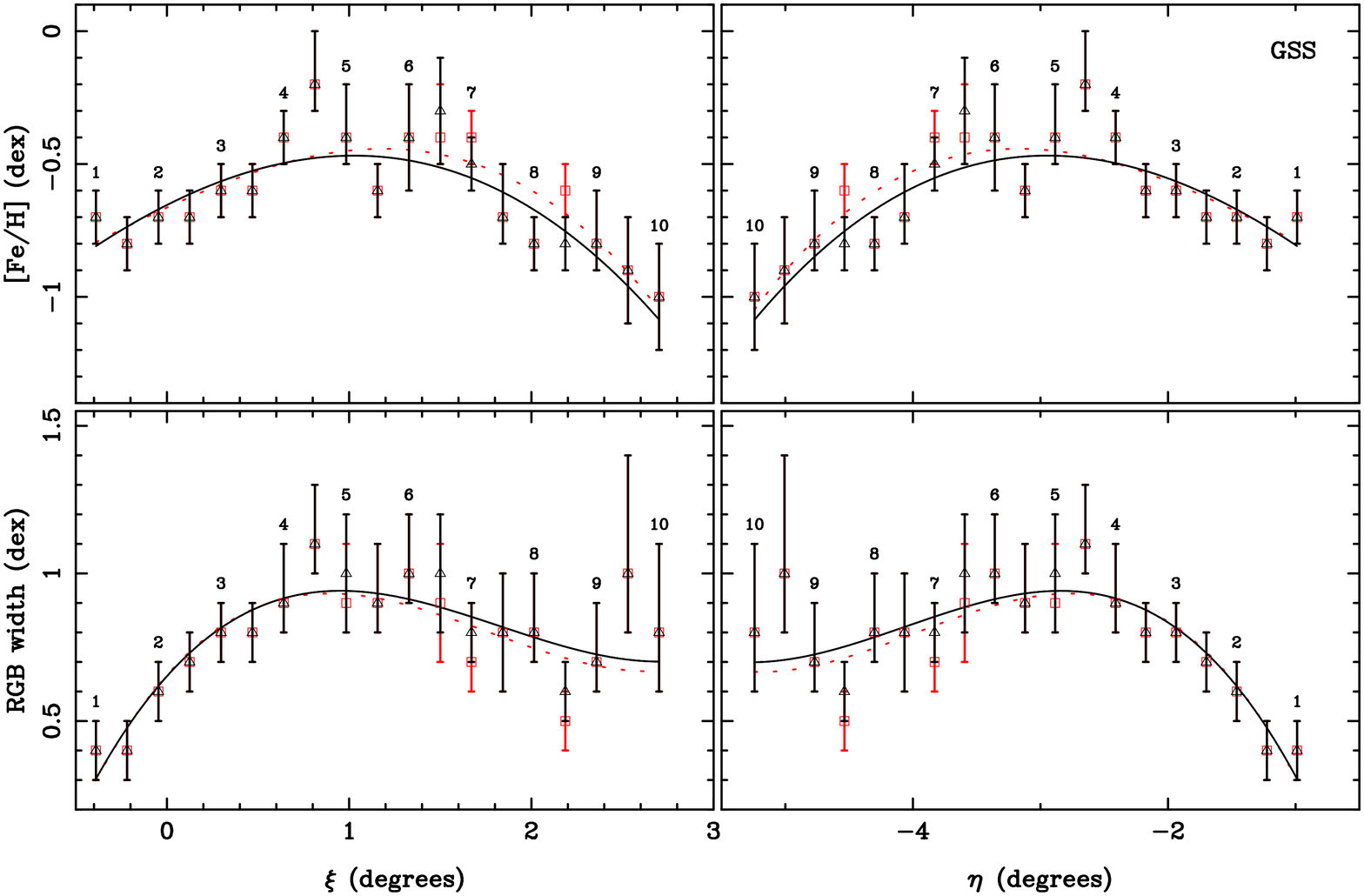}
\caption{Metallicity and RGB width as a function of both $\xi$ and $\eta$ for the GSS subfields. The symbols used are the same as for Fig. \ref{GSSDist}.}
\label{GSSMet}
\end{center}

\end{figure*}

\begin{figure*} 
\begin{center}
\includegraphics[width = 0.70\textwidth]{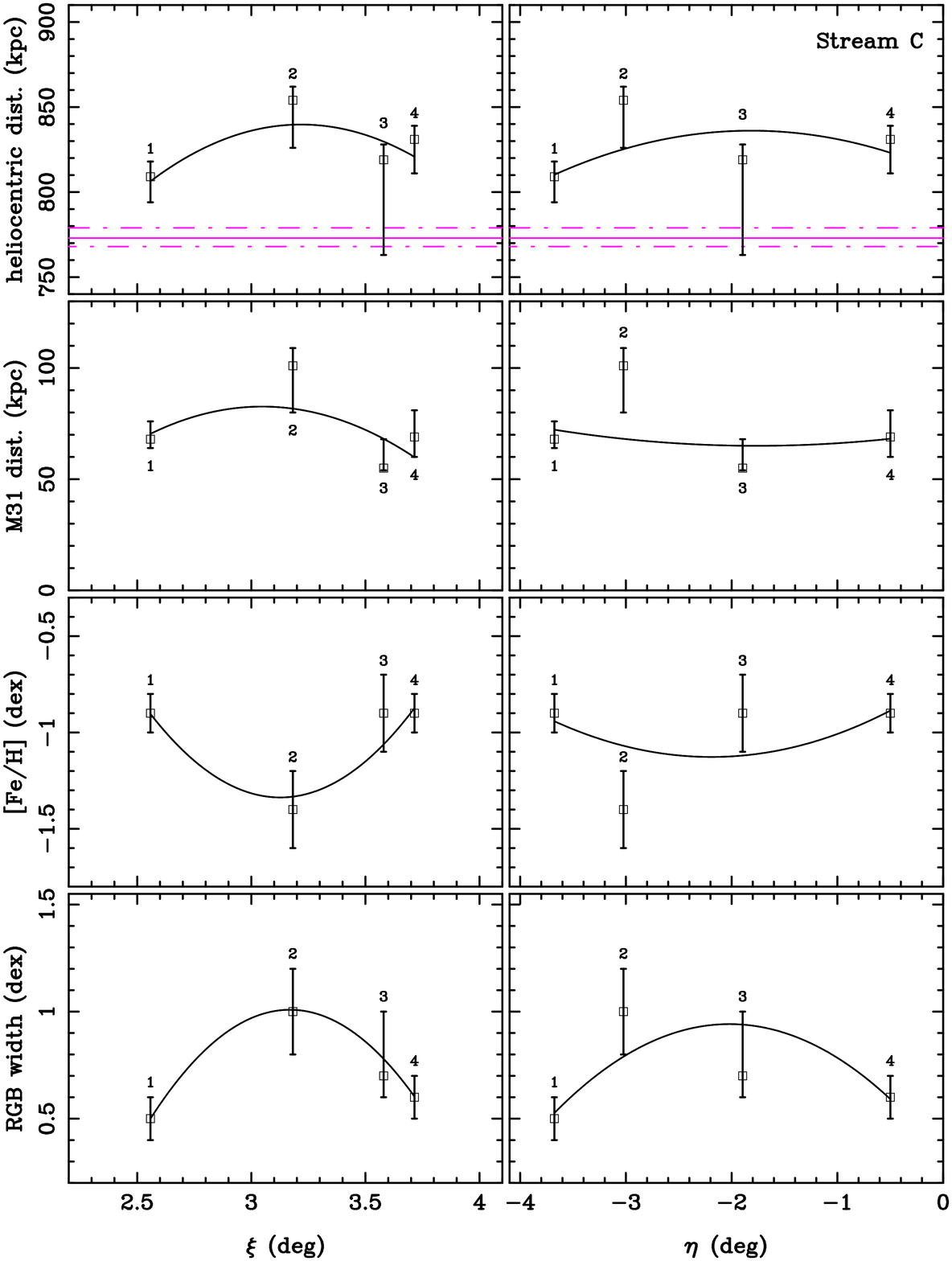}
\caption{Heliocentric distance, M31 distance, metallicity and RGB width as a function of both $\xi$ and $\eta$ for the Stream C subfields. \emph{Solid} and \emph{dashed} horizontal purple lines in the top two panels denote the heliocentric distance to M31 and its associated uncertainties respectively.}
\label{Stream_C_parameters}
\end{center}

\end{figure*}

\begin{figure*} 
\begin{center}
\includegraphics[width = 0.70\textwidth]{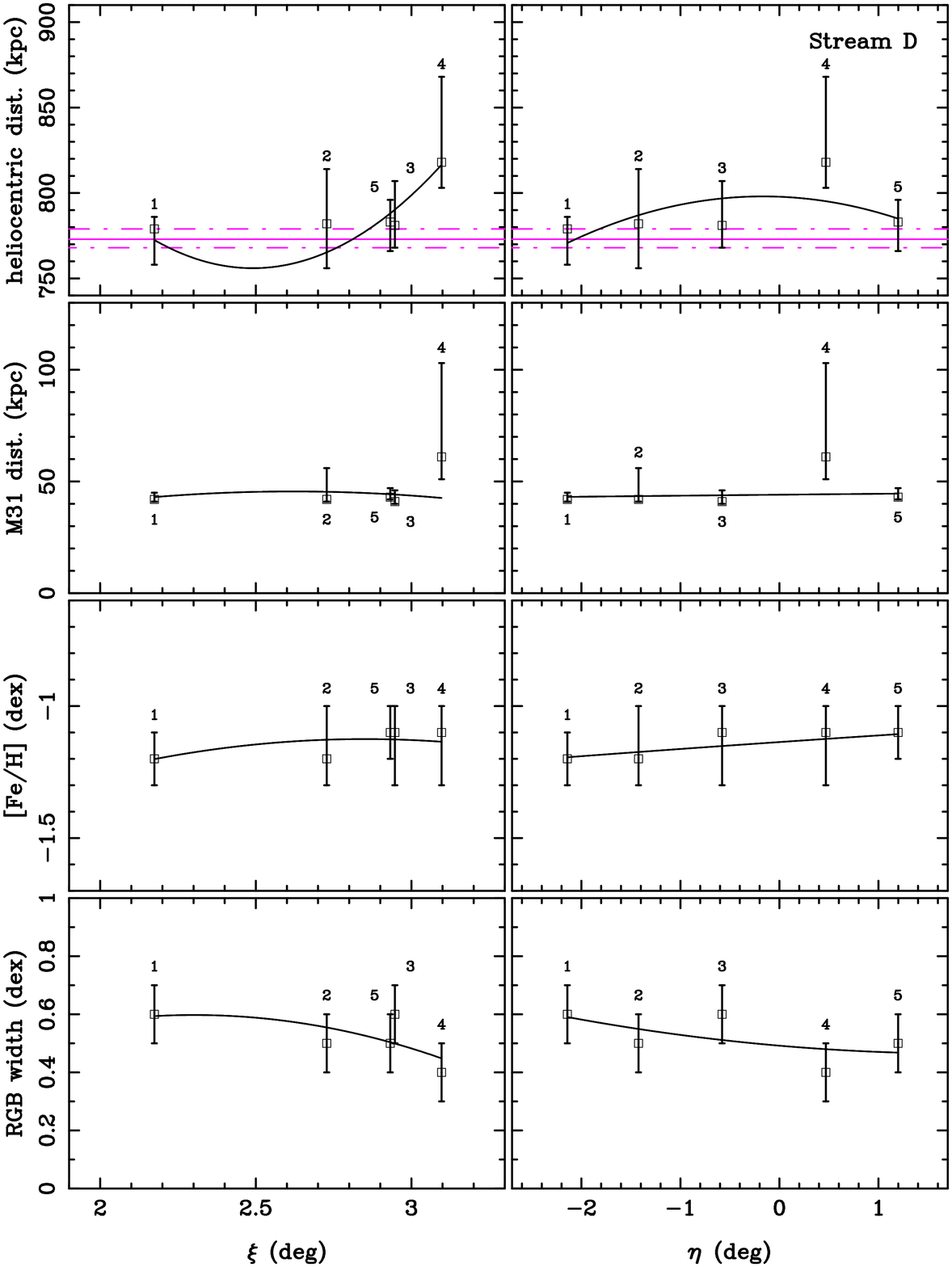}
\caption{Heliocentric distance, M31 distance, metallicity and RGB width as a function of both $\xi$ and $\eta$ for the Stream D subfields. \emph{Solid} and \emph{dashed} horizontal purple lines in the top two panels denote the heliocentric distance to M31 and its associated uncertainties respectively.}
\label{Stream_D_parameters}
\end{center}

\end{figure*}

As discussed in \S \ref{method}, for each subfield we obtain estimates of the heliocentric distance, the metallicity [$Fe/H$] and the RGB width ($w_{RGB}$), as well as the contamination fraction from Milky Way stars ($f_{cont}$). These are quantified in Tables \ref{par_table1} and \ref{par_table2}, as are the distance modulus, extinction ($E(B-V)$) and M31 distance for each subfield. Distances along the GSS (both heliocentric and M31-centric) are plotted as a function of their M31-centric tangent plane coordinates $\xi$ and $\eta$ in Figure \ref{GSSDist}. Metallicities and RGB widths for the GSS are plotted as a function of $\xi$ and $\eta$ in Figure \ref{GSSMet}. Figures \ref{Stream_C_parameters} and \ref{Stream_D_parameters} present the distances (heliocentric and M31-centric), metallicities and RGB widths for Stream C and Stream D respectively. All data points are plotted together with their one-sigma ($68.2 \%$) uncertainties. Note that for the GSS, an overlapping system of fields was implemented such that a given field GSS$X.5$ contains the stars from the lower half of field GSS$X$ and the upper half of field GSS$X+1$. For this reason, data points are shown in between the numbered fields in Figure \ref{GSSDist} and Figure \ref{GSSMet}. In each of the Figures \ref{GSSDist} through \ref{Stream_D_parameters}, basis splines are over plotted on each structure to aid the eye - they are not intended as a fit to the data. The splines are simply a smoothing function weighted by the errors in each data point - they are not constrained to pass through any specific data point. Each combination of parameters is smoothed separately and smoothing does not take into account the full three dimensions ($\xi$, $\eta$, $<$parameter$>$). Cubic splines are used for our GSS measurements whilst quadratic splines are used for all other measurements.

For the derivation of the M31 distance for each subfield, a new distance to M31 of $773^{+6}_{-5}$ kpc was determined via our new method, by fitting to stars within an elliptical annulus centered on M31 and defined by inner and outer ellipses with ellipticities of $0.68$, position angles of $39.8^{\circ}$ and semi-major axes of $2.45^{\circ}$ and $2.55^{\circ}$ respectively (as indicated in Fig. \ref{FieldMaps}). This distance is a little smaller than the $779^{+19}_{-18}$ kpc determined by the 1D predecessor of our current method \citep{Conn2012} and larger than the $752 \pm 27$ kpc determined from Cepheid Variables \citep{Riess2012} or the $744 \pm 33$ kpc determined from eclipsing binaries \citep{Vilardell2010} but nevertheless well within the uncertainties of each of these measurements.      

It is immediately clear, both from the large error bars in Figures \ref{GSSDist} through \ref{Stream_D_parameters} and in particular from the last column ($f_{cont}$) of Tables \ref{par_table1} and \ref{par_table2}, that our parameter estimates for most subfields are derived from heavily contaminated structures. Nevertheless, on closer inspection, much can be inferred from the estimates returned by our algorithm. Our results support the same general distance gradient reported by \citet{McConn2003}, as can be seen in Fig. \ref{GSSDist}, although we note a slightly greater increase in distance as a function of angular separation from M31.  We also find no evidence of the sudden distance increase between fields $7$ and $8$ of that study, and importantly, we note that the stream appears to emerge from a small distance in front of the M31 disk center. It should be noted that the results reported in \citet{McConn2003} determine distance shifts of each field with respect to field $8$ - taken as the M31 distance - whereas our estimates are independent of any inter-field correlations. Our data is also the product of a different imager to that used in this earlier study and of a different photometric calibration. We also stress that the technique used in the earlier contribution did not take metallicity changes into account on a field-by-field basis. The mid GSS fields are in fact slightly more metal rich than the inner most fields (see Fig. \ref{GSSMet}) which would yield inflated distance estimates for those fields.

 It is evident from Fig. \ref{GSSDist} that our distance estimates appear to depart markedly from the general trend between subfields GSS4 and GSS5.5 as well as between GSS6 and GSS7.5. These subfields coincide with the intersection (on the sky) between the GSS and streams D and C respectively. With the exception of subfield GSS4.5, each of these anomalous subfields contain parameter probability distributions that are double peaked, with the second peak more in keeping with the GSS trend and thus presumably attributable to the GSS. In the case of subfield GSS5, Stream D would appear to be consistent with the additional peak in so far as distance is concerned, but the same cannot be said for either the metallicity or the RGB width. In the case of subfields GSS6.5 and GSS7, the additional peak is roughly consistent with the secondary peak derived for subfield C3 in terms of distance and RGB width but the metallicity is different. For all fields where a restriction on the TRGB probability distribution proved informative (namely subfields GSS5, GSS6.5, GSS7, GSS8.5 and GSS9), parameter estimates are provided for both the restricted and unrestricted case. The fields are denoted in the restricted case with the symbol ${\dagger}$$^*$ in Table \ref{par_table1} and in Appendix A and Appendix B, whilst ${\dagger}$ is used in the unrestricted case. Fields denoted ${\dagger}$$^*$ will be represented as black triangle symbols in Figs \ref{GSSDist} and \ref{GSSMet} whilst those denoted ${\dagger}$ will be represented as red square symbols. We note that even when the GSS subfield distances are determined from the full parameter distributions, they remain in general keeping with the trend when the full uncertainties are considered. 
 
 Moving on to the outer most portion of the GSS, it is interesting to observe that the distance seems to plateau and even diminish beyond the brightest portion of the stream covered in \citet{McConn2003}, although caution must be exercised with inferences made from the outermost subfields, due to the extremely low signal available. 

For streams C and D, we find average distances of $\sim 828^{+9}_{-30}$ kpc and $\sim 789^{+26}_{-18}$ kpc respectively. We are unable to determine any reliable distance gradient along either of these structures. In addition to Streams C and D, consideration had been given to the possibility of an arching segment of the GSS, extending outward from subfields GSS8, GSS9 and GSS10 and falling back onto the M31 disk in the vicinity of subfields C4 and D4/ D5. Despite the conceivable existence of such a feature based on visual inspection of stellar density plots, no distinct population could be reliably determined in any of the fitted parameters. If such a continuation of the GSS exists, it is heavily contaminated by the much brighter Stream C and Stream D and beyond the reach of our method in its present form.  

When we examine the metallicity and RGB width estimates returned by our algorithm (see Figure \ref{GSSMet}), we observe an unusual trend as we move out along the main part of the GSS. Closest to the M31 disk, the stream is found to be moderately metal poor, with metallicities in the range $-0.7 > [Fe/H] > -0.8$ whilst midway along the stream we find more metal rich stars with $[Fe/H] > -0.5$. Then, as we move out still further, the metallicity diminishes again, falling below the levels in the inner part of the stream with $[Fe/H] \approx -1$ at the furthest reaches in subfield GSS10. A similar trend is observed for the RGB width. This would suggest that the range of metallicities present is relatively small in the inner part of the stream, whilst increasing significantly as we move toward the middle part of the stream. Once again, in the outer most parts of the stream, we observe a return to lower values, although not to the same degree as we observed for the metallicity. Once again, we must stress however that the contamination fraction is exceedingly high in the outermost subfields and thus the metallicity and RGB width estimates for these subfields should be treated with caution. We find streams C and D to be consistently more metal poor than the GSS, with average metallicities of $-1.0^{+0.1}_{-0.1}$ dex and $-1.1^{+0.1}_{-0.1}$ dex respectively. They are also generally less diverse in terms of the range of metallicities present. 

When we compare our halo fields to our GSS and Stream C and D fields, we find a clear indication that we are indeed picking up the signal of the intended structures. When we examine the contour plots in Appendix C, we find distributions that are markedly different from those of our target structures presented in Appendices A and B. These fields were carefully chosen to be of comparable size to our target fields, and to traverse the approximate M31 halo radii spanned by our target structures. The lack of any clear structure to fit to in fields H1 and H2 is clear from the breadth of the distributions in all parameters, whereas clearly such poor parameter constraints are not observed for any of our target fields. Likewise, we find little correlation between the location of the distribution maxima. Halo field H3 is somewhat different to fields H1 and H2 in that it is expected to be heavily contaminated by the M31 disk. More overlap in the distributions is found between the H3 field and our target fields (the Stream D subfields for instance), particularly in tip magnitude and metallicity, but  the signal-to-noise ratio is much higher for our inner fields, suggesting that any correlations are real and not merely the result of contamination. We should also note that we expect any parameter gradients across the halo to be diffuse and unsuited to our method which works most favorably with sharply defined structure boundaries along the line of sight. This is indeed exemplified by the plots in Appendix C.

\section{Discussion}
\label{discussion}

The key findings of our method lie in the spatially resolved metallicities and distances along the main inner-halo structures around M31. Our metallicity measurements are consistent with all prior published measurements. Whilst these measurements utilize data from a variety of instruments, we note that our method was not tuned to be consistent with any of these prior results. 

The initial discovery of the GSS by \citet{Ibata2001} in the Issac Newton Telescope (INT) Survey measured a metallicity of slightly higher than $[Fe/H]=-0.71$ at a position consistent with our innermost GSS subfields (GSS1 to GSS3).
Of the 16 Hubble Space Telescope (HST) WFPC2 fields analyzed by \citet{Bellazzini2003}, those overlapping our fields correspond to our innermost GSS subfields (GSS1 to GSS3), and have metallicity measurements in the range $[Fe/H]=-0.7$ to $-0.5$, with a tendency towards increasing metallicity moving South-East, in the same sense as our results.

Further out, at a location consistent with our GSS subfield GSS4, Keck DEIMOS spectra analyzed by \citet{Guhat2006} gave a higher mean metallicity measurement of $[Fe/H]=-0.51$, matching our findings.
A detailed analysis by \citet{Ibata2014} is in broad agreement with our results, with the GSS dominating the inner halo down to a metallicity of $[Fe/H]=-1.1$, the lowest metallicity we find for the GSS.

\citet{Ibata2014} also found the inner halo streams (including Streams C and D) to be dominant in the metallicity range $[Fe/H]=-1.7$ to $-1.1$, where our results for Stream D and one subfield of Stream C are situated, although there are also signs of a significant population of Stream C members in the range $[Fe/H]=-1.1$ to $-0.6$, where the bulk of our Stream C results lie.
This lends support to the suggestion by \citet{Chapman2008} that there are two, potentially completely separate populations that make up Stream C. These populations are found separable by their velocity measurements, and also by their metallicities of $[Fe/H]=-1.3$ and $-0.7$ in the aforesaid publication, which match our findings for subfield C2, and the rest of Stream C respectively. Indeed, \citet{Gilbert2009} also find evidence of two populations in Stream C, separable into a more metal rich component ($[Fe/H]_{mean} = -0.79 \pm 0.12$ dex) and metal poor component ($[Fe/H]_{mean} = -1.31 \pm 0.18$ dex).  We caution however that our detection of two populations is tentative and independent velocity measurements for our field locations are warranted if a clear distinction is to be confirmed.  
\citet{Chapman2008} additionally measured the metallicity of Stream D to be $[Fe/H]=-1.1\pm 0.3$, in good agreement with our results.

A key finding of this paper is the extraordinary extent of the GSS to the South-East, reaching a full degree further away from M31 in projection than previously measured, at a $5.5$ degree separation for subfield GSS10.

\citet{Fardal2008} was able to find a model for the GSS which sufficiently matched observations of some of the inner structures, however the low velocity dispersions, physical thickness and narrow metallicity ranges of Streams C and D found by \citet{Chapman2008} suggest that a single accretion event is unlikely to be sufficient to form both of these structures as well as the GSS. One possible scenario that might explain the difference in metallicity between Streams C and D and the main GSS structure is a spinning disk galaxy progenitor with a strong metallicity gradient following a radial plunging orbit into M31 resulting in the outer portion ending up on a counter orbit with a lower metallicity \citep{Chapman2008}. Although each new observation makes explanations such as this increasingly contrived.

None of the current simulations of this system predict or include an extension of the GSS as far out as we find it, or the existence of any arching segment to the GSS (\citealt{Fardal2008}; \citealt{Fardal2013}; \citealt{Sadoun2014}). Although the latest simulations of \citet{Fardal2013} include distances for the main GSS, which while consistent with the distances presented by \citet{McConn2003}, are also highly consistent with the distances presented here, particularly for the innermost and outermost portions of the GSS. This suggests that finding a simulation consistent with our much more restrictive distance constraints for the GSS may only require minor alterations.

Whilst our method has been very successful fitting these structures, particularly considering the high levels of contamination in this region (over 85 per cent for most subfields), it is in some instances difficult to resolve all the populations, especially for the fainter structures. Some additional information will be gleaned by running a full multi-population fit (Martin et al in prep), but to fully uncover the history of this system, we will need detailed simulations of the formation and evolution of the GSS and associated structures. These simulations should take into account realistic gas physics, combined with next generation observations including wide field kinematic surveys.



\begin{table*} 

\caption{Stream Parameters (Field A subfields).}{} 

\vspace{5 mm}

\centering      
\begin{tabular}{c c c c c c c c c c}  
\hline\hline                        
Subfield & Xi & Eta & Distance Modulus & $E(B-V)$ & Distance (kpc) & M31 Distance (kpc) & $Fe/H$ (dex) & RGB width (dex) & $f_{cont}$  \\ [0.5ex] 
\hline                    
GSS1 & $-0.390$ & $-0.988$ & $24.39^{+0.01}_{-0.01}$ & $0.076$ & $756.^{+5.}_{-5.}$ & $21.^{+7.}_{-4.}$ & $-0.7^{+0.1}_{-0.1}$ & $0.4^{+0.1}_{-0.1}$ & $0.170^{+0.003}_{-0.001}$ \\ [1.0ex]
 \hline
 GSS1.5 & $-0.219$ & $-1.225$ & $24.41^{+0.01}_{-0.01}$ & $0.073$ & $762.^{+5.}_{-5.}$ & $17.^{+6.}_{-1.}$ & $-0.8^{+0.1}_{-0.1}$ & $0.4^{+0.1}_{-0.1}$ & $0.263^{+0.004}_{-0.004}$ \\ [1.0ex]
 \hline
 GSS2 & $-0.047$ & $-1.462$ & $24.40^{+0.03}_{-0.02}$ & $0.070$ & $760.^{+10.}_{-6.}$ & $20.^{+5.}_{-1.}$ & $-0.7^{+0.1}_{-0.1}$ & $0.6^{+0.1}_{-0.1}$ & $0.416^{+0.008}_{-0.006}$ \\ [1.0ex]
 \hline
 GSS2.5 & $0.125$ & $-1.699$ & $24.45^{+0.02}_{-0.02}$ & $0.058$ & $778.^{+6.}_{-7.}$ & $23.^{+2.}_{-1.}$ & $-0.7^{+0.1}_{-0.1}$ & $0.7^{+0.1}_{-0.1}$ & $0.523^{+0.008}_{-0.008}$ \\ [1.0ex]
 \hline
 GSS3 & $0.297$ & $-1.937$ & $24.48^{+0.01}_{-0.02}$ & $0.053$ & $787.^{+5.}_{-7.}$ & $27.^{+4.}_{-1.}$ & $-0.6^{+0.1}_{-0.1}$ & $0.8^{+0.1}_{-0.1}$ & $0.572^{+0.008}_{-0.008}$ \\ [1.0ex]
 \hline
 GSS3.5 & $0.469$ & $-2.174$ & $24.50^{+0.01}_{-0.01}$ & $0.050$ & $795.^{+5.}_{-5.}$ & $36.^{+5.}_{-3.}$ & $-0.6^{+0.1}_{-0.1}$ & $0.8^{+0.1}_{-0.1}$ & $0.617^{+0.006}_{-0.008}$ \\ [1.0ex]
 \hline
 GSS4 & $0.641$ & $-2.411$ & $24.52^{+0.02}_{-0.02}$ & $0.050$ & $800.^{+8.}_{-7.}$ & $43.^{+7.}_{-5.}$ & $-0.4^{+0.1}_{-0.1}$ & $0.9^{+0.2}_{-0.1}$ & $0.628^{+0.008}_{-0.009}$ \\ [1.0ex]
 \hline
 GSS4.5 & $0.812$ & $-2.648$ & $24.45^{+0.02}_{-0.02}$ & $0.054$ & $776.^{+6.}_{-6.}$ & $38.^{+1.}_{-1.}$ & $-0.2^{+0.2}_{-0.1}$ & $1.1^{+0.2}_{-0.1}$ & $0.622^{+0.009}_{-0.009}$ \\ [1.0ex]
 \hline
  GSS5${\dagger}$$^*$ & $0.984$ & $-2.885$ & $24.57^{+0.02}_{-0.02}$ & $0.058$ & $821.^{+7.}_{-9.}$ & $63.^{+7.}_{-7.}$ & $-0.4^{+0.2}_{-0.1}$ & $1.0^{+0.1}_{-0.1}$ & $0.665^{+0.008}_{-0.009}$ \\ [1.0ex]
 \hline
 GSS5.5 & $1.156$ & $-3.121$ & $24.61^{+0.02}_{-0.02}$ & $0.057$ & $836.^{+7.}_{-9.}$ & $77.^{+8.}_{-8.}$ & $-0.6^{+0.1}_{-0.1}$ & $0.9^{+0.2}_{-0.1}$ & $0.693^{+0.008}_{-0.008}$ \\ [1.0ex]
 \hline
 GSS6 & $1.328$ & $-3.358$ & $24.67^{+0.02}_{-0.05}$ & $0.051$ & $859.^{+7.}_{-21.}$ & $99.^{+8.}_{-18.}$ & $-0.4^{+0.2}_{-0.2}$ & $1.0^{+0.2}_{-0.1}$ & $0.728^{+0.008}_{-0.009}$ \\ [1.0ex]
 \hline
 GSS6.5${\dagger}$$^*$ & $1.500$ & $-3.594$ & $24.58^{+0.09}_{-0.02}$ & $0.053$ & $825.^{+35.}_{-8.}$ & $74.^{+28.}_{-6.}$ & $-0.3^{+0.2}_{-0.2}$ & $1.0^{+0.2}_{-0.2}$ & $0.762^{+0.009}_{-0.010}$ \\ [1.0ex]
 \hline
  GSS7${\dagger}$$^*$ & $1.671$ & $-3.830$ & $24.58^{+0.05}_{-0.02}$ & $0.052$ & $826.^{+18.}_{-8.}$ & $79.^{+14.}_{-6.}$ & $-0.5^{+0.1}_{-0.1}$ & $0.8^{+0.1}_{-0.1}$ & $0.780^{+0.009}_{-0.009}$ \\ [1.0ex]
 \hline
 GSS7.5 & $1.843$ & $-4.066$ & $24.71^{+0.01}_{-0.03}$ & $0.053$ & $873.^{+6.}_{-12.}$ & $117.^{+8.}_{-11.}$ & $-0.7^{+0.2}_{-0.1}$ & $0.8^{+0.2}_{-0.2}$ & $0.812^{+0.009}_{-0.008}$ \\ [1.0ex]
 \hline
 GSS8 & $2.015$ & $-4.302$ & $24.70^{+0.01}_{-0.02}$ & $0.054$ & $871.^{+6.}_{-7.}$ & $118.^{+7.}_{-7.}$ & $-0.8^{+0.1}_{-0.1}$ & $0.8^{+0.2}_{-0.1}$ & $0.827^{+0.008}_{-0.008}$ \\ [1.0ex]
 \hline
  GSS8.5${\dagger}$$^*$ & $2.186$ & $-4.537$ & $24.65^{+0.03}_{-0.02}$ & $0.055$ & $853.^{+10.}_{-9.}$ & $108.^{+8.}_{-8.}$ & $-0.8^{+0.1}_{-0.1}$ & $0.6^{+0.1}_{-0.1}$ & $0.841^{+0.008}_{-0.009}$ \\ [1.0ex]
 \hline
  GSS9${\dagger}$$^*$ & $2.358$ & $-4.772$ & $24.63^{+0.02}_{-0.02}$ & $0.050$ & $844.^{+8.}_{-7.}$ & $103.^{+7.}_{-6.}$ & $-0.8^{+0.2}_{-0.1}$ & $0.7^{+0.2}_{-0.1}$ & $0.875^{+0.008}_{-0.009}$ \\ [1.0ex]
 \hline
 GSS9.5 & $2.530$ & $-5.007$ & $24.64^{+0.05}_{-0.06}$ & $0.047$ & $847.^{+18.}_{-24.}$ & $107.^{+14.}_{-15.}$ & $-0.9^{+0.2}_{-0.2}$ & $1.0^{+0.4}_{-0.2}$ & $0.900^{+0.008}_{-0.009}$ \\ [1.0ex]
 \hline
 GSS10 & $2.701$ & $-5.242$ & $24.70^{+0.06}_{-0.10}$ & $0.047$ & $870.^{+25.}_{-41.}$ & $128.^{+21.}_{-29.}$ & $-1.0^{+0.2}_{-0.2}$ & $0.8^{+0.3}_{-0.2}$ & $0.924^{+0.008}_{-0.008}$ \\ [1.0ex]
 \hline\hline
 GSS5${\dagger}$ & $0.984$ & $-2.885$ & $24.46^{+0.05}_{-0.02}$ & $0.058$ & $780.^{+19.}_{-7.}$ & $42.^{+4.}_{-1.}$ & $-0.4^{+0.2}_{-0.1}$ & $0.9^{+0.2}_{-0.1}$ & $0.652^{+0.011}_{-0.009}$ \\ [1.0ex]
 \hline
 GSS6.5${\dagger}$ & $1.500$ & $-3.594$ & $24.41^{+0.22}_{-0.03}$ & $0.053$ & $762.^{+83.}_{-10.}$ & $53.^{+31.}_{-1.}$ & $-0.4^{+0.2}_{-0.1}$ & $0.9^{+0.2}_{-0.2}$ & $0.753^{+0.013}_{-0.011}$ \\ [1.0ex]
 \hline
 GSS7${\dagger}$ & $1.671$ & $-3.830$ & $24.36^{+0.23}_{-0.04}$ & $0.052$ & $744.^{+83.}_{-12.}$ & $61.^{+18.}_{-3.}$ & $-0.4^{+0.1}_{-0.1}$ & $0.7^{+0.2}_{-0.1}$ & $0.763^{+0.015}_{-0.010}$ \\ [1.0ex]
 \hline
 GSS8.5${\dagger}$ & $2.186$ & $-4.537$ & $24.37^{+0.04}_{-0.08}$ & $0.055$ & $749.^{+15.}_{-26.}$ & $68.^{+11.}_{-1.}$ & $-0.6^{+0.1}_{-0.1}$ & $0.5^{+0.1}_{-0.1}$ & $0.824^{+0.009}_{-0.010}$ \\ [1.0ex]
 \hline
 GSS9${\dagger}$ & $2.358$ & $-4.772$ & $24.63^{+0.02}_{-0.04}$ & $0.050$ & $845.^{+8.}_{-15.}$ & $103.^{+7.}_{-12.}$ & $-0.8^{+0.2}_{-0.1}$ & $0.7^{+0.2}_{-0.1}$ & $0.872^{+0.009}_{-0.010}$ \\ [1.0ex]
 \hline
\\

\end{tabular} 

\label{par_table1}  

\begin{flushleft} This table quantifies the MCMC-fitted parameter estimates for the Giant Stellar Stream subfields - i.e. labelled `GSS$X$'. Parameters are given with their one-sigma (68.2\%) uncertainties. Field boundaries are illustrated in Fig. \ref{FieldMaps}. Note that subfields labelled GSS$X.5$ include the lower half of Subfield GSS$X$ and the upper half of Subfield GSS$X+1$. Subfields with probability peaks omitted for the determination of their best fit parameter estimates (due to the presence of prominent peaks that are inconsistent with the overwhelming trend) are denoted $\dagger$$^*$. The alternative estimates derived from the unrestricted distributions are denoted $\dagger$ and appear at the bottom of the table below the double line. For fields external to the GSS, see Table \ref{par_table2}. \end{flushleft}

\end{table*} 




\newpage



\begin{table*} 

\caption{Stream Parameters (Stream C, Stream D and Halo comparison fields). See Table \ref{par_table1} caption for explanation.}{} 

\vspace{1 mm}

\centering      
\begin{tabular}{c c c c c c c c c c}  
\hline\hline                        
Subfield & Xi & Eta & Distance Modulus & $E(B-V)$ & Distance (kpc) & M31 Distance (kpc) & $Fe/H$ (dex) & RGB width (dex) & $f_{cont}$  \\ [0.5ex] 
\hline                    
 C1 & $2.558$ & $-3.676$ & $24.54^{+0.02}_{-0.04}$ & $0.050$ & $809.^{+9.}_{-15.}$ & $68.^{+8.}_{-4.}$ & $-0.9^{+0.1}_{-0.1}$ & $0.5^{+0.1}_{-0.1}$ & $0.846^{+0.011}_{-0.013}$ \\ [1.0ex]
 \hline
 C2 & $3.182$ & $-3.023$ & $24.66^{+0.02}_{-0.07}$ & $0.050$ & $854.^{+8.}_{-28.}$ & $101.^{+8.}_{-21.}$ & $-1.4^{+0.2}_{-0.2}$ & $1.0^{+0.2}_{-0.2}$ & $0.889^{+0.008}_{-0.011}$ \\ [1.0ex]
 \hline
 C3 & $3.580$ & $-1.896$ & $24.57^{+0.02}_{-0.15}$ & $0.048$ & $819.^{+9.}_{-56.}$ & $55.^{+13.}_{-1.}$ & $-0.9^{+0.2}_{-0.2}$ & $0.7^{+0.3}_{-0.1}$ & $0.871^{+0.010}_{-0.013}$ \\ [1.0ex]
 \hline
 C4 & $3.715$ & $-0.499$ & $24.60^{+0.02}_{-0.05}$ & $0.054$ & $831.^{+8.}_{-20.}$ & $69.^{+12.}_{-9.}$ & $-0.9^{+0.1}_{-0.1}$ & $0.6^{+0.1}_{-0.1}$ & $0.889^{+0.009}_{-0.009}$ \\ [1.0ex]
 \hline
 D1 & $2.174$ & $-2.142$ & $24.46^{+0.02}_{-0.06}$ & $0.049$ & $779.^{+7.}_{-21.}$ & $42.^{+3.}_{-1.}$ & $-1.2^{+0.1}_{-0.1}$ & $0.6^{+0.1}_{-0.1}$ & $0.867^{+0.010}_{-0.013}$ \\ [1.0ex]
 \hline
 D2 & $2.728$ & $-1.423$ & $24.47^{+0.09}_{-0.07}$ & $0.057$ & $782.^{+32.}_{-26.}$ & $42.^{+14.}_{-1.}$ & $-1.2^{+0.2}_{-0.1}$ & $0.5^{+0.1}_{-0.1}$ & $0.902^{+0.009}_{-0.013}$ \\ [1.0ex]
 \hline
 D3 & $2.947$ & $-0.579$ & $24.46^{+0.07}_{-0.04}$ & $0.055$ & $781.^{+26.}_{-13.}$ & $41.^{+5.}_{-1.}$ & $-1.1^{+0.1}_{-0.2}$ & $0.6^{+0.1}_{-0.1}$ & $0.884^{+0.010}_{-0.013}$ \\ [1.0ex]
 \hline
 D4 & $3.097$ & $0.469$ & $24.56^{+0.13}_{-0.04}$ & $0.056$ & $818.^{+50.}_{-15.}$ & $61.^{+42.}_{-10.}$ & $-1.1^{+0.1}_{-0.2}$ & $0.4^{+0.1}_{-0.1}$ & $0.907^{+0.010}_{-0.011}$ \\ [1.0ex]
 \hline
 D5 & $2.932$ & $1.198$ & $24.47^{+0.04}_{-0.05}$ & $0.081$ & $783.^{+13.}_{-17.}$ & $43.^{+4.}_{-1.}$ & $-1.1^{+0.1}_{-0.1}$ & $0.5^{+0.1}_{-0.1}$ & $0.869^{+0.009}_{-0.010}$ \\ [1.0ex]
 \hline
 H1 & $4.8$ & $-4.5$ & $24.54^{+0.14}_{-0.27}$ & $0.048$ & $809.^{+53.}_{-96.}$ & $89.^{+29.}_{-1.}$ & $-1.5^{+0.5}_{-0.4}$ & $0.9^{+0.3}_{-0.2}$ & $0.967^{+0.010}_{-0.014}$ \\ [1.0ex]
 \hline
 H2 & $5.2$ & $0.2$ & $24.43^{+0.21}_{-0.15}$ & $0.062$ & $768.^{+77.}_{-51.}$ & $70.^{+26.}_{-1.}$ & $-0.9^{+0.2}_{-1.0}$ & $1.0^{+0.5}_{-0.3}$ & $0.971^{+0.008}_{-0.010}$ \\ [1.0ex]
 \hline
 H3 & $1.586$ & $-0.823$ & $24.50^{+0.01}_{-0.04}$ & $0.052$ & $795.^{+5.}_{-13.}$ & $25.^{+9.}_{-1.}$ & $-1.3^{+0.1}_{-0.1}$ & $0.8^{+0.1}_{-0.1}$ & $0.792^{+0.009}_{-0.008}$ \\ [1.0ex]
 \hline
\\

\end{tabular} 

\label{par_table2}  

\end{table*} 




\section{Conclusions}
\label{conclusions}

We have presented the distances and metallicities for the major inner-halo streams of M31 using the highest quality data currently available. There is a great deal of overlap between many of these features, making clear measurements troublesome, however the new method we developed to fit populations to the data have allowed some details to be revealed.

There is a clear need for a wide field kinematic survey of the stellar substructure within the halo of M31, which combined with the superb PAndAS photometric data, would allow for a complete decomposition of these structures. This would bring a much greater understanding of the current and past accretion history of our nearest neighbour analogue, and would represent a great leap forward in galactic archaeology.

The conclusion of this work then, is that the GSS, Stream C, and Stream D, are in general extremely faint, and can not be completely separated using the currently available photometric data. Our method however, allows for even the lowest contrast structures to be partially resolved into separate populations, providing both distance and metallicity probability distributions. These values will be invaluable for future simulations of the M31 system, placing much stronger constraints on the three dimensional present day positions of the major inner-halo structures. A full population fit based on this data, will lead to a deeper understanding, and will be the subject of a future contribution.

\section*{Acknowledgments}

ARC thanks the University of Sydney for funding via a 2014 Laffan Fellowship. BM acknowledges the support of an Australian Postgraduate Award. NFB and GFL thank the Australian Research Council (ARC) for support through Discovery Project (DP110100678). GFL also gratefully acknowledges financial support through his ARC Future Fellowship (FT100100268). PJE is supported by the SSimPL programme and the Sydney Institute for Astronomy (SIfA), and {\it Australian Research Council} (ARC) grants DP130100117 and DP140100198.

\bsp

\label{lastpage}

\newpage


\begin{figure*}
\begin{center} 
$ \begin{array}{c}
\begin{overpic}[width = 0.23\textwidth,angle=-90,clip=true,trim={0cm 0cm 1.6cm 0cm}]{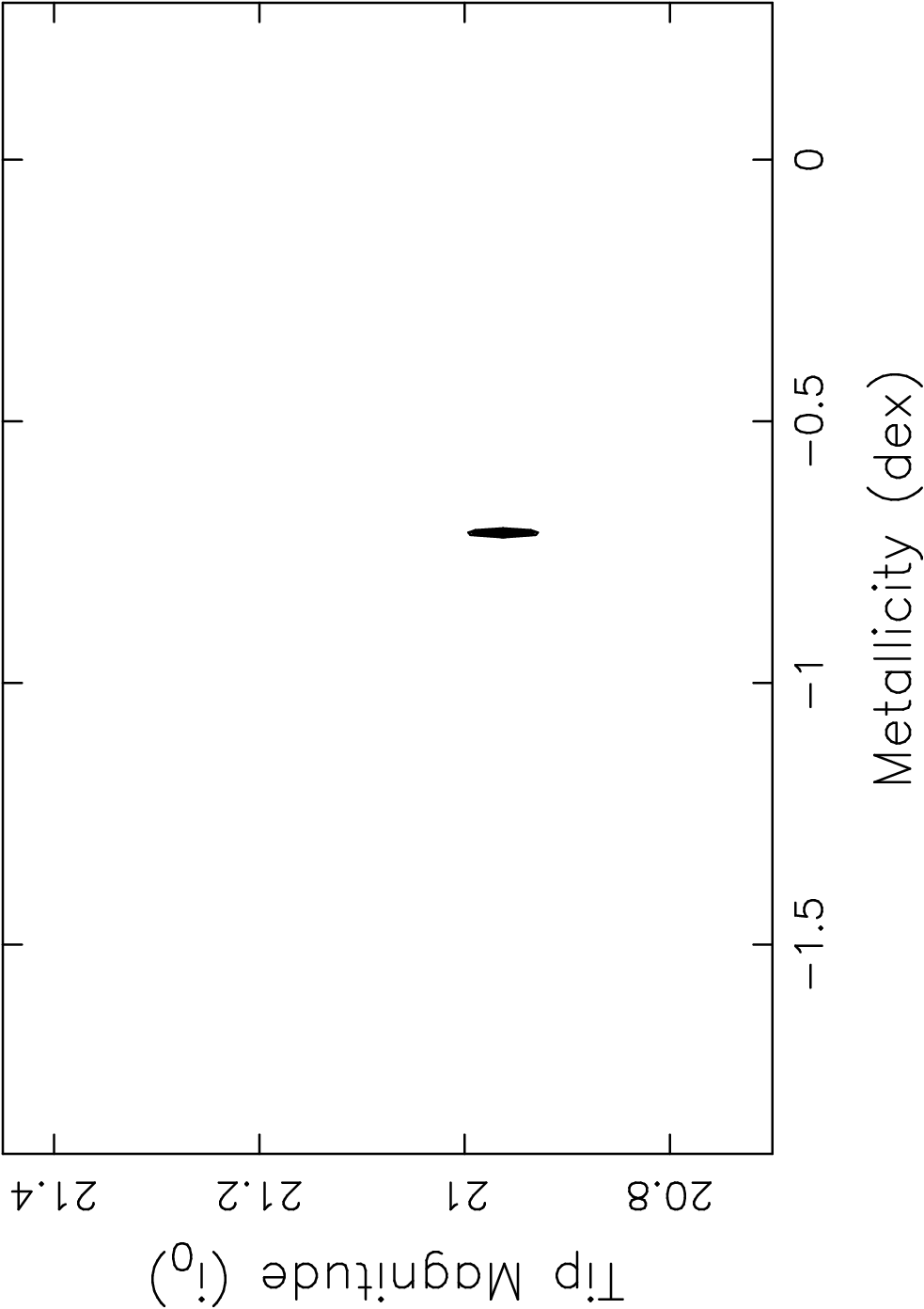}
\put(15,56){\small GSS1}
\end{overpic}
\begin{overpic}[width = 0.23\textwidth,angle=-90,clip=true,trim={0cm 1.1cm 1.6cm 0cm}]{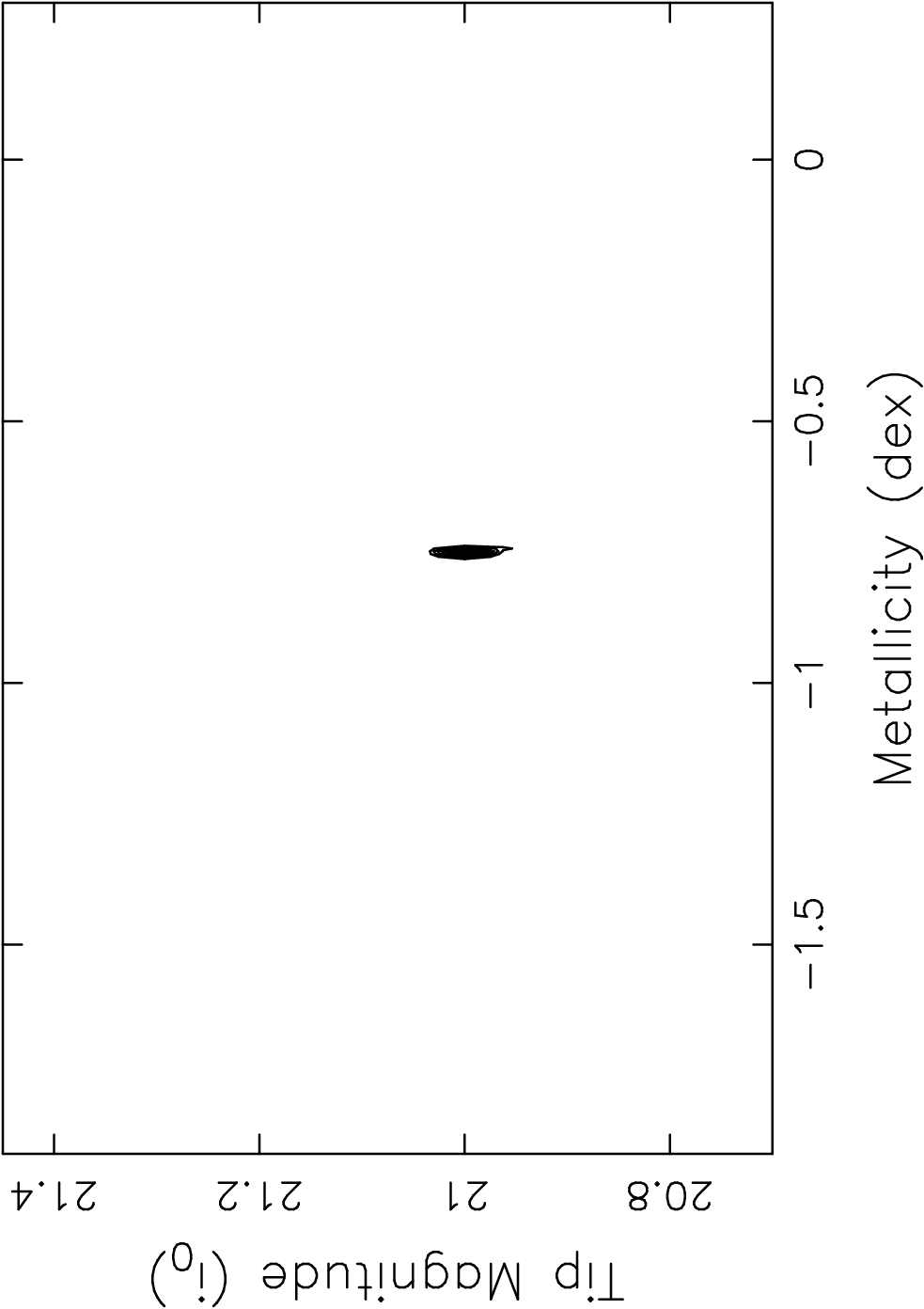}
\put(15,58){\small GSS1.5}
\end{overpic}
\begin{overpic}[width = 0.23\textwidth,angle=-90,clip=true,trim={0cm 1.1cm 1.6cm 0cm}]{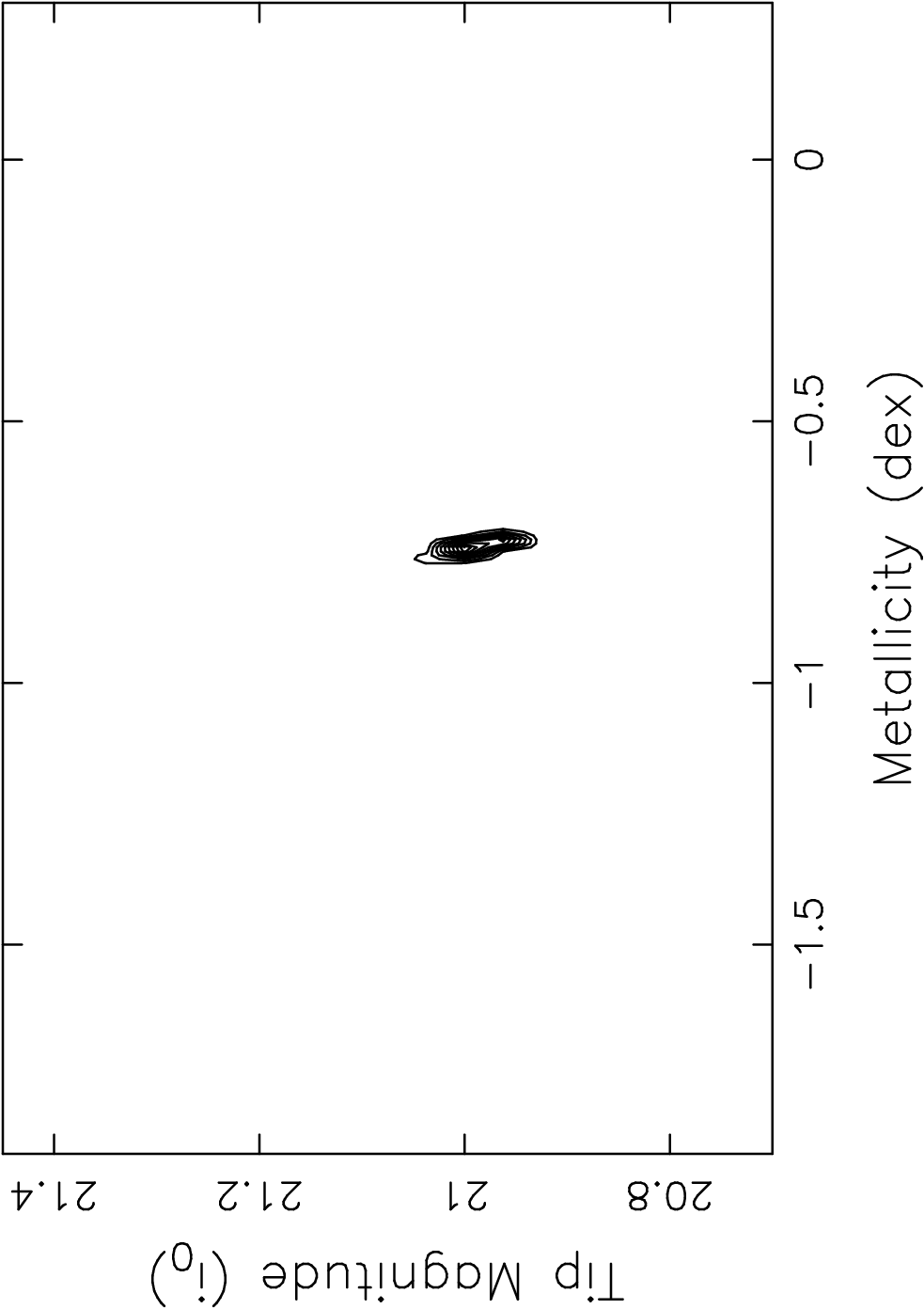}
\put(15,58){\small GSS2}
\end{overpic}
\\ \begin{overpic}[width = 0.23\textwidth,angle=-90,clip=true,trim={0cm 0cm 1.6cm 0cm}]{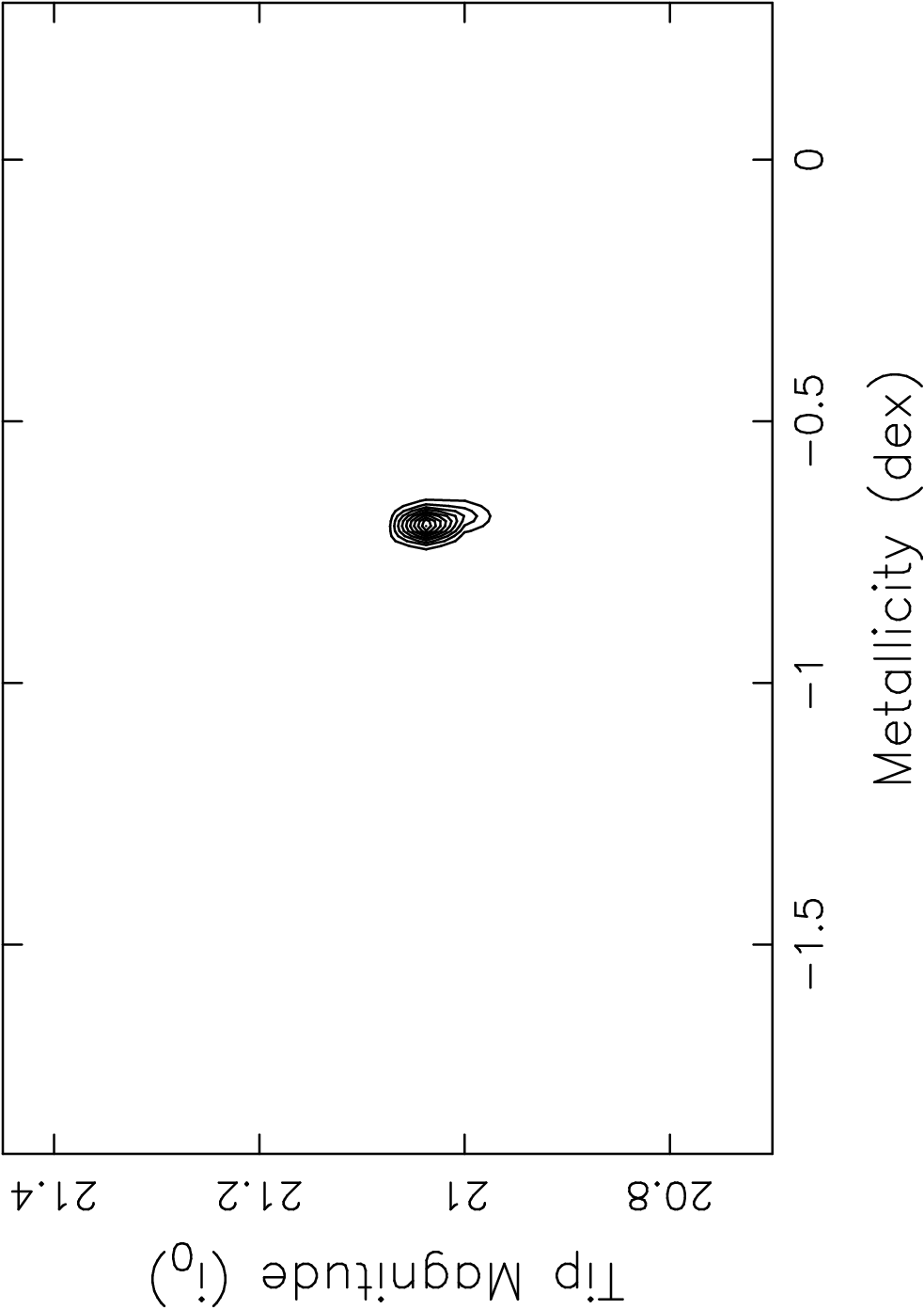}
\put(15,55){\small GSS2.5}
\end{overpic}
\begin{overpic}[width = 0.23\textwidth,angle=-90,clip=true,trim={0cm 1.1cm 1.6cm 0cm}]{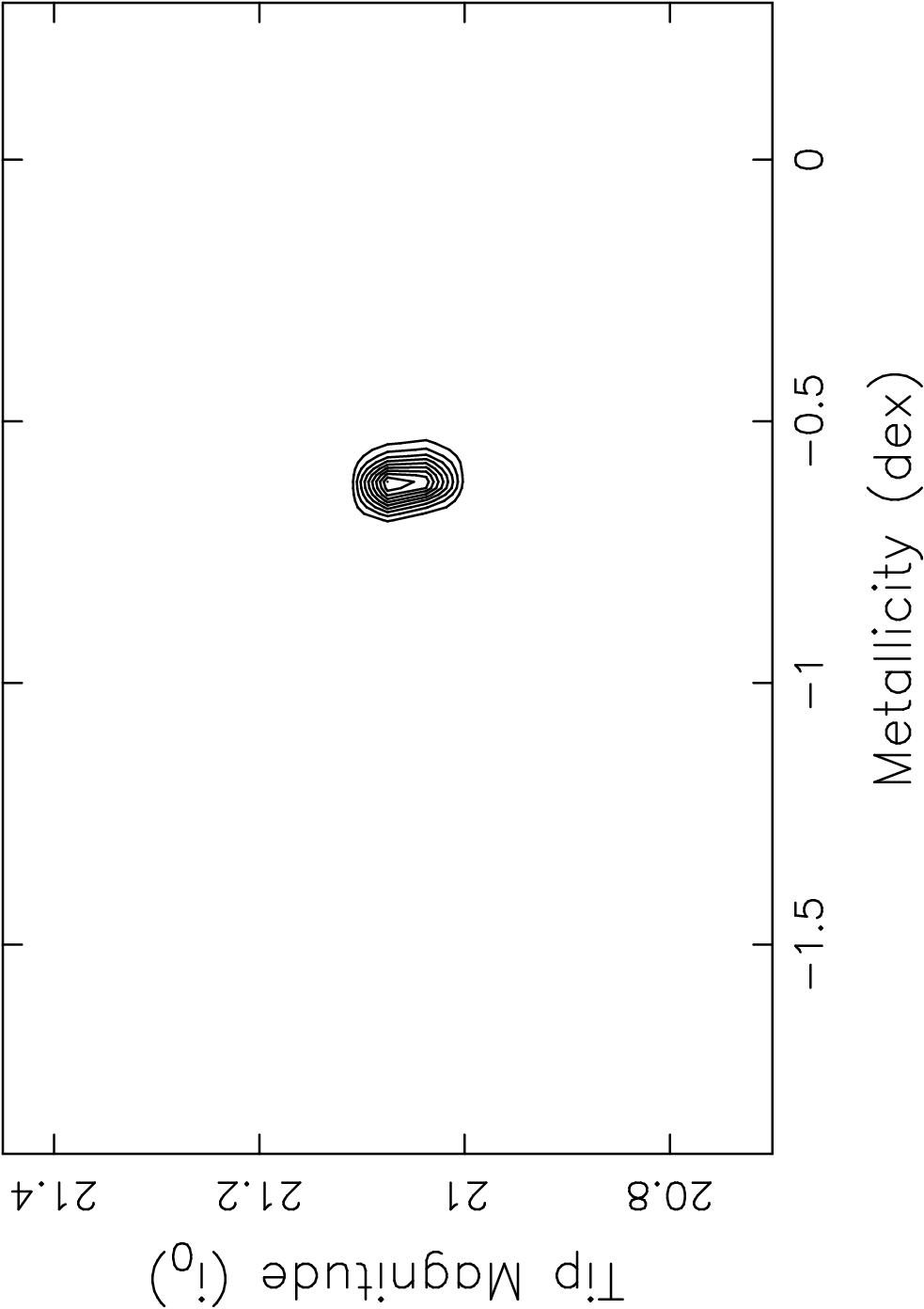}
\put(15,58){\small GSS3}
\end{overpic}
\begin{overpic}[width = 0.23\textwidth,angle=-90,clip=true,trim={0cm 1.1cm 1.6cm 0cm}]{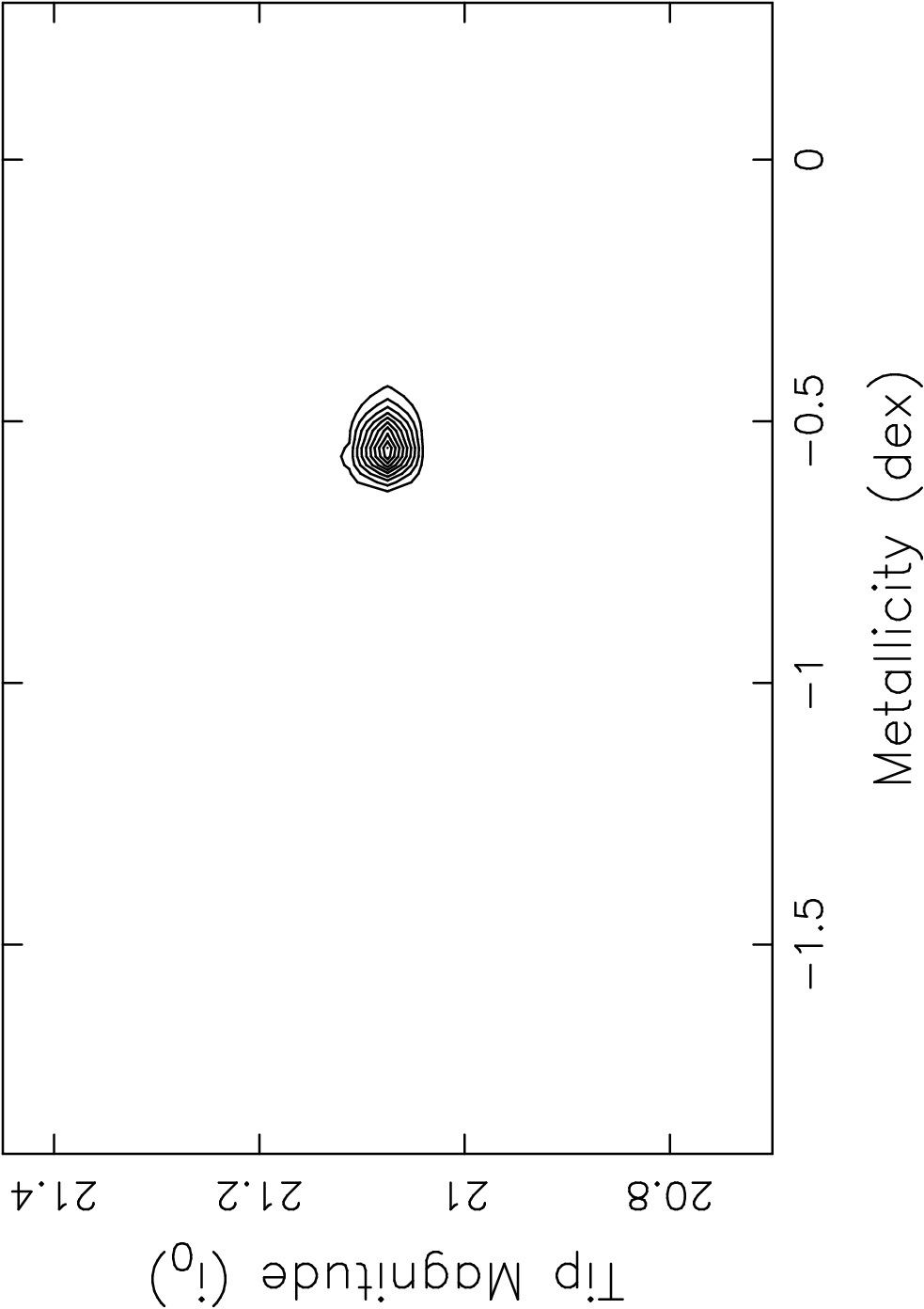}
\put(15,58){\small GSS3.5}
\end{overpic}
\\ \begin{overpic}[width = 0.23\textwidth,angle=-90,clip=true,trim={0cm 0cm 1.6cm 0cm}]{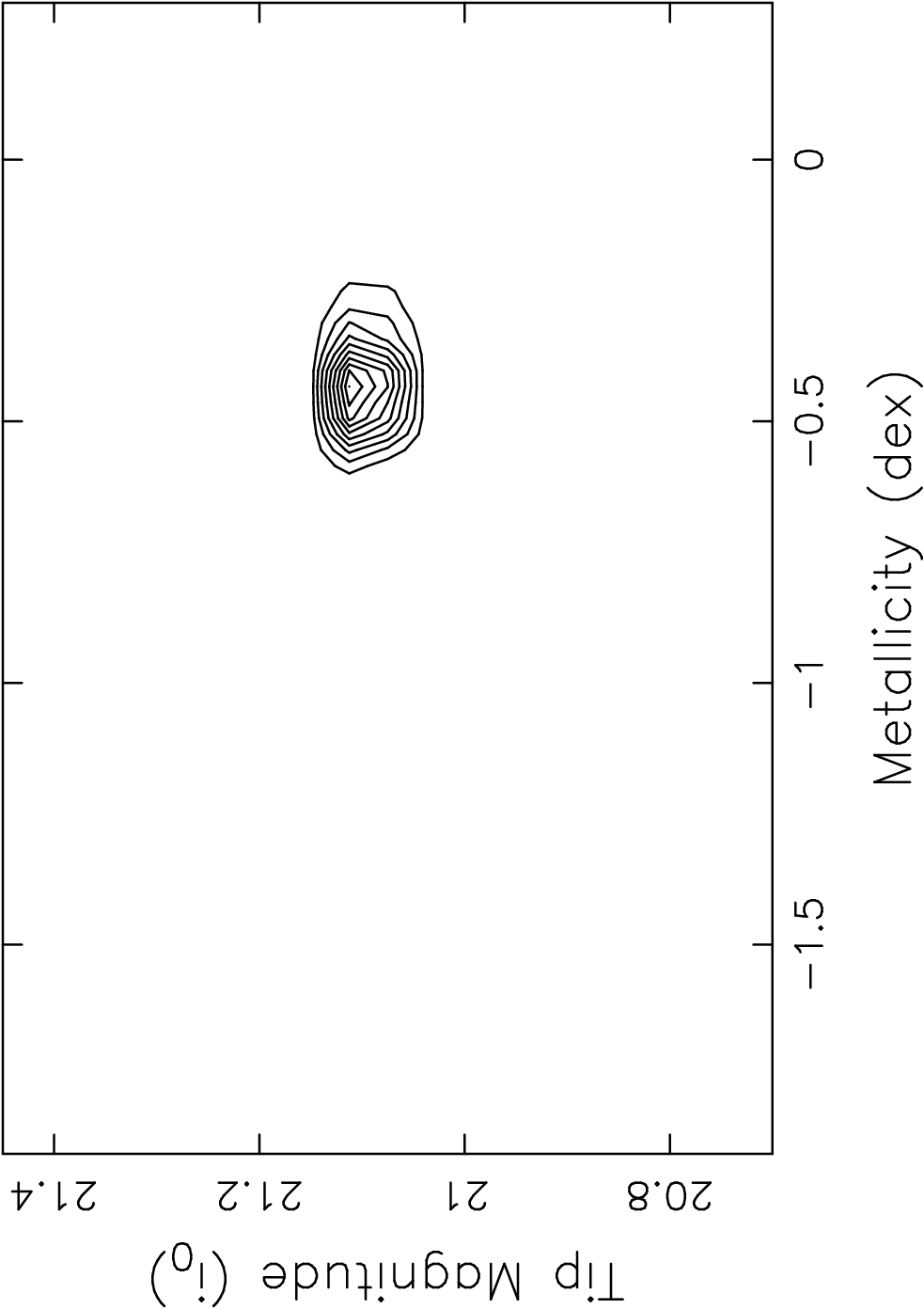}
\put(15,55){\small GSS4}
\end{overpic}
\begin{overpic}[width = 0.23\textwidth,angle=-90,clip=true,trim={0cm 1.1cm 1.6cm 0cm}]{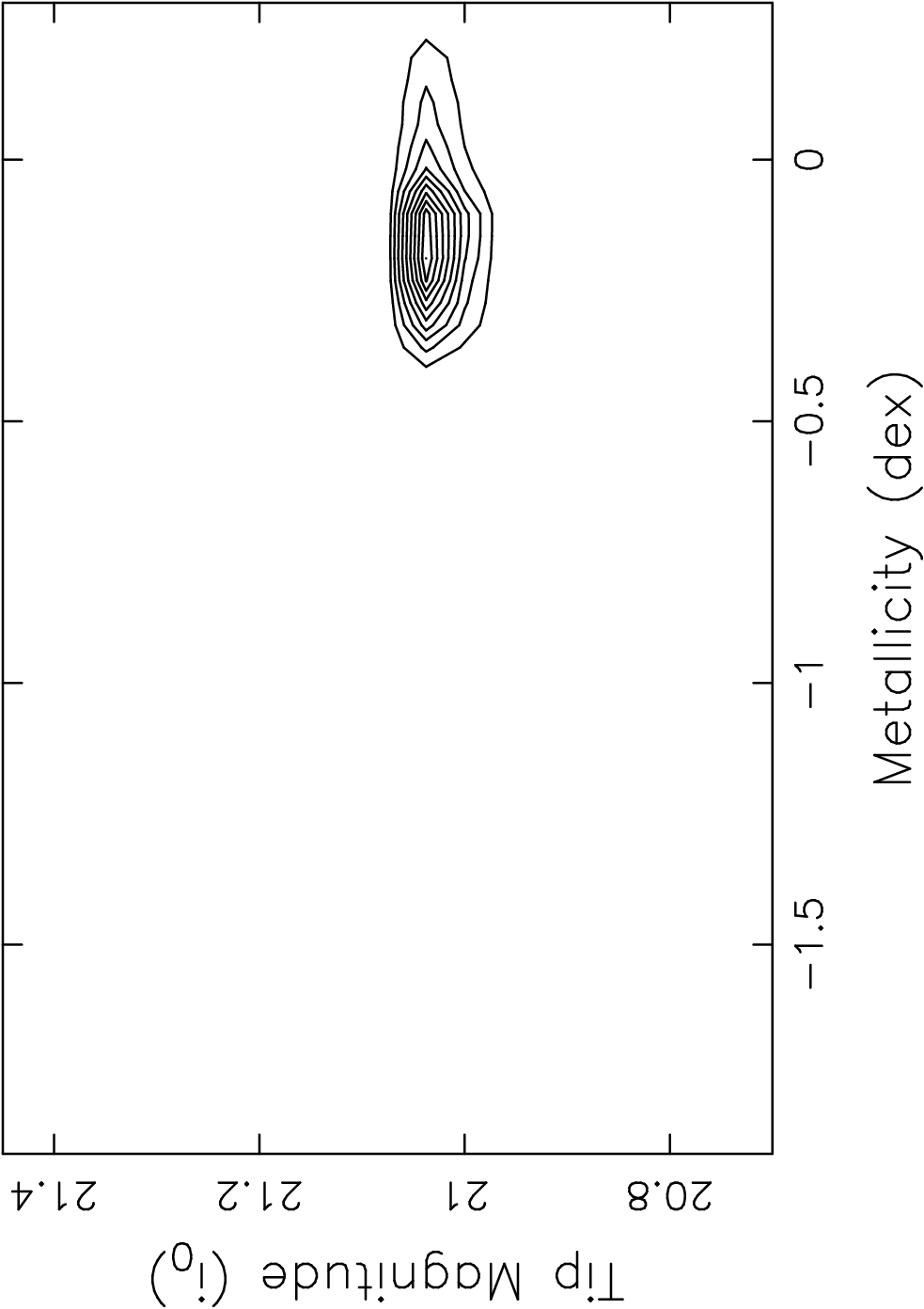}
\put(15,58){\small GSS4.5}
\end{overpic}
\begin{overpic}[width = 0.23\textwidth,angle=-90,clip=true,trim={0cm 1.1cm 1.6cm 0cm}]{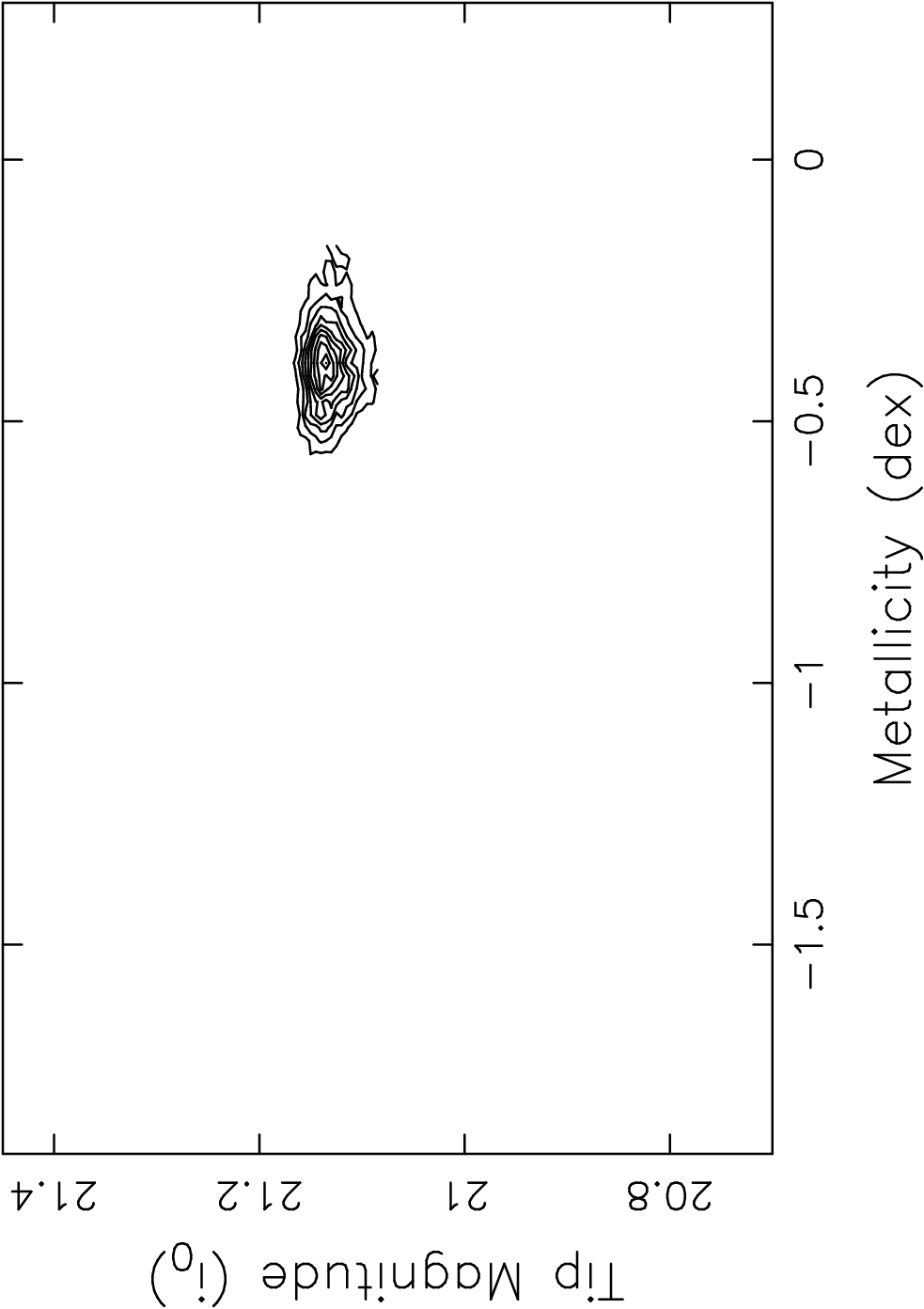}
\put(15,58){\small GSS5$\dagger$$^*$}
\end{overpic}
\\ \begin{overpic}[width = 0.23\textwidth,angle=-90,clip=true,trim={0cm 0cm 1.6cm 0cm}]{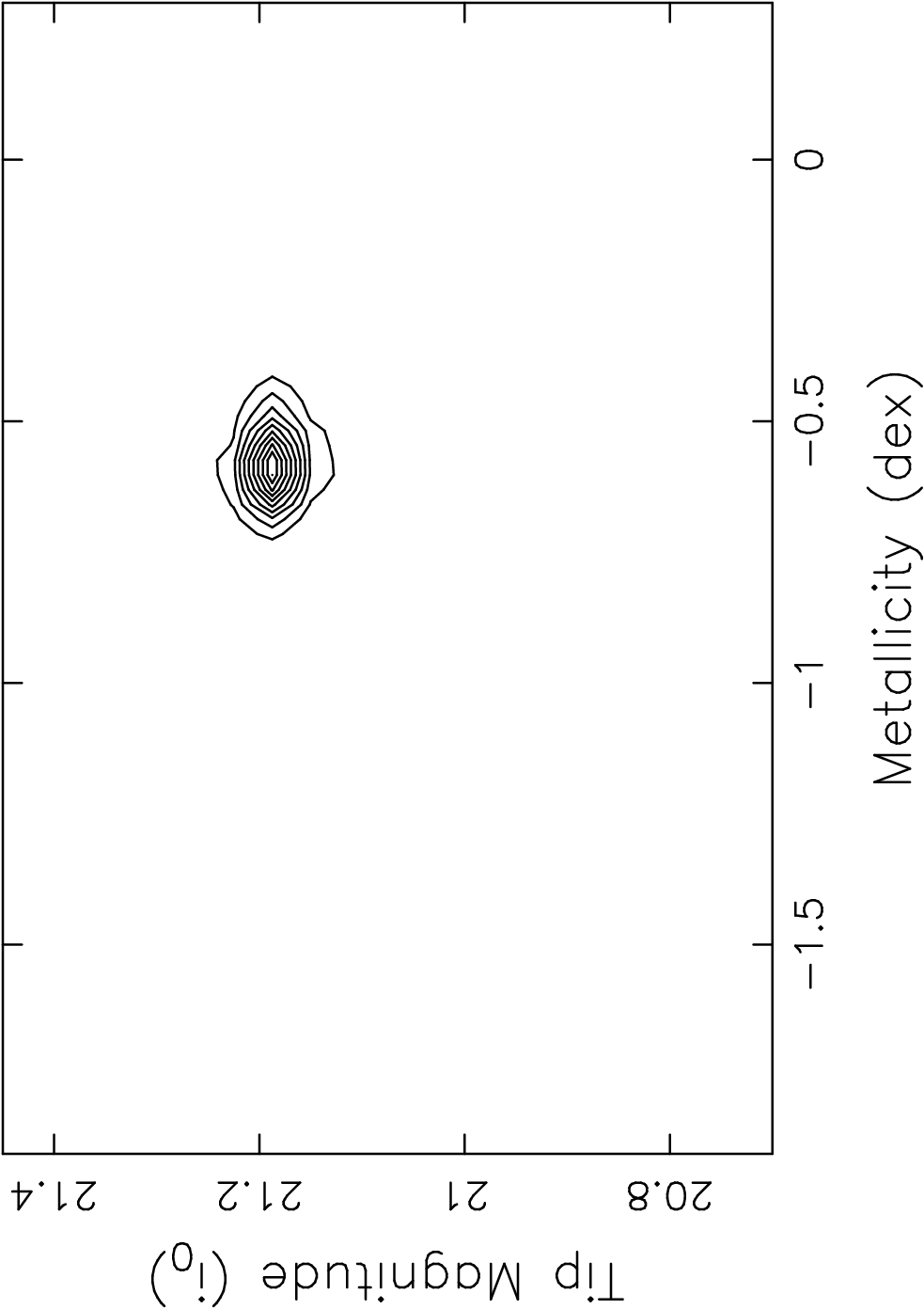}
\put(15,55){\small GSS5.5}
\end{overpic}
\begin{overpic}[width = 0.23\textwidth,angle=-90,clip=true,trim={0cm 1.1cm 1.6cm 0cm}]{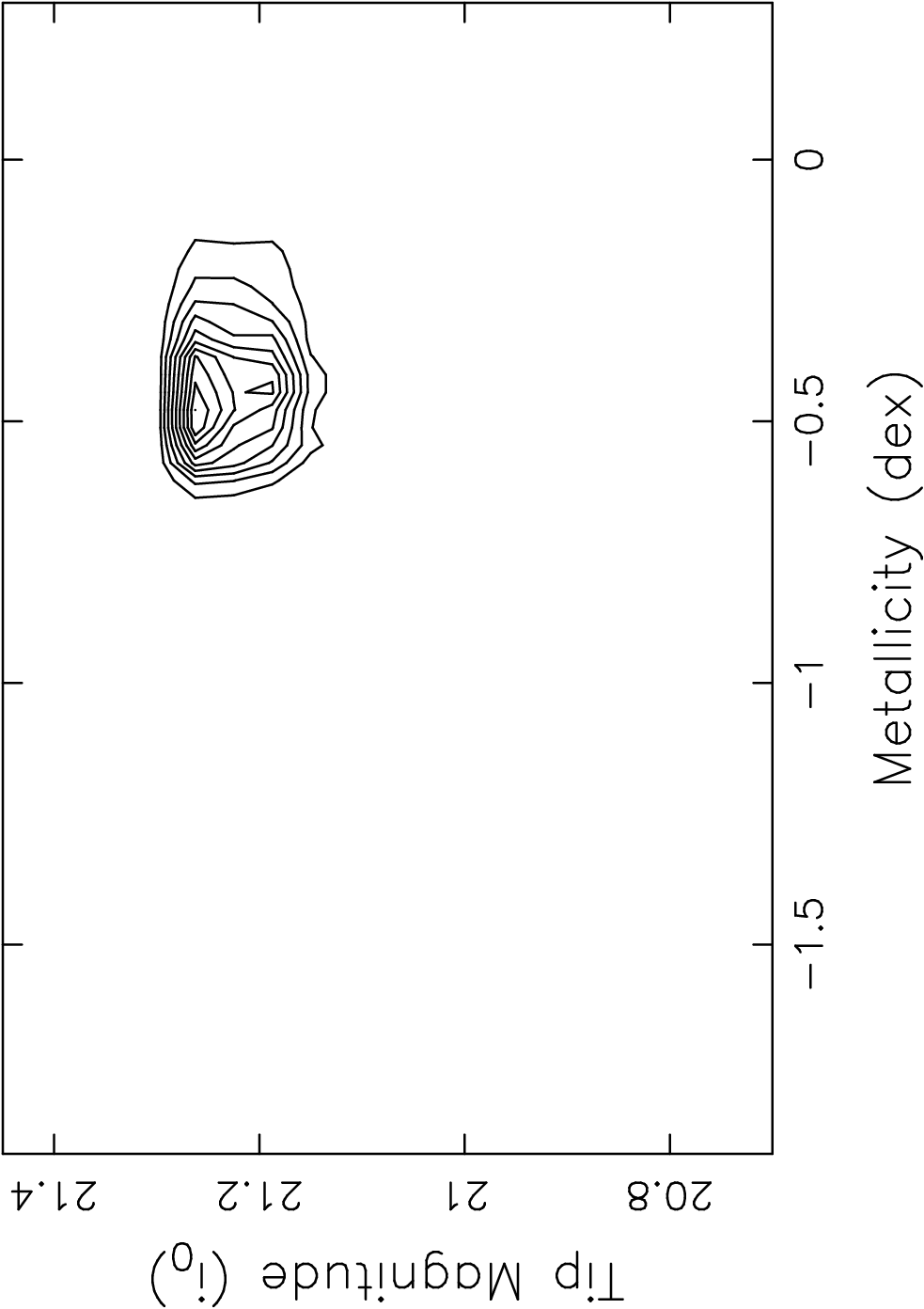}
\put(15,58){\small GSS6}
\end{overpic}
\begin{overpic}[width = 0.23\textwidth,angle=-90,clip=true,trim={0cm 1.1cm 1.6cm 0cm}]{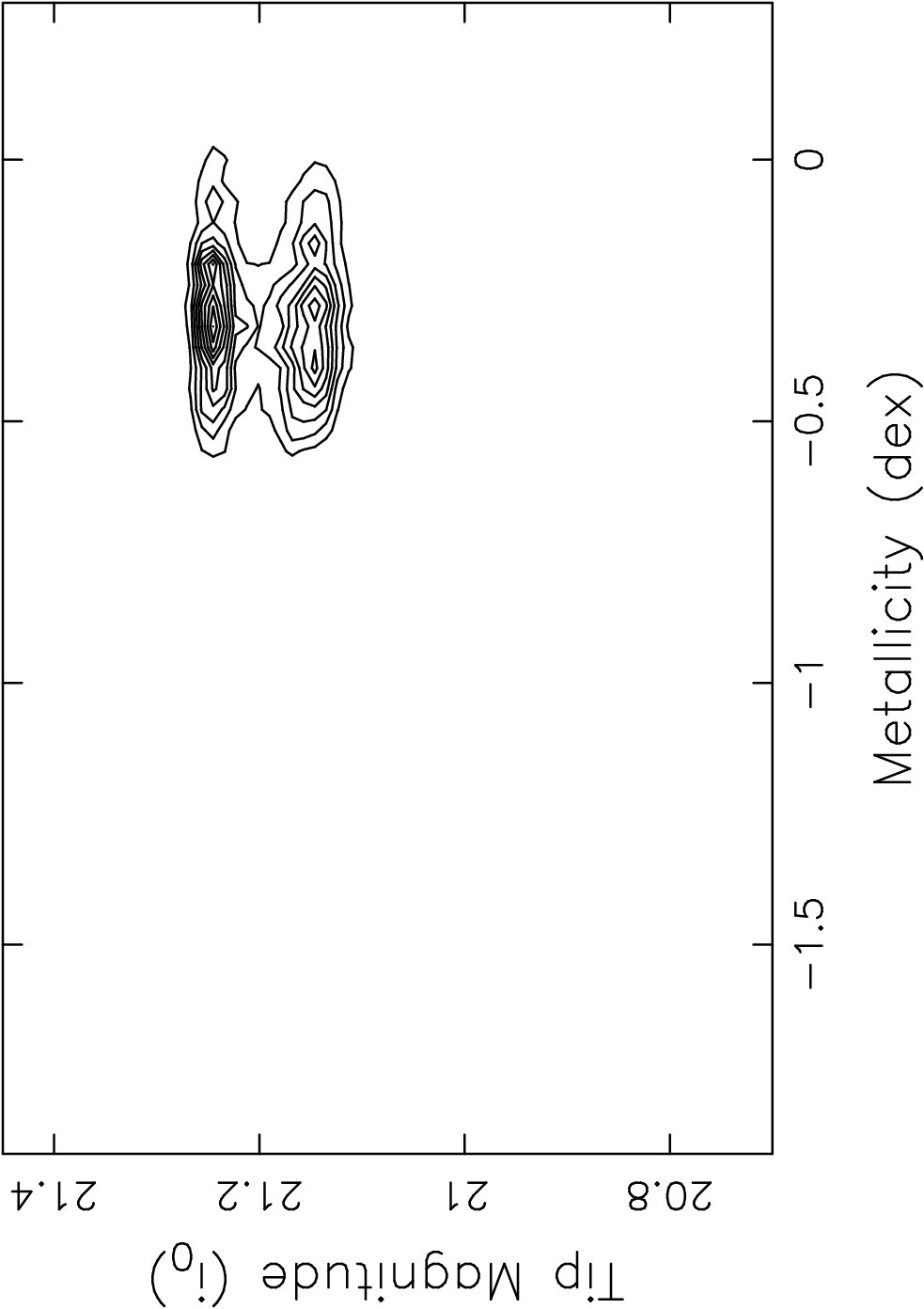}
\put(15,58){\small GSS6.5$\dagger$$^*$}
\end{overpic}
\\ \begin{overpic}[width = 0.258\textwidth,angle=-90,clip=true,trim={0cm 0cm 0cm 0cm}]{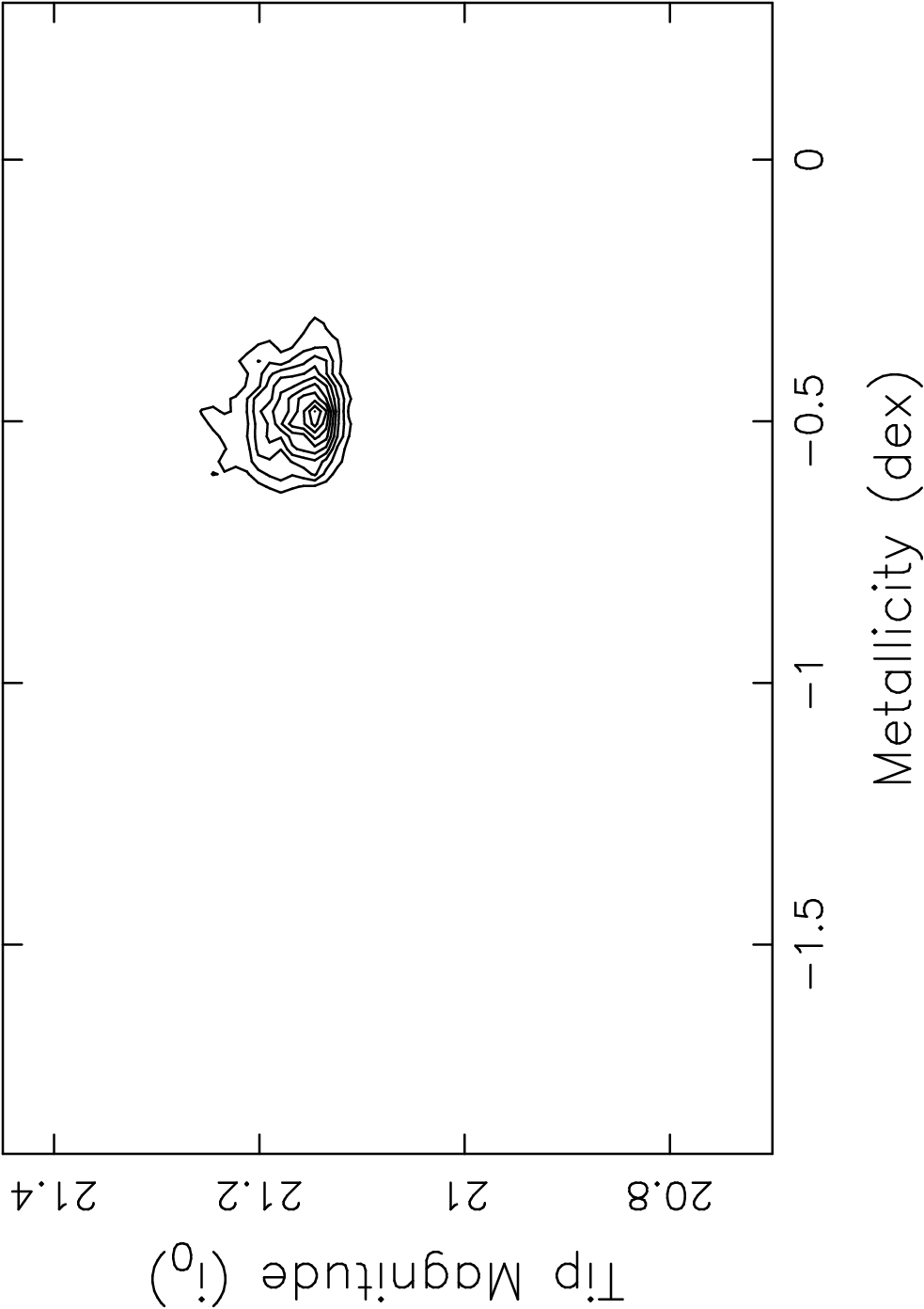}
\put(15,63){\small GSS7$\dagger$$^*$}
\end{overpic}
\begin{overpic}[width = 0.258\textwidth,angle=-90,clip=true,trim={0cm 1.1cm 0cm 0cm}]{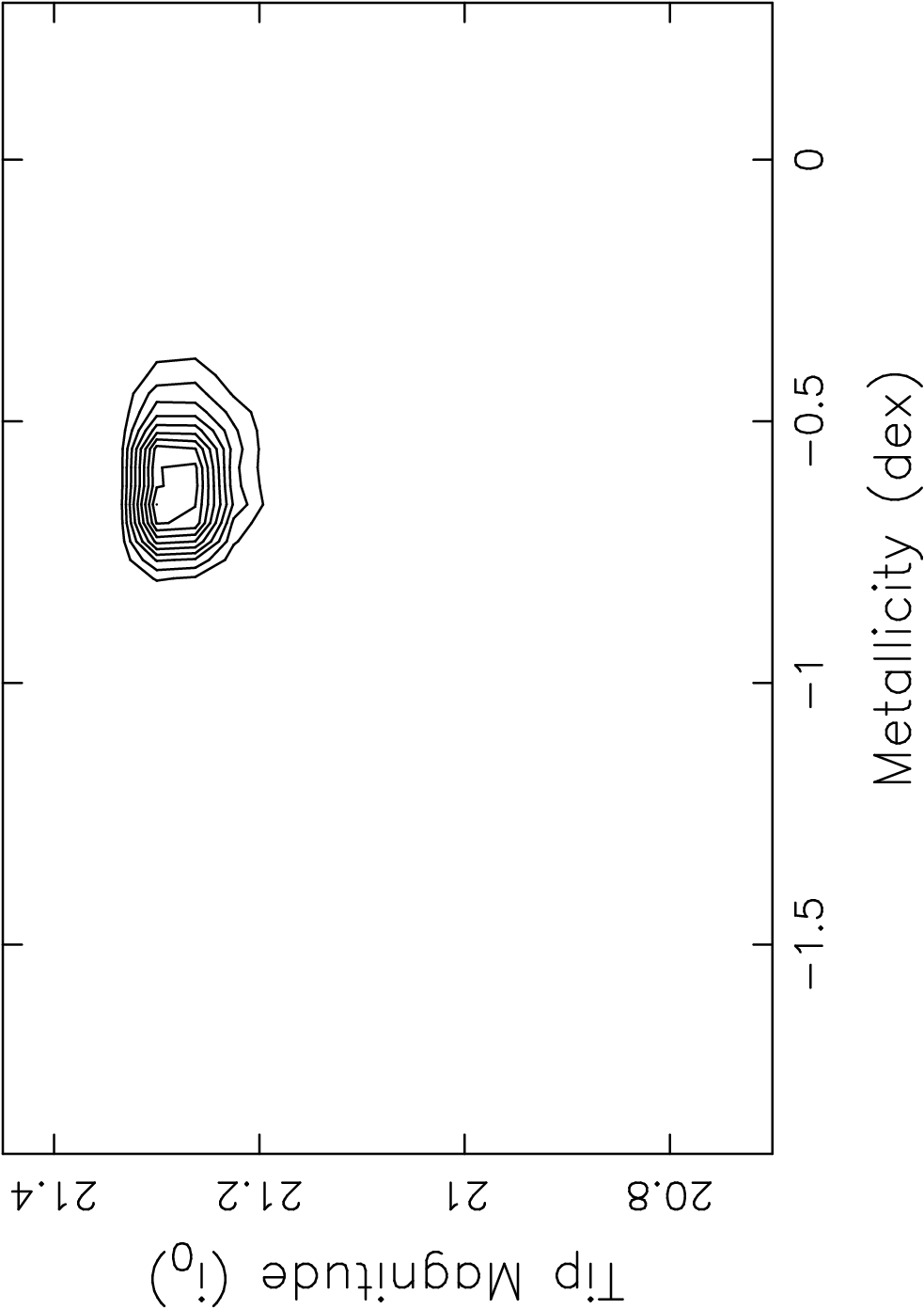}
\put(15,66){\small GSS7.5}
\end{overpic}
\begin{overpic}[width = 0.258\textwidth,angle=-90,clip=true,trim={0cm 1.1cm 0cm 0cm}]{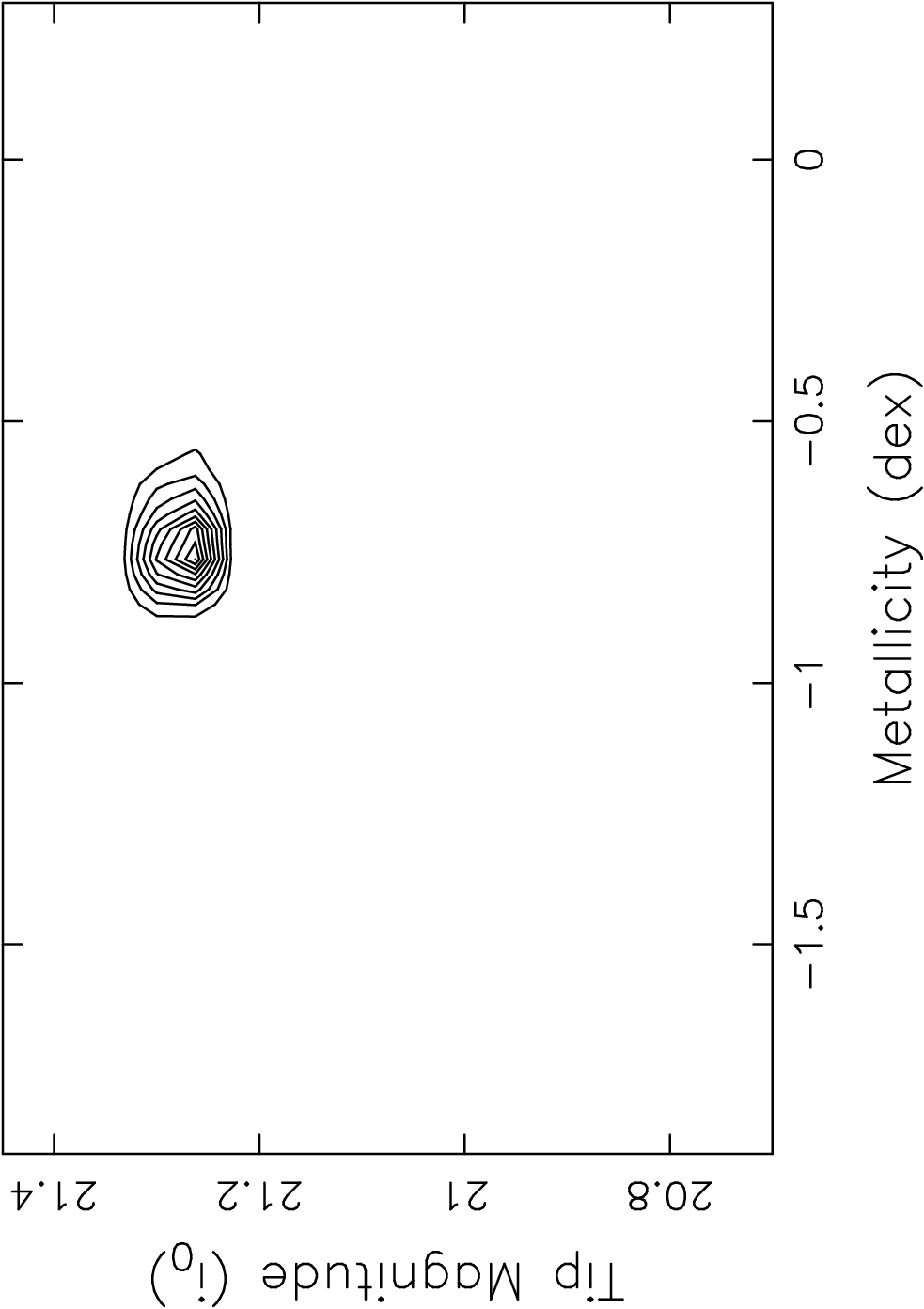}
\put(15,67){\small GSS8}
\end{overpic}

\end{array}$
\end{center}

\textbf{Appendix A Part I:} Contour plots illustrating the correlation between tip magnitude and metallicity probability distributions for the fields listed in Table \ref{par_table1}. \\ Contours are drawn at $10\%$ intervals (as is the case for all subsequent Appendix plots). Fields GSS1 through GSS8 are represented here. \\ Plots denoted $\dagger$$^*$ are generated by sampling only the parameter values consistent with a restricted TRGB range. The full, unrestricted versions denoted $\dagger$ are shown on the next page. The restricted ranges are: GSS5$\dagger$$^*$, $21.08 \le TRGB \le 21.18$; GSS6.5$\dagger$$^*$, $21.08 \le TRGB \le 21.30$; GSS7$\dagger$$^*$, $21.08 \le TRGB \le 21.30$.  

\label{AppAI}

\end{figure*}


\newpage


\begin{figure*}
\begin{center} 
$ \begin{array}{c}
\begin{overpic}[width = 0.23\textwidth,angle=-90,clip=true,trim={0cm 0cm 1.6cm 0cm}]{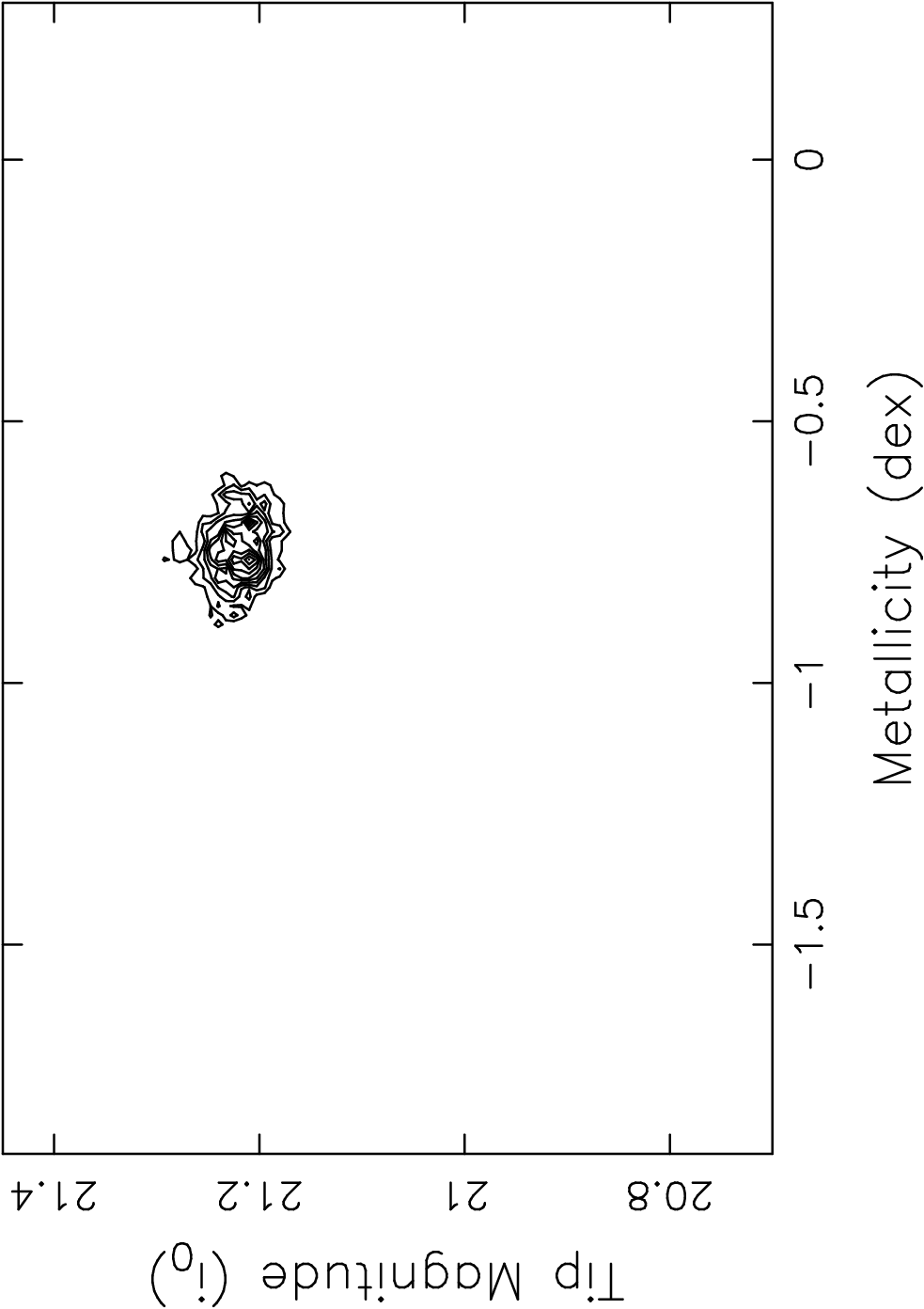}
\put(15,54){\small GSS8.5$\dagger$$^*$}
\end{overpic}
\begin{overpic}[width = 0.23\textwidth,angle=-90,clip=true,trim={0cm 1.1cm 1.6cm 0cm}]{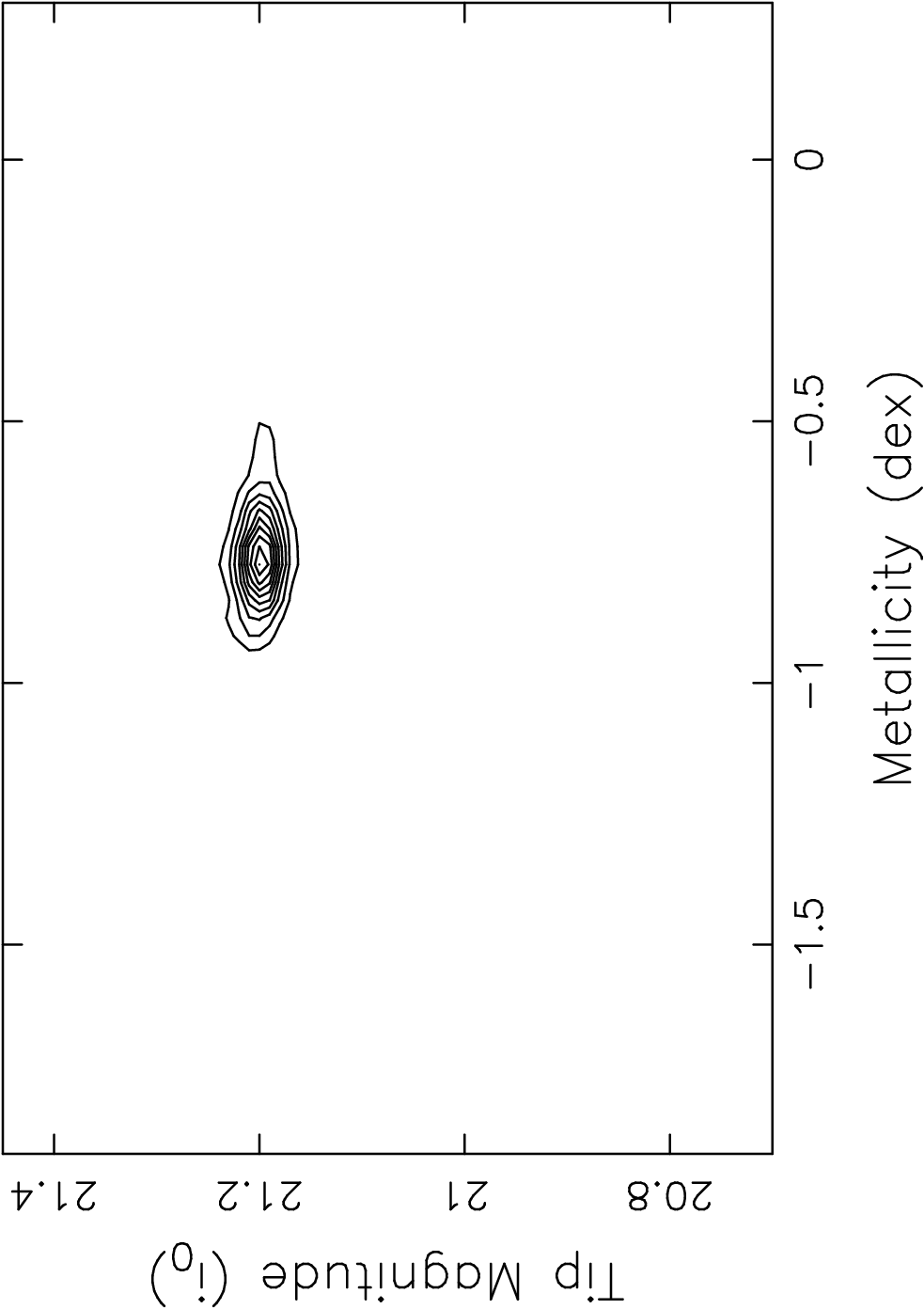}
\put(15,57){\small GSS9$\dagger$$^*$}
\end{overpic}
\begin{overpic}[width = 0.23\textwidth,angle=-90,clip=true,trim={0cm 1.1cm 1.6cm 0cm}]{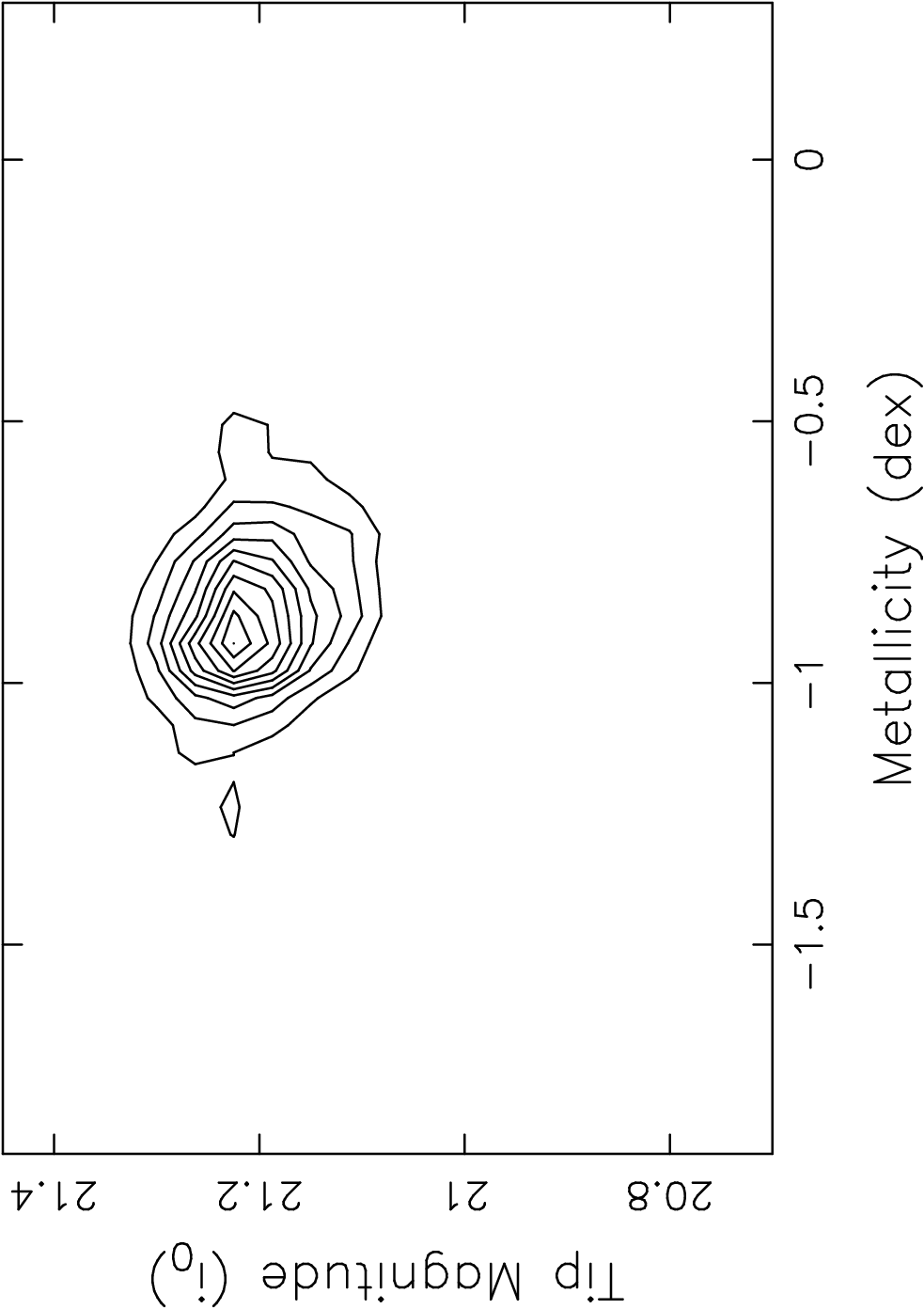}
\put(15,57){\small GSS9.5}
\end{overpic}
\\ \begin{overpic}[width = 0.23\textwidth,angle=-90,clip=true,trim={0cm 0cm 1.6cm 0cm}]{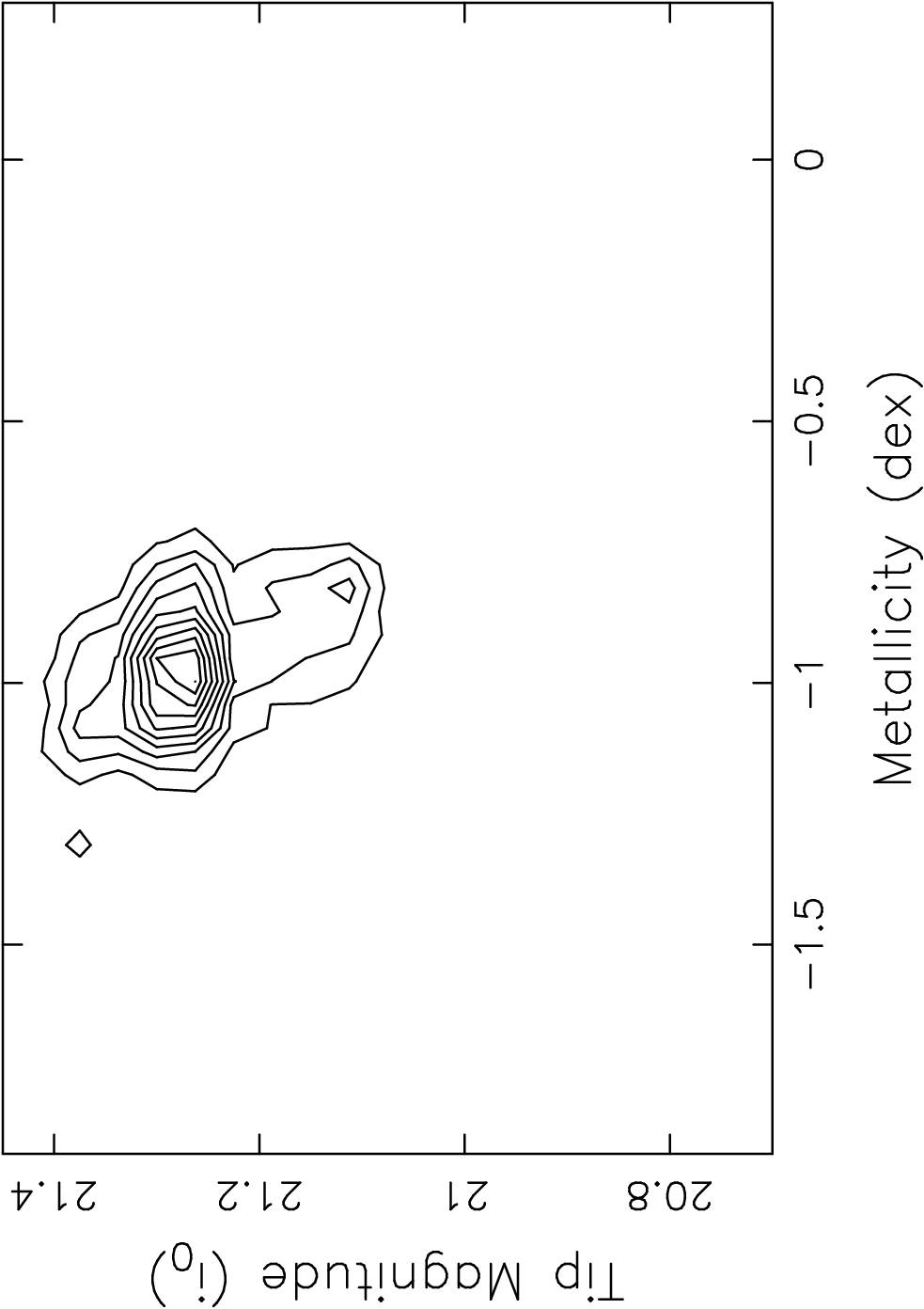}
\put(15,54){\small GSS10}
\end{overpic}
\begin{overpic}[width = 0.23\textwidth,angle=-90,clip=true,trim={0cm 1.1cm 1.6cm 0cm}]{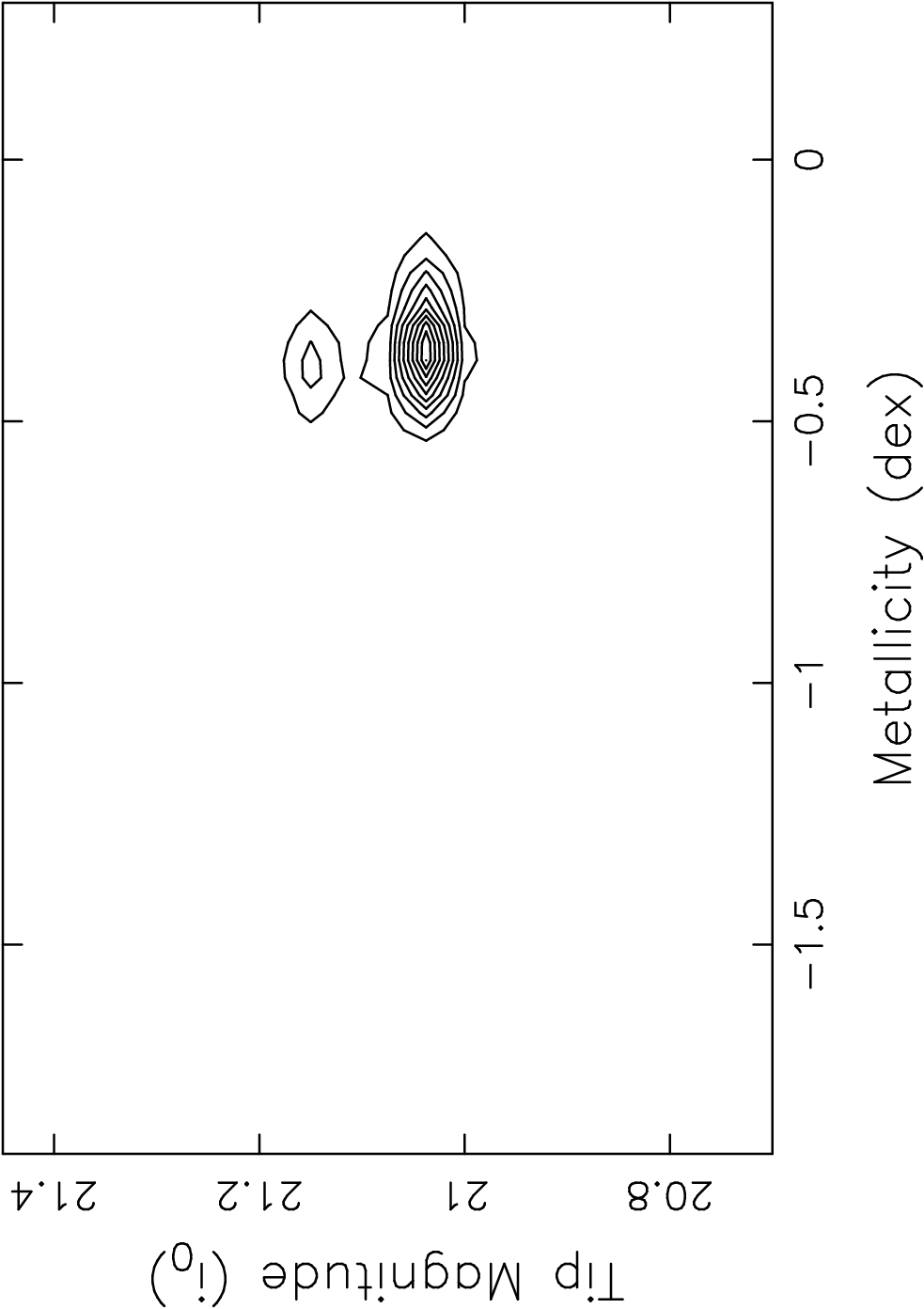}
\put(15,57){\small GSS5$\dagger$}
\end{overpic}
\begin{overpic}[width = 0.23\textwidth,angle=-90,clip=true,trim={0cm 1.1cm 1.6cm 0cm}]{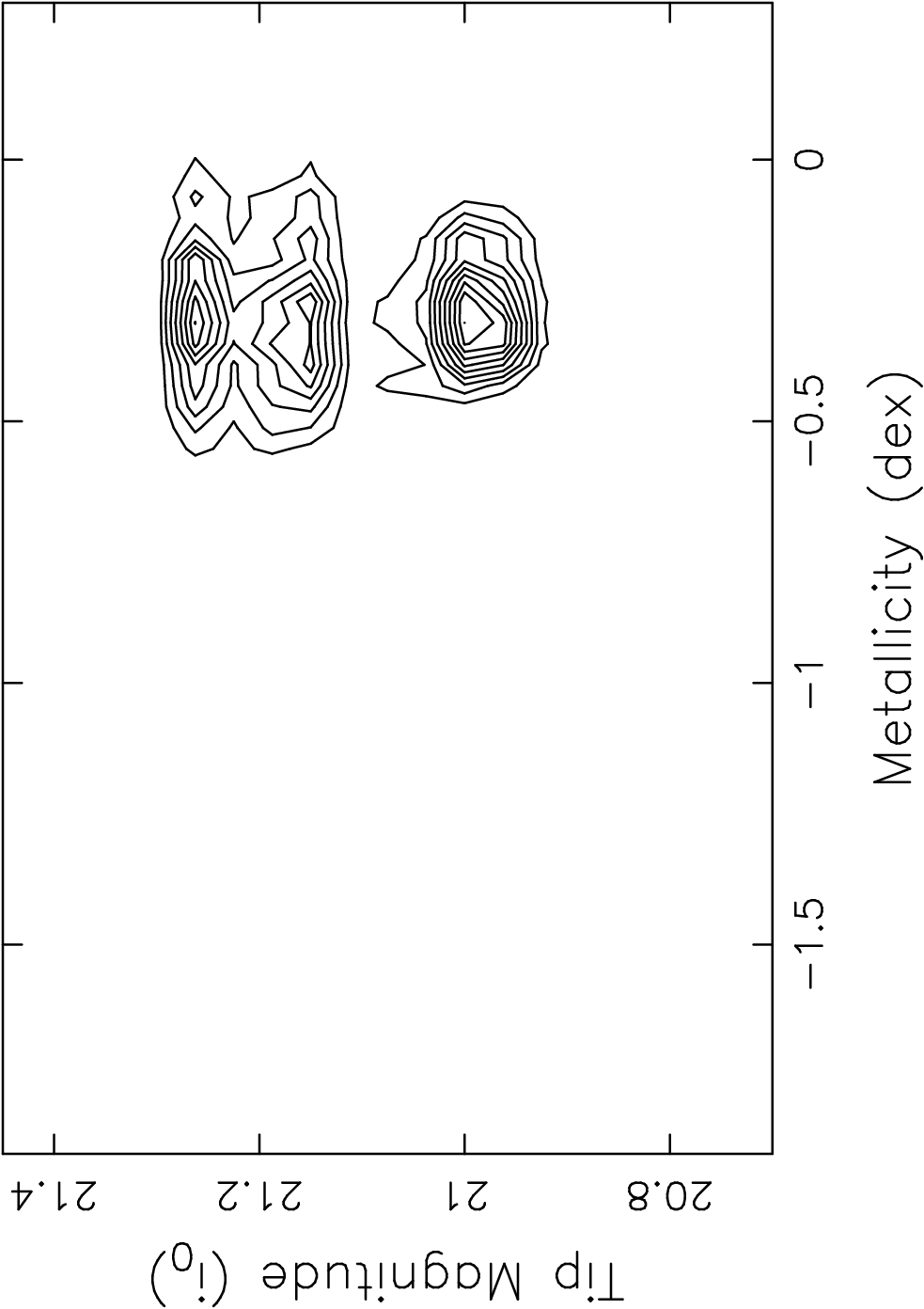}
\put(15,57){\small GSS6.5$\dagger$}
\end{overpic}
\\ \begin{overpic}[width = 0.258\textwidth,angle=-90,clip=true,trim={0cm 0cm 0cm 0cm}]{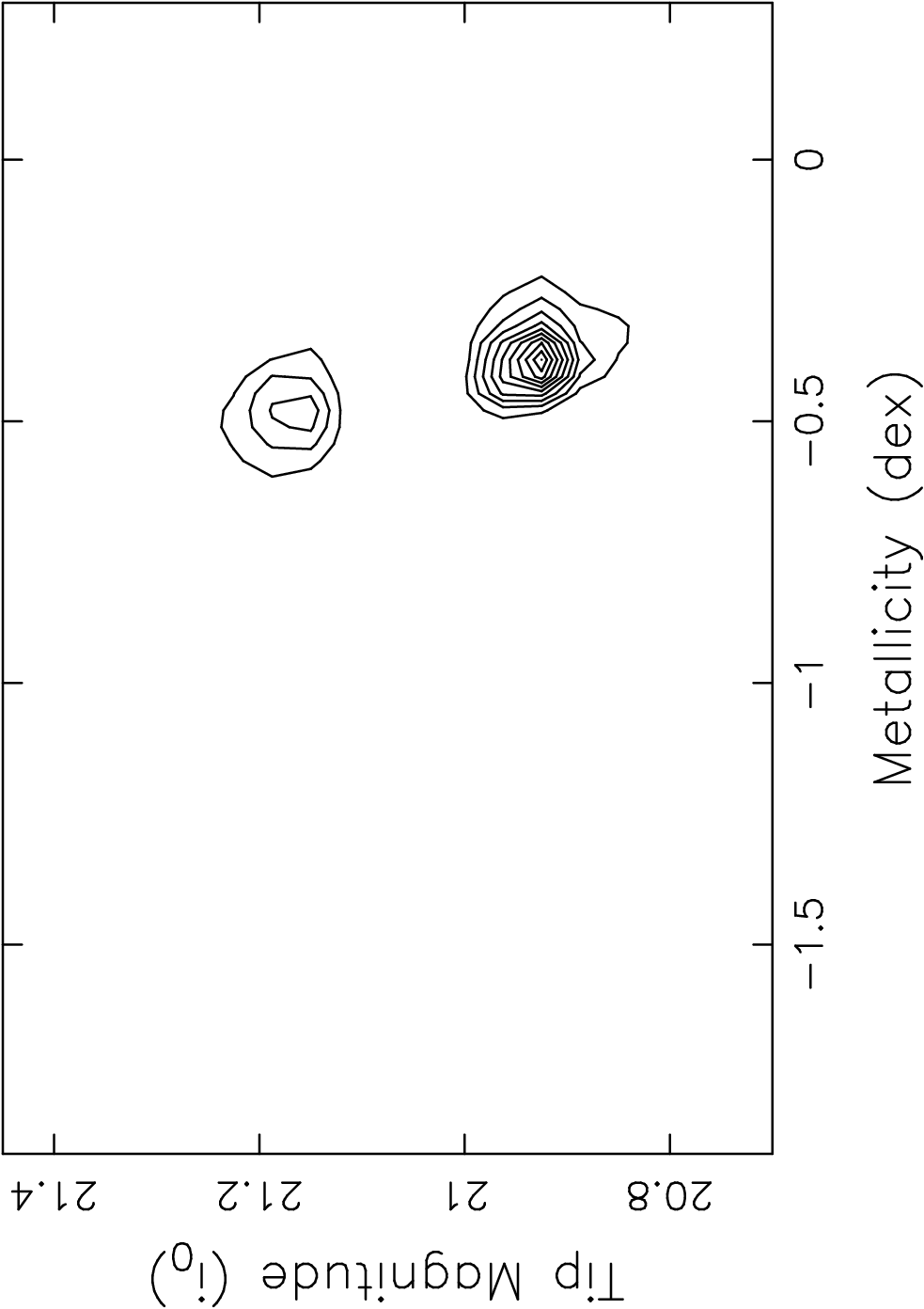}
\put(15,60){\small GSS7$\dagger$}
\end{overpic}
\begin{overpic}[width = 0.258\textwidth,angle=-90,clip=true,trim={0cm 1.1cm 0cm 0cm}]{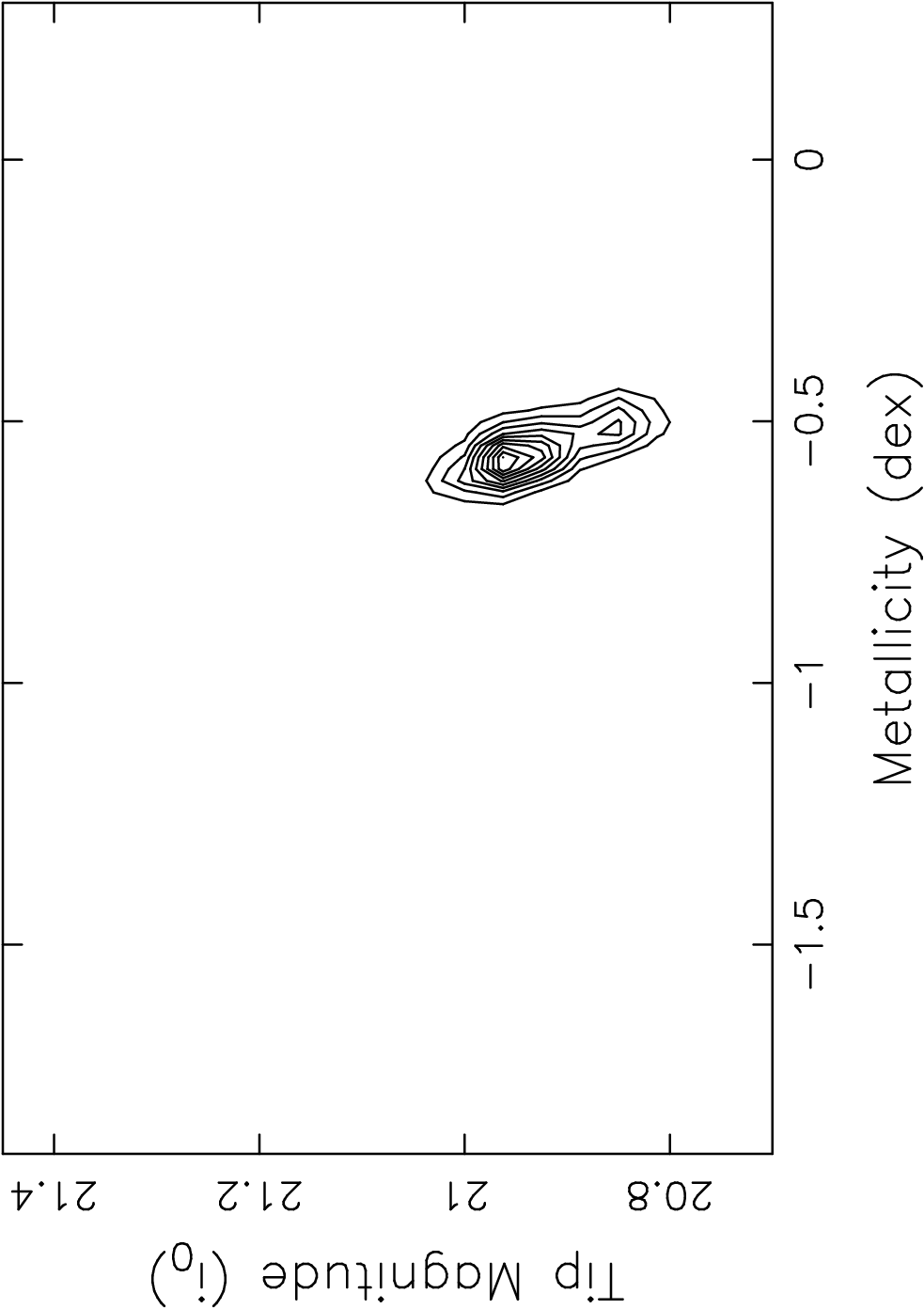}
\put(15,63){\small GSS8.5$\dagger$}
\end{overpic}
\begin{overpic}[width = 0.258\textwidth,angle=-90,clip=true,trim={0cm 1.1cm 0cm 0cm}]{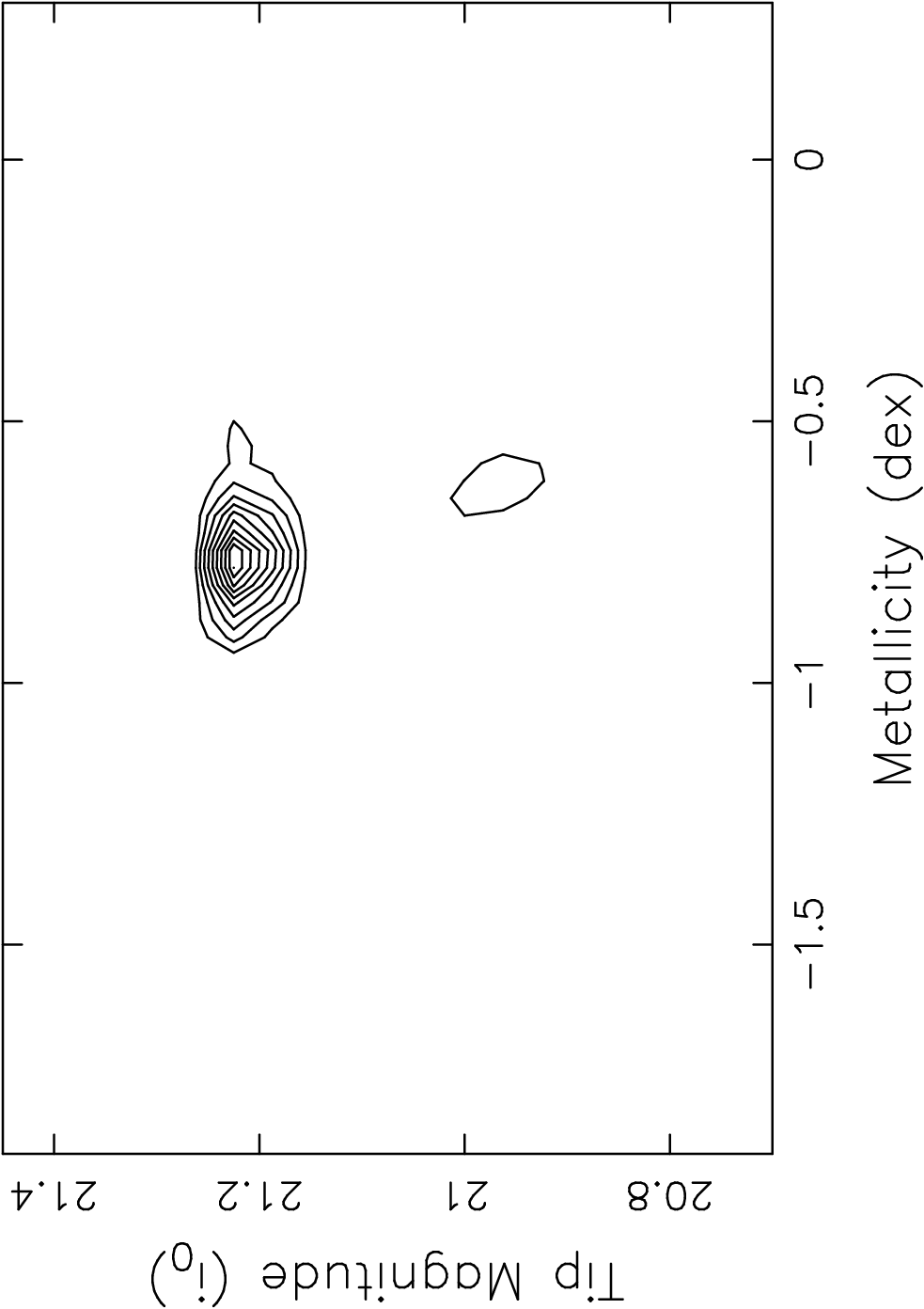}
\put(15,63){\small GSS9$\dagger$}
\end{overpic}

\end{array}$
\end{center}

\textbf{Appendix A Part II:} Contour plots illustrating the correlation between tip magnitude and metallicity for the fields listed in Table \ref{par_table1}. \\ Fields GSS8.5 through GSS10 are represented here. Plots denoted $\dagger$$^*$ are generated by sampling only the parameter values consistent with a restricted TRGB range. The full, unrestricted versions (for both Appendix A Parts I and II) are displayed here also and are denoted $\dagger$. \\ The restricted range plots are generated with the following limits: GSS8.5$\dagger$$^*$, $21.15 \le TRGB \le 21.30$; GSS9$\dagger$$^*$, $21.10 \le TRGB \le 21.30$. 

\label{AppAII}

\end{figure*}


\newpage


\begin{figure*}
\begin{center} 
$ \begin{array}{c}
\begin{overpic}[width = 0.23\textwidth,angle=-90,clip=true,trim={0cm 0cm 1.6cm 0cm}]{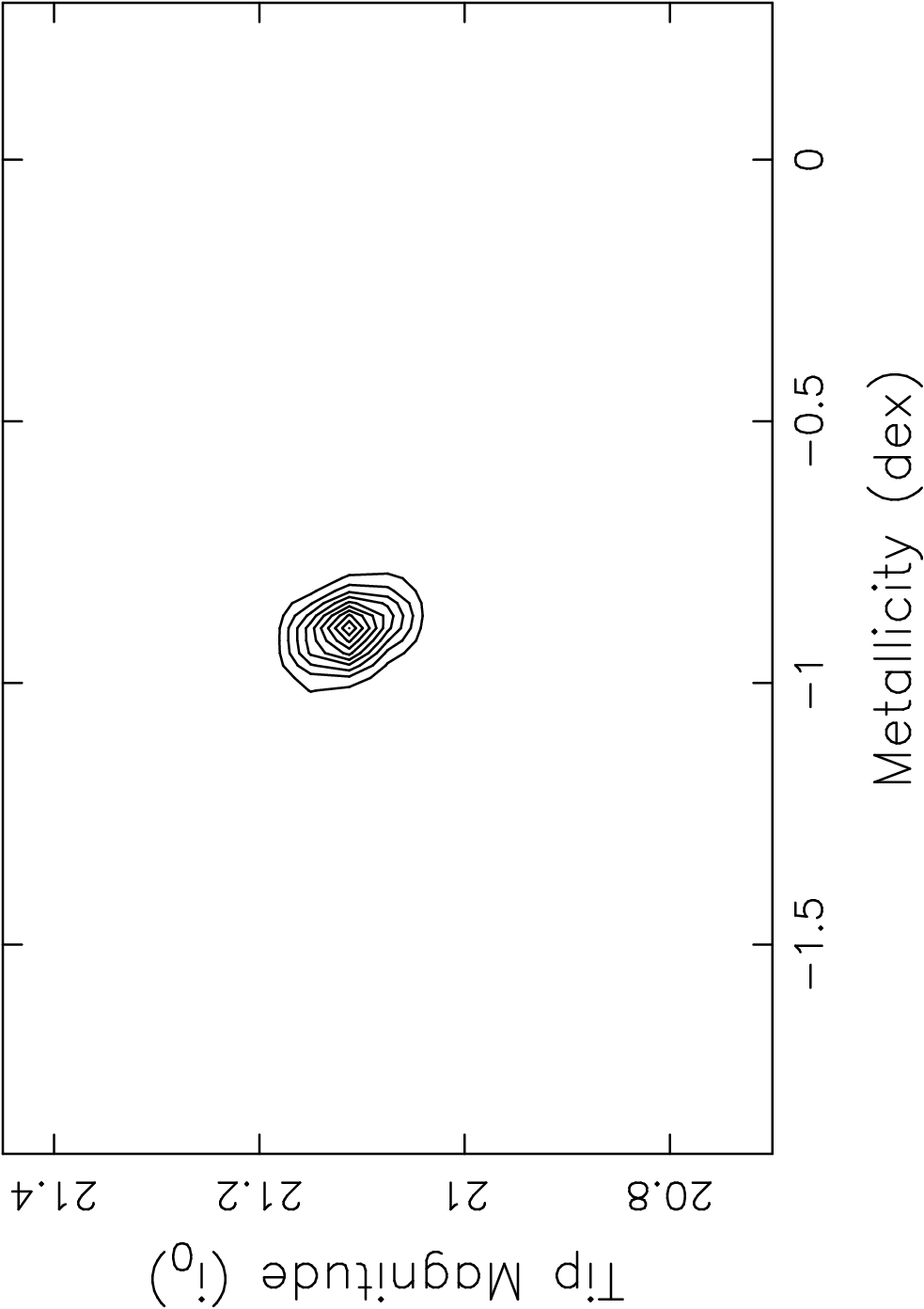}
\put(15,55){\small C1}
\end{overpic}
\begin{overpic}[width = 0.23\textwidth,angle=-90,clip=true,trim={0cm 1.1cm 1.6cm 0cm}]{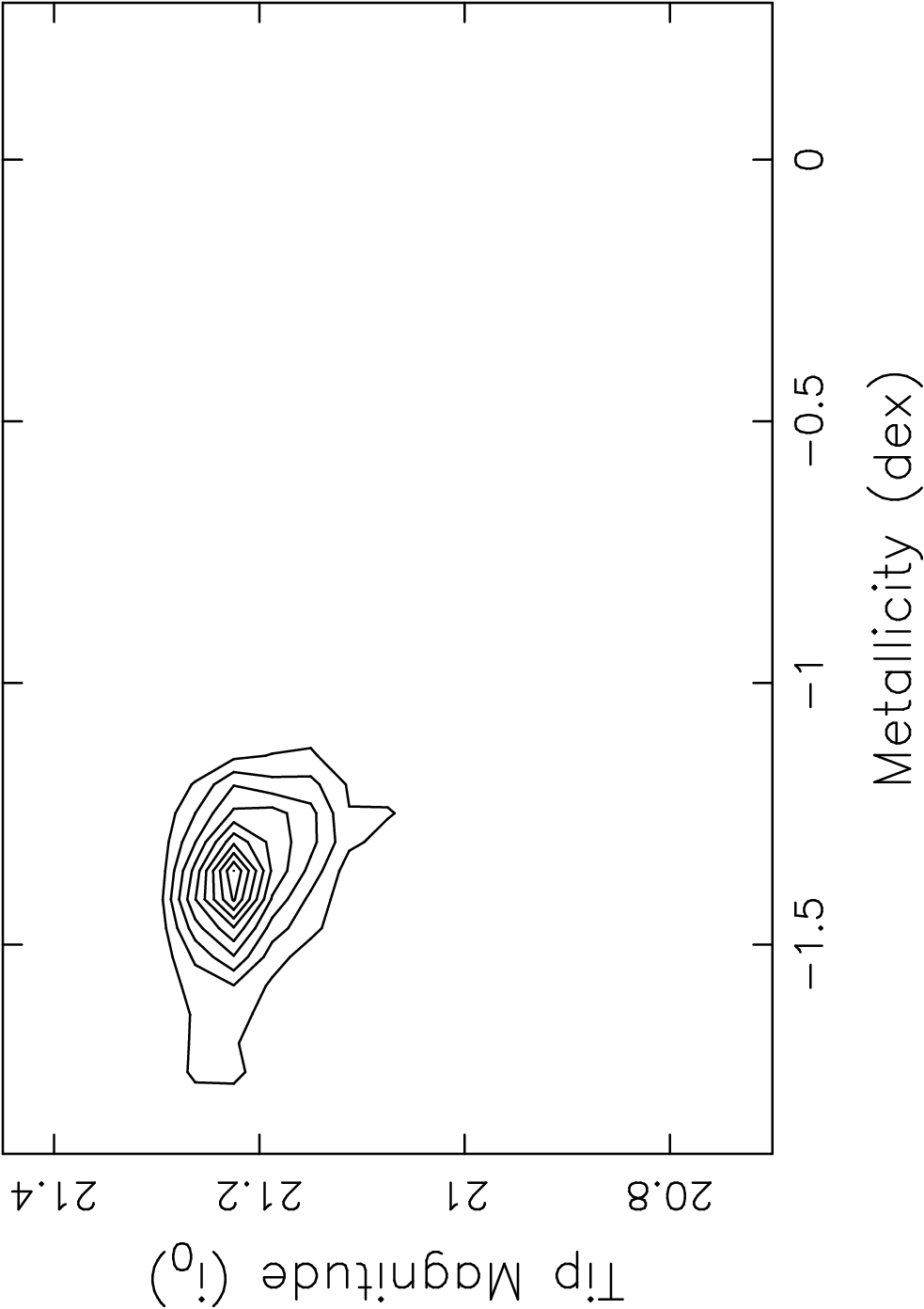}
\put(15,58){\small C2}
\end{overpic}
\\ \begin{overpic}[width = 0.23\textwidth,angle=-90,clip=true,trim={0cm 0cm 1.6cm 0cm}]{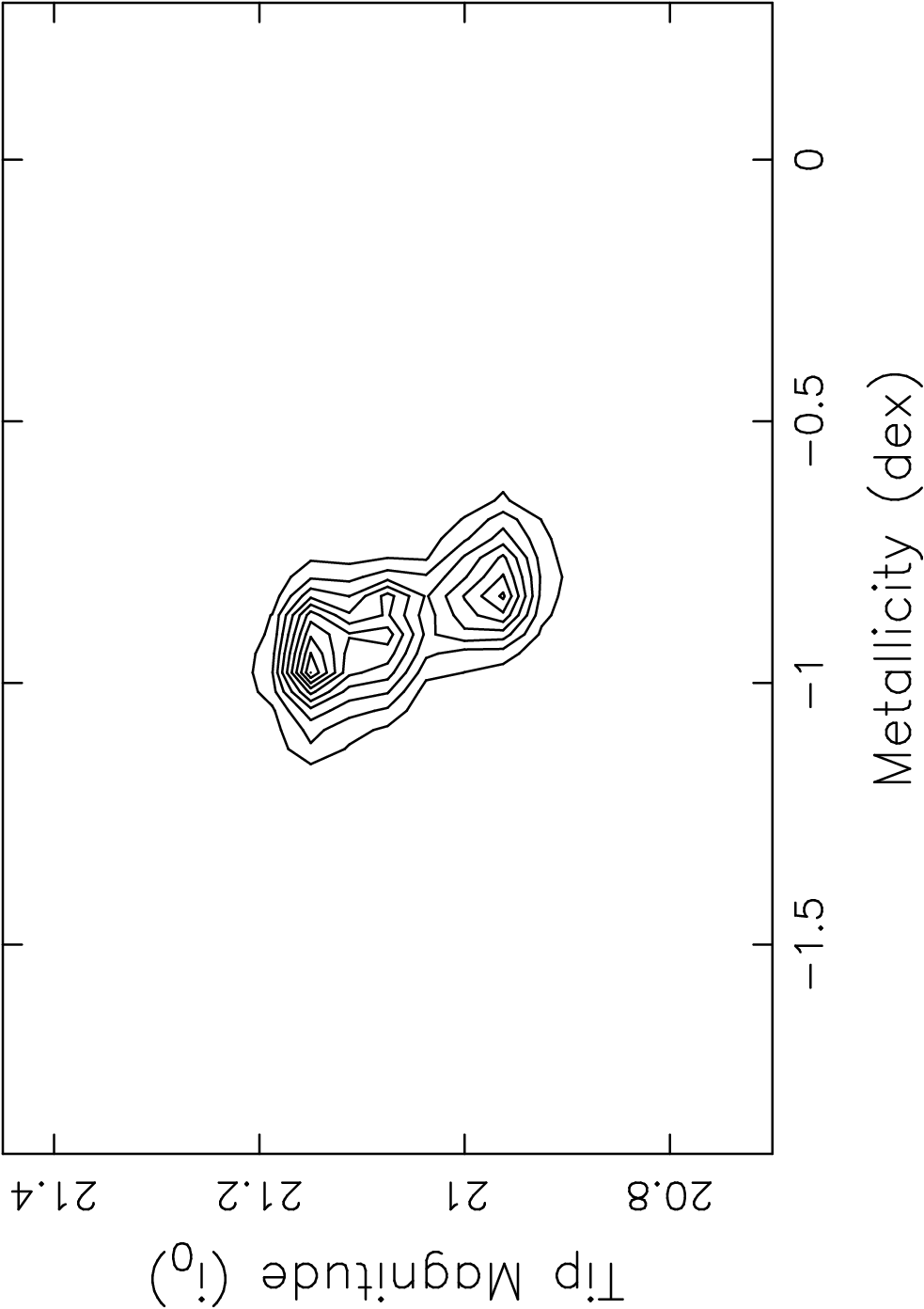}
\put(15,55){\small C3}
\end{overpic}
\begin{overpic}[width = 0.23\textwidth,angle=-90,clip=true,trim={0cm 1.1cm 1.6cm 0cm}]{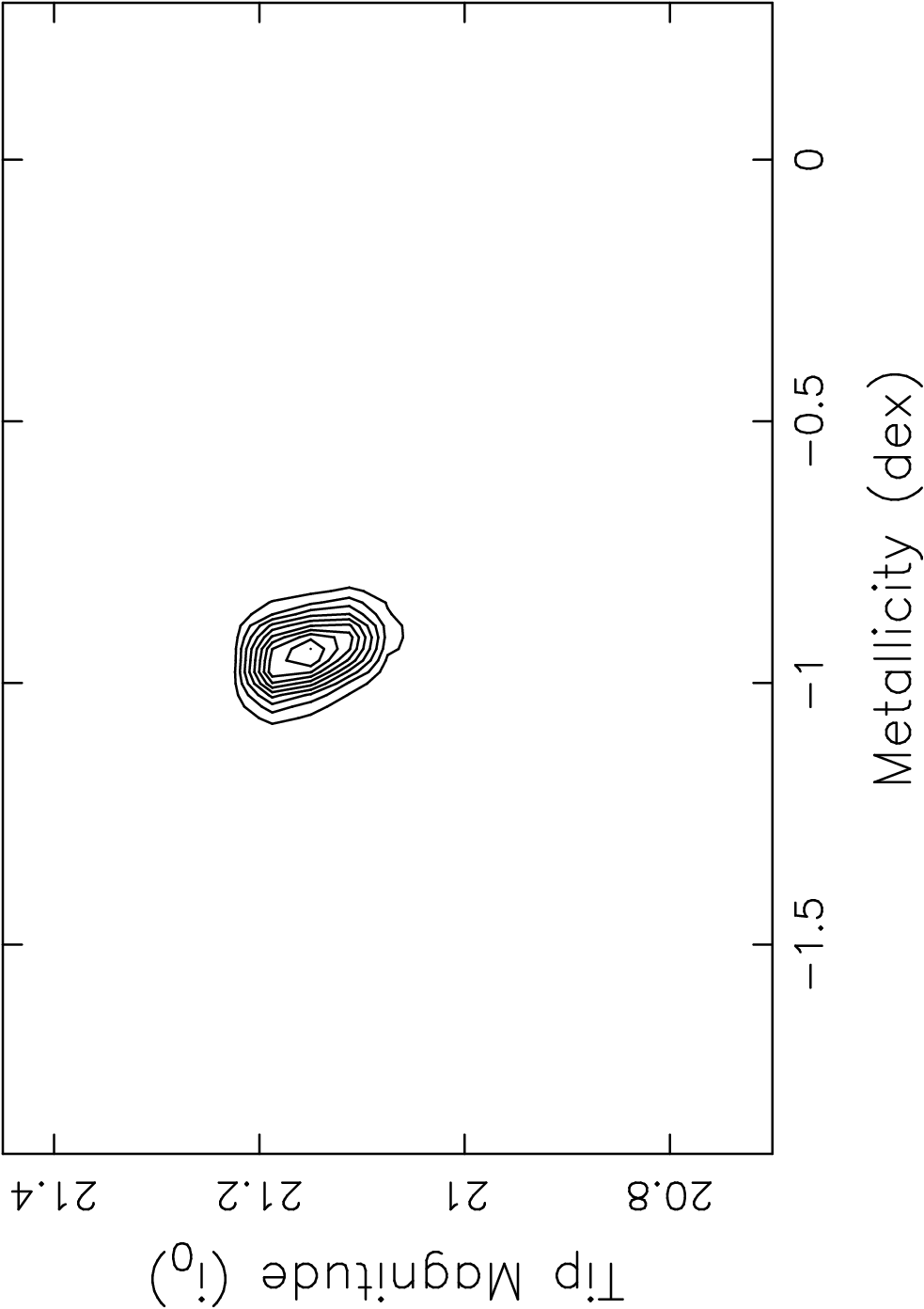}
\put(15,58){\small C4}
\end{overpic}
\\ \begin{overpic}[width = 0.23\textwidth,angle=-90,clip=true,trim={0cm 0cm 1.6cm 0cm}]{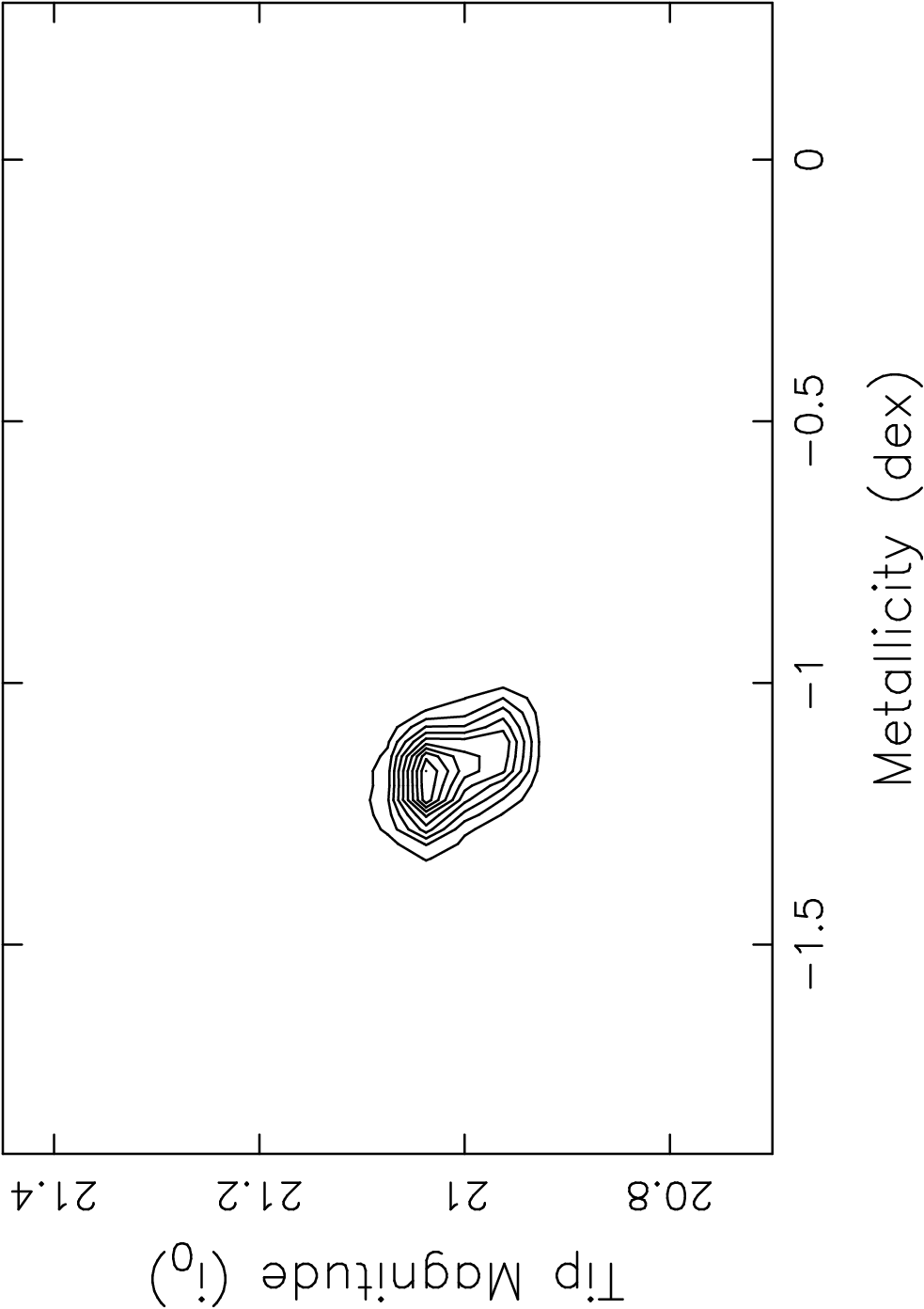}
\put(15,55){\small D1}
\end{overpic}
\begin{overpic}[width = 0.23\textwidth,angle=-90,clip=true,trim={0cm 1.1cm 1.6cm 0cm}]{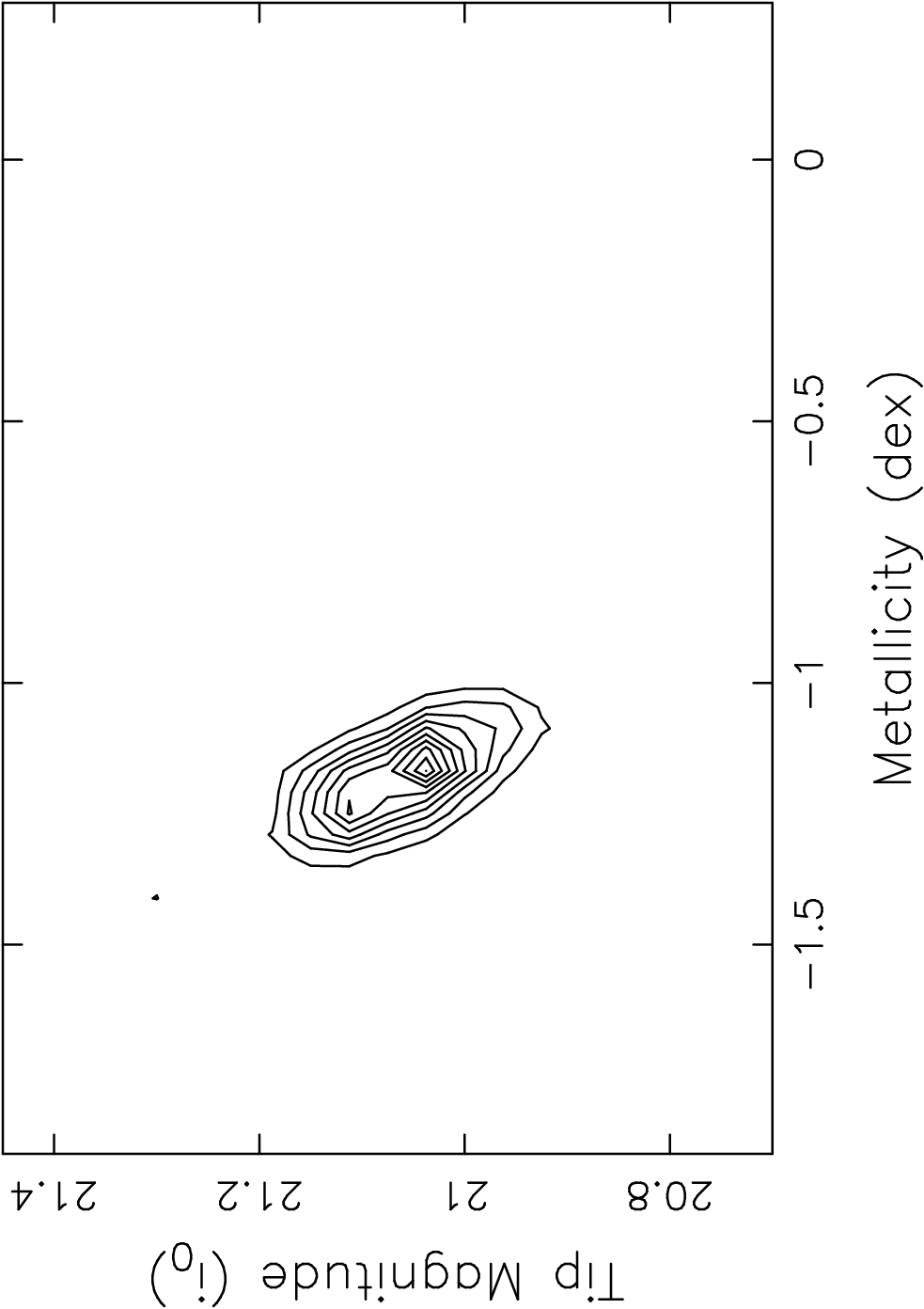}
\put(15,58){\small D2}
\end{overpic}
\begin{overpic}[width = 0.23\textwidth,angle=-90,clip=true,trim={0cm 1.1cm 1.6cm 0cm}]{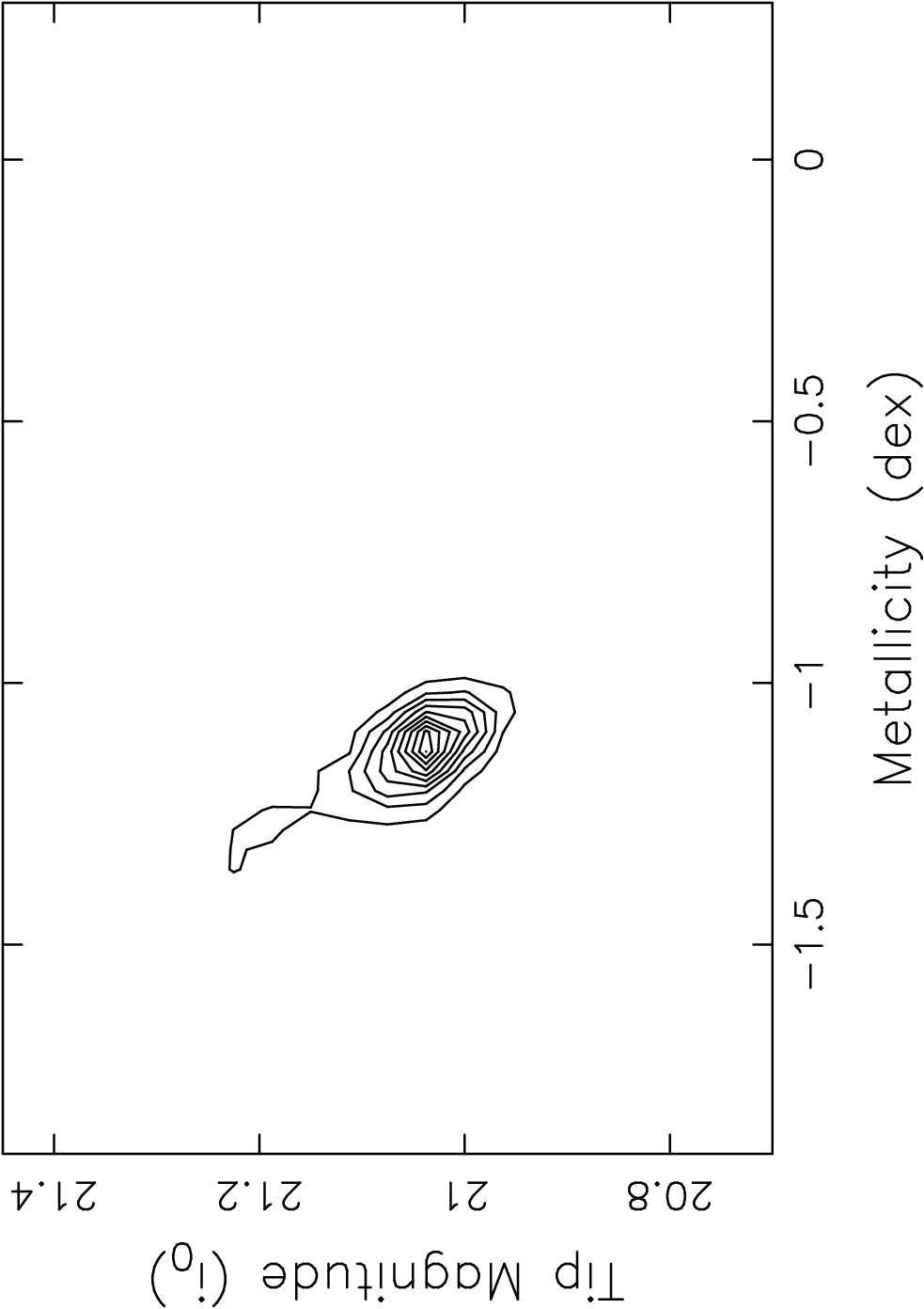}
\put(15,58){\small D3}
\end{overpic}
\\ \begin{overpic}[width = 0.252\textwidth,angle=-90,clip=true,trim={0cm 0cm 0cm 0cm}]{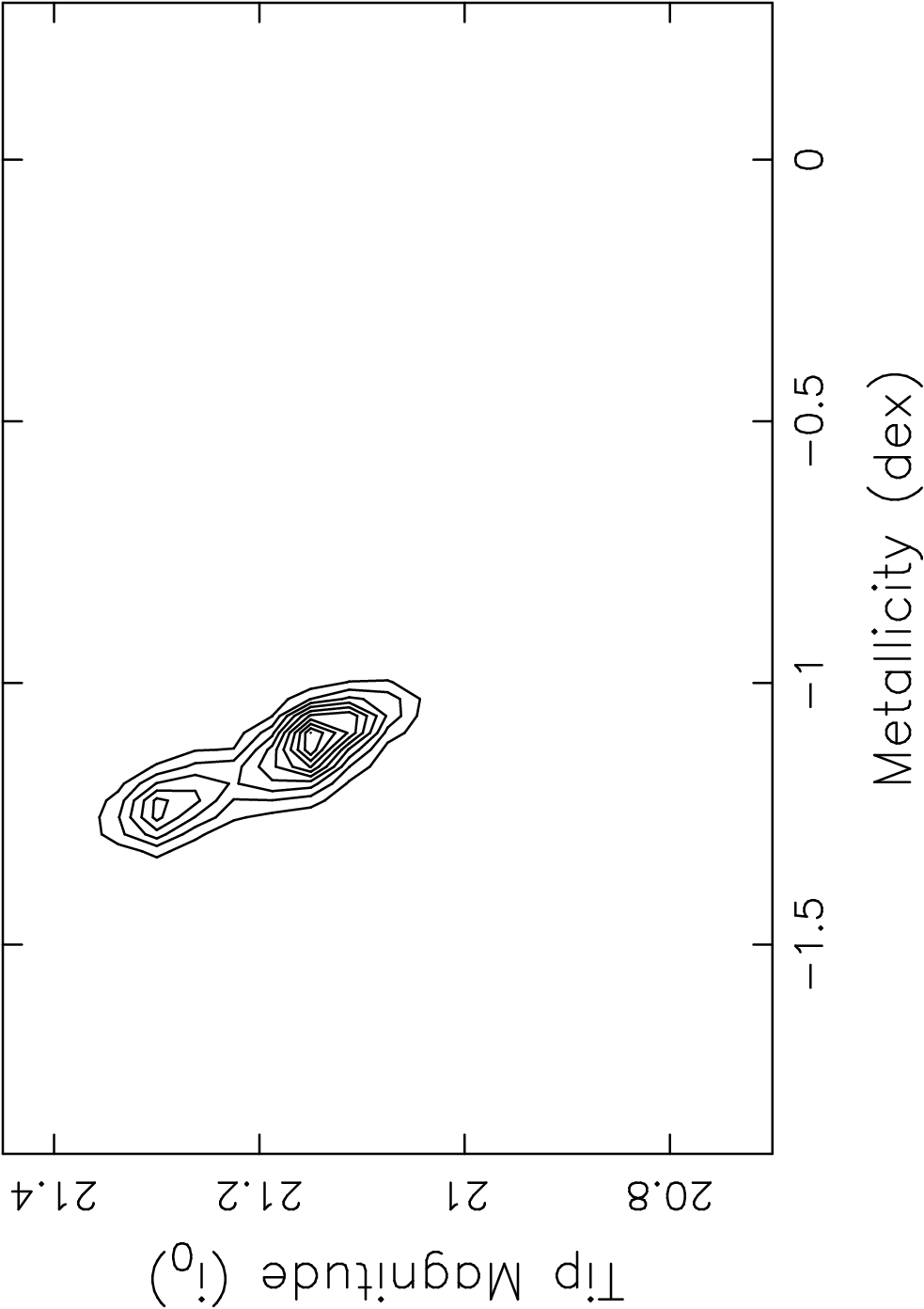}
\put(15,61){\small D4}
\end{overpic}
\begin{overpic}[width = 0.252\textwidth,angle=-90,clip=true,trim={0cm 1.1cm 0cm 0cm}]{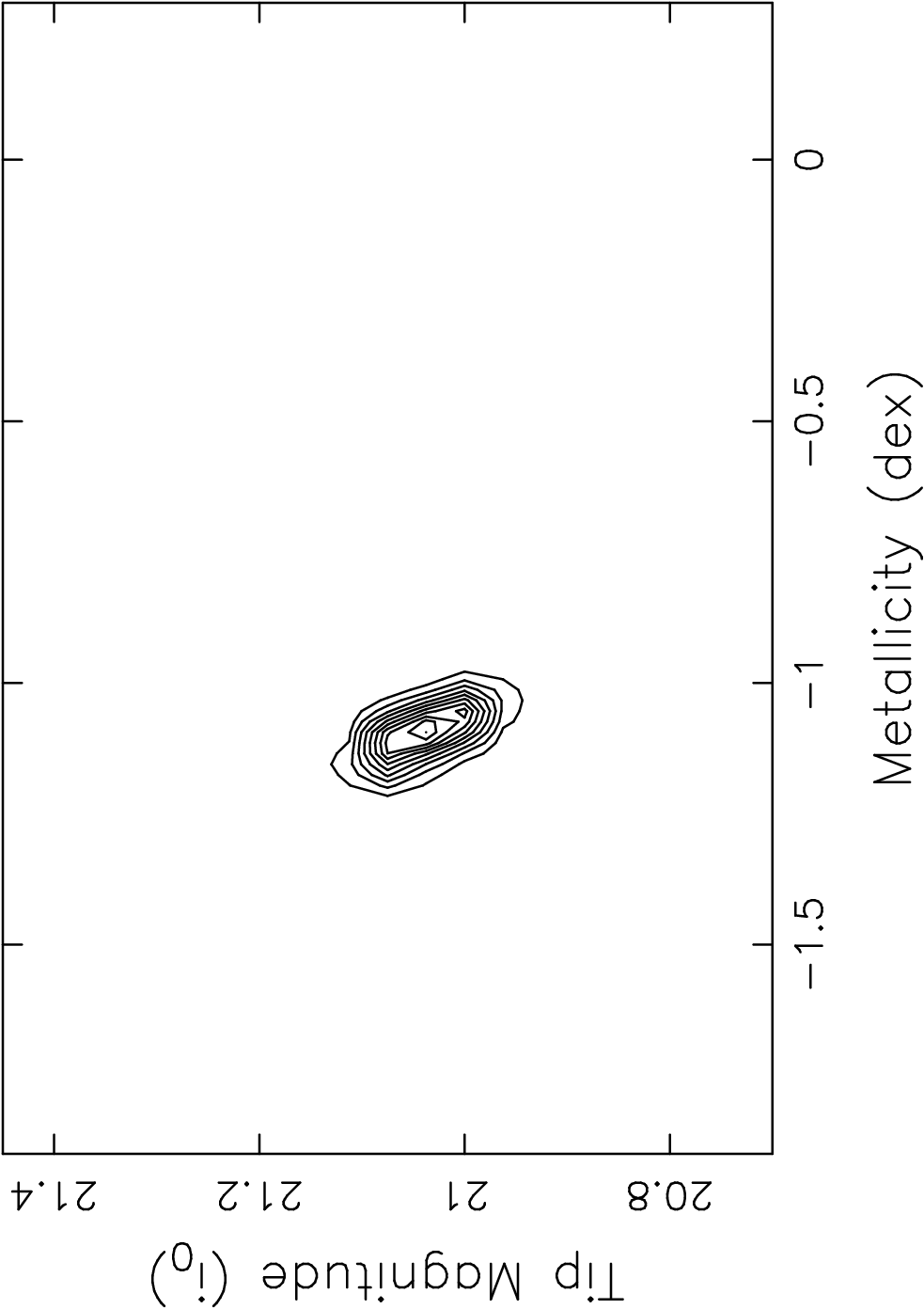}
\put(15,64){\small D5}
\end{overpic}

\end{array}$
\end{center}

\textbf{Appendix A Part III:} Contour plots illustrating the correlation between tip magnitude and metallicity for the fields listed in Table \ref{par_table2}. \\ Fields from Streams C \& D are represented here.

\label{AppAIII}

\end{figure*}

\newpage


\begin{figure*}
\begin{center} 
$ \begin{array}{c}
\begin{overpic}[width = 0.23\textwidth,angle=-90,clip=true,trim={0cm 0cm 1.6cm 0cm}]{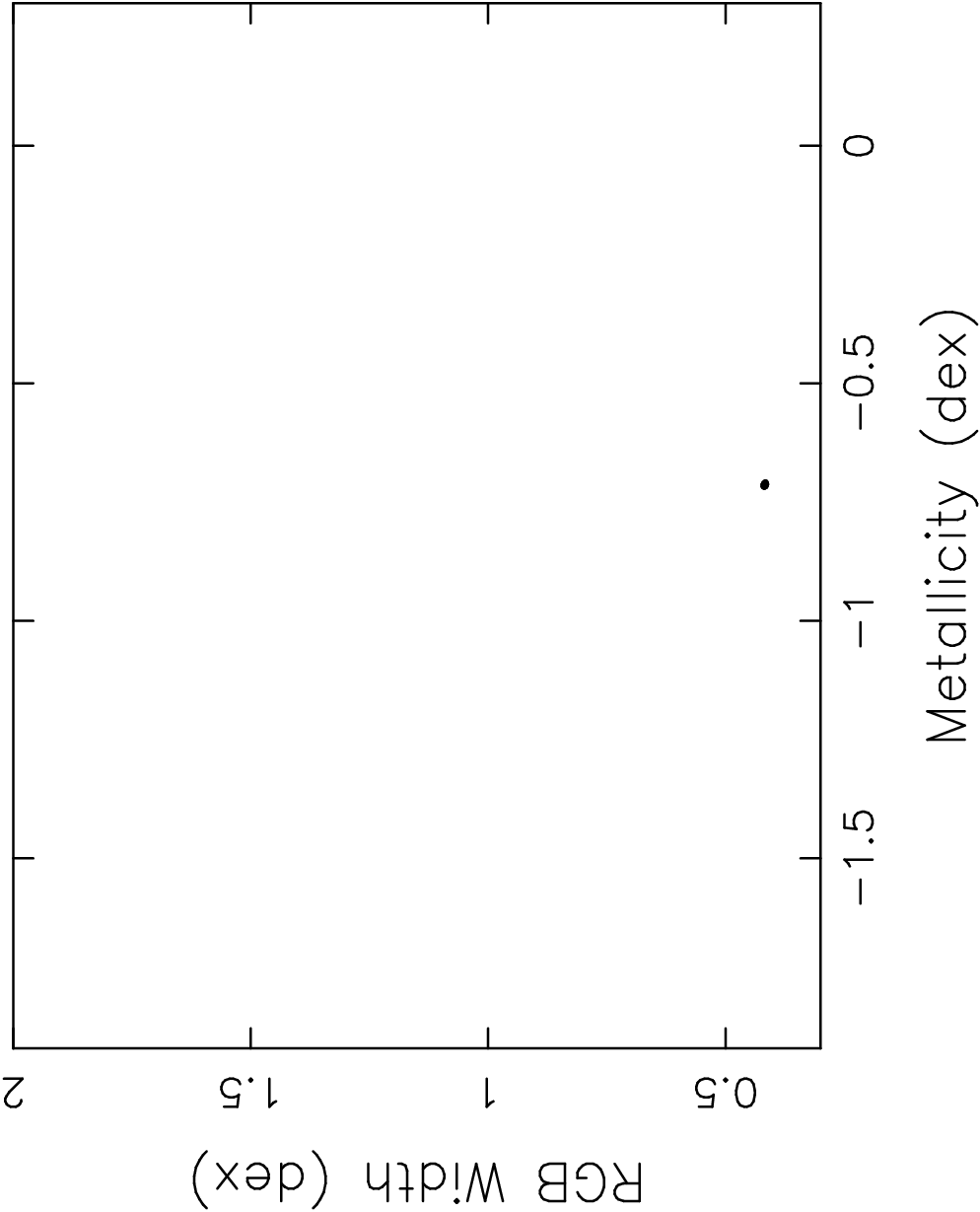}
\put(20,62){\small GSS1}
\end{overpic}
\begin{overpic}[width = 0.23\textwidth,angle=-90,clip=true,trim={0cm 1.1cm 1.6cm 0cm}]{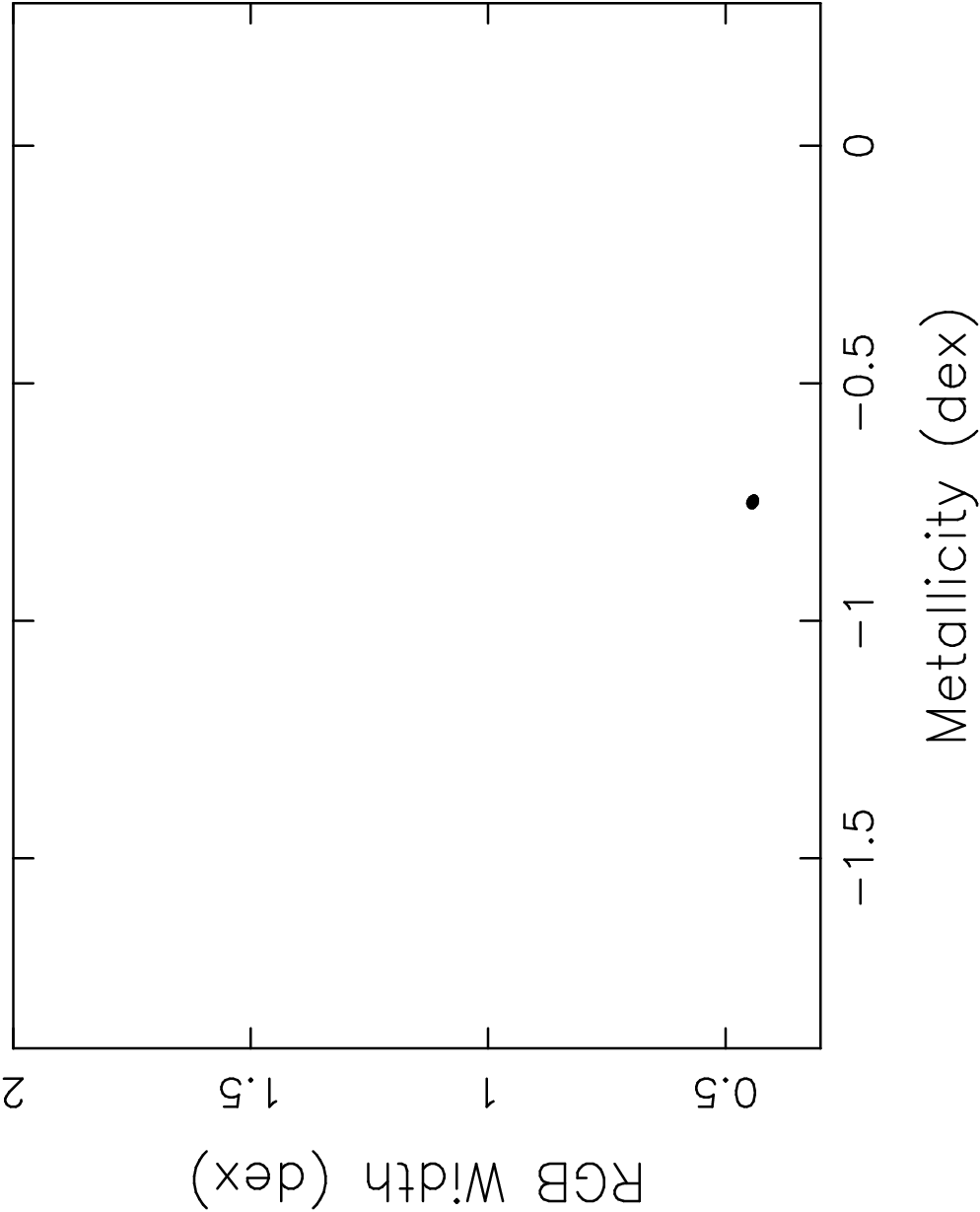}
\put(15,65){\small GSS1.5}
\end{overpic}
\begin{overpic}[width = 0.23\textwidth,angle=-90,clip=true,trim={0cm 1.1cm 1.6cm 0cm}]{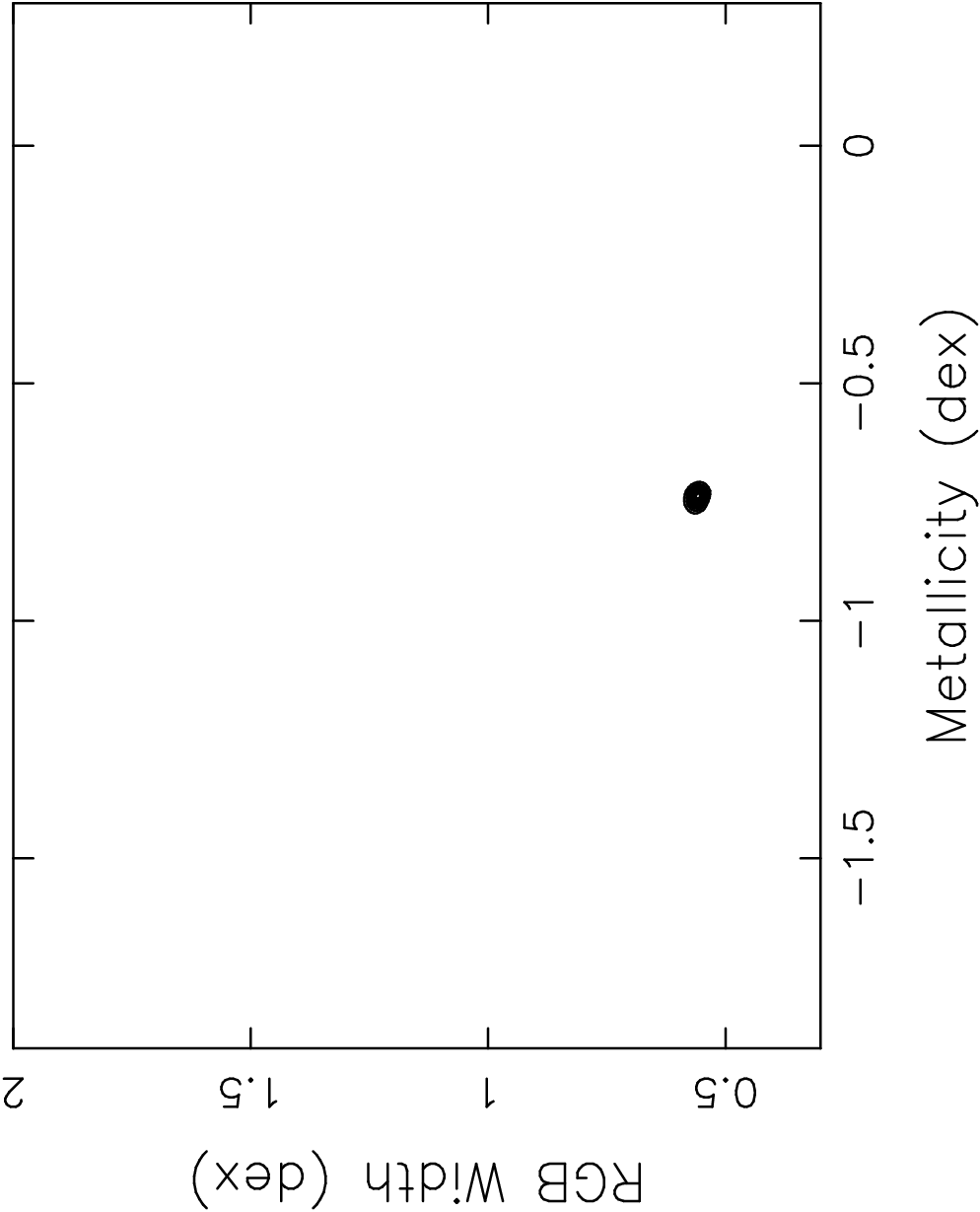}
\put(15,65){\small GSS2}
\end{overpic}
\\ \begin{overpic}[width = 0.23\textwidth,angle=-90,clip=true,trim={0cm 0cm 1.6cm 0cm}]{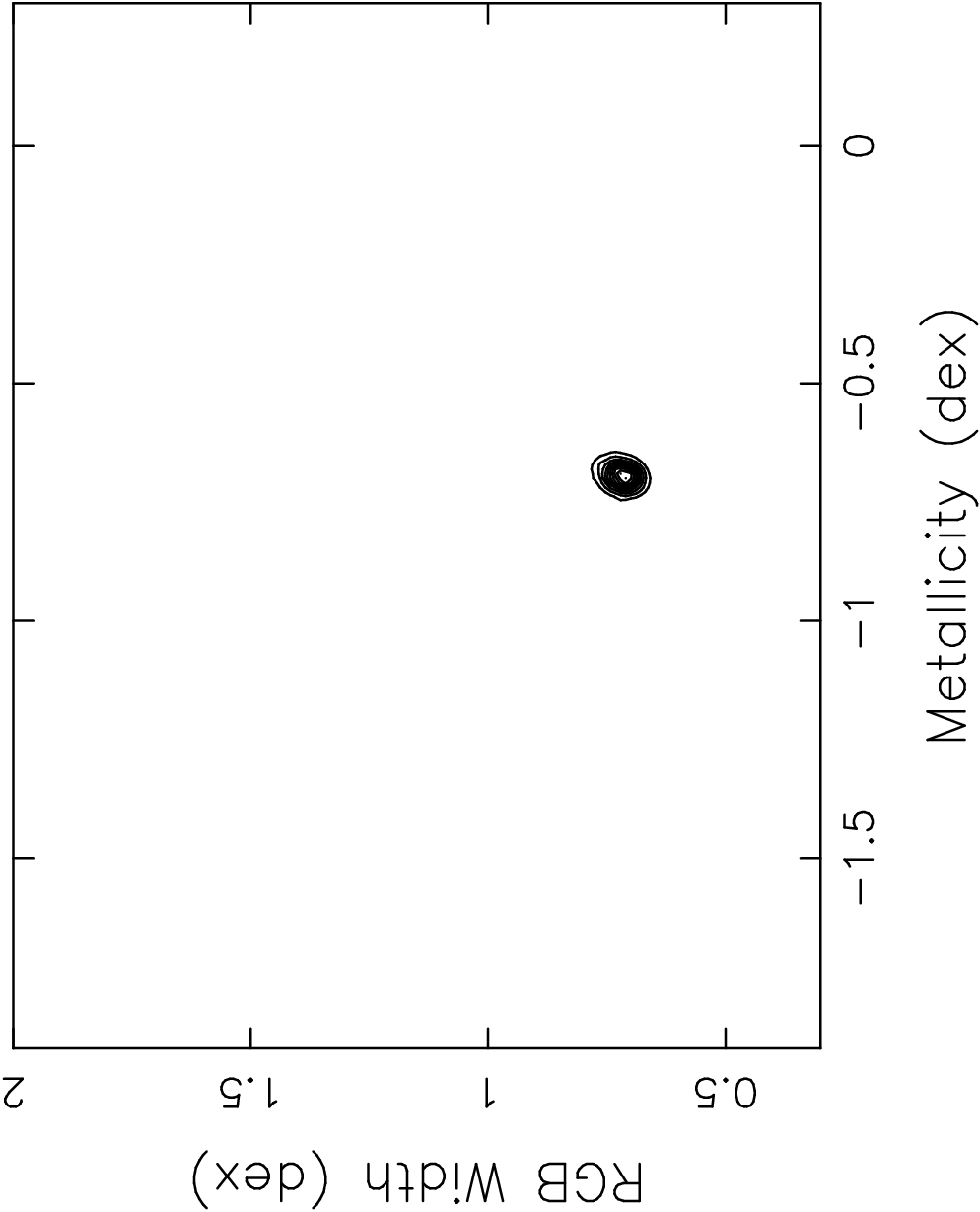}
\put(20,62){\small GSS2.5}
\end{overpic}
\begin{overpic}[width = 0.23\textwidth,angle=-90,clip=true,trim={0cm 1.1cm 1.6cm 0cm}]{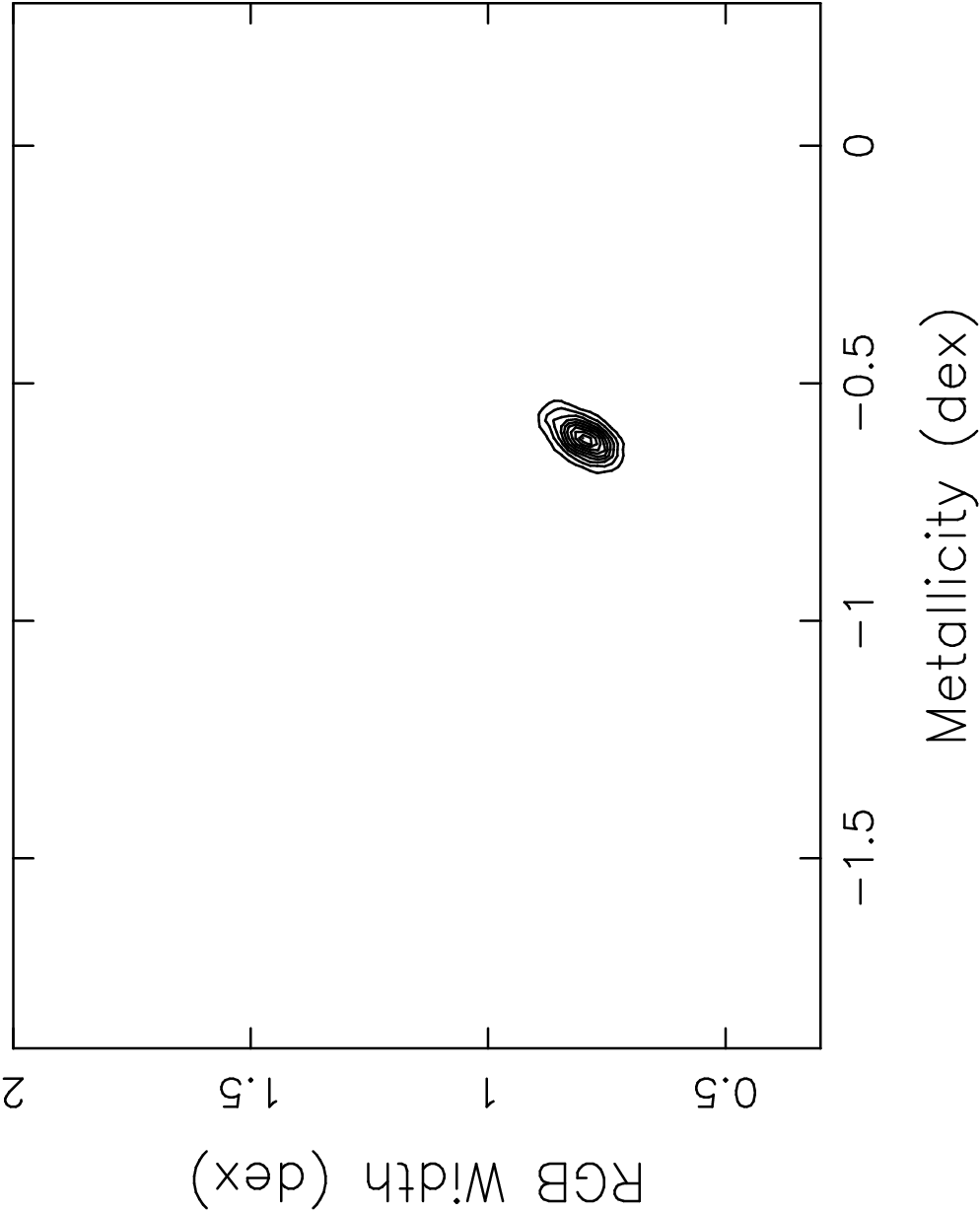}
\put(15,65){\small GSS3}
\end{overpic}
\begin{overpic}[width = 0.23\textwidth,angle=-90,clip=true,trim={0cm 1.1cm 1.6cm 0cm}]{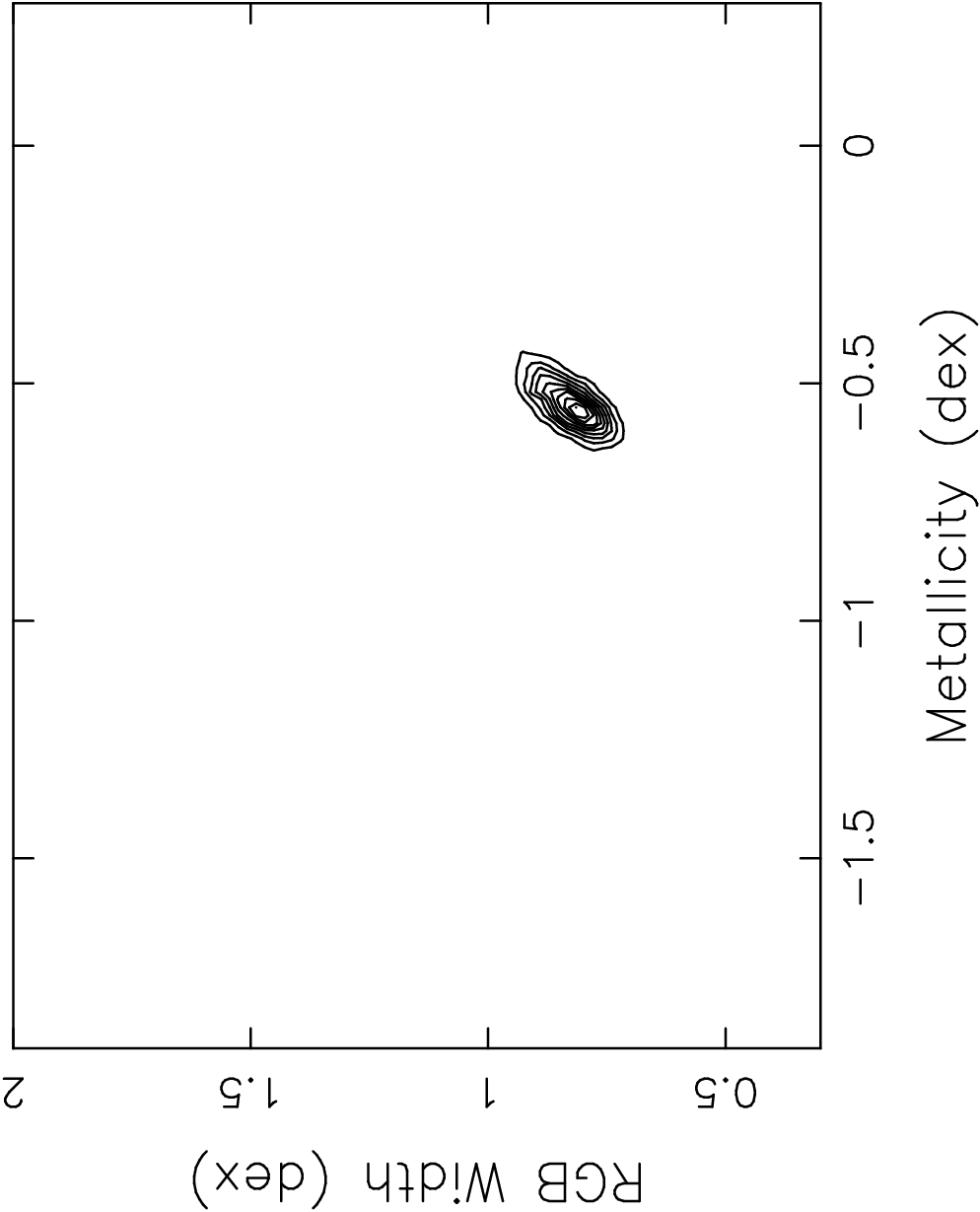}
\put(15,65){\small GSS3.5}
\end{overpic}
\\ \begin{overpic}[width = 0.23\textwidth,angle=-90,clip=true,trim={0cm 0cm 1.6cm 0cm}]{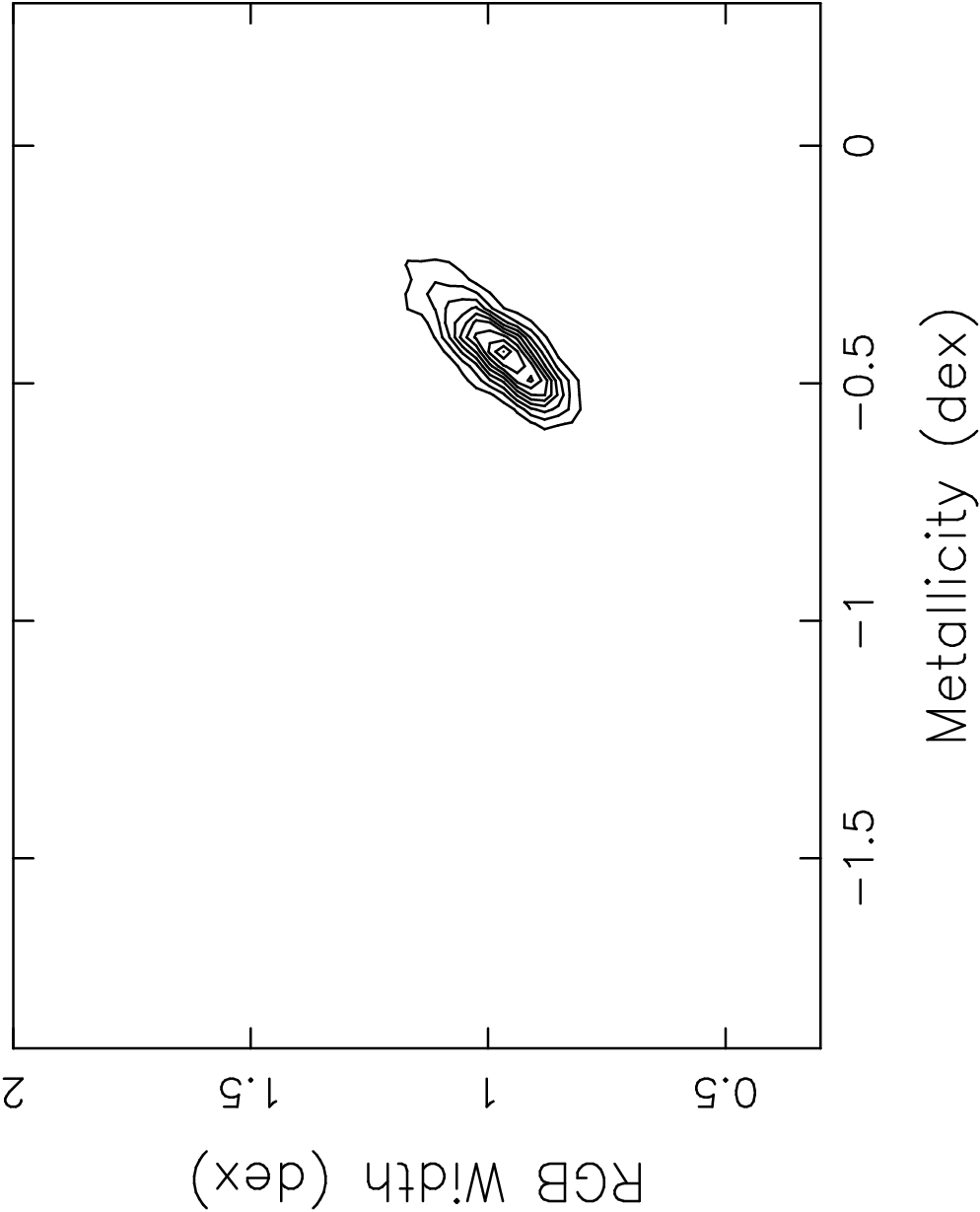}
\put(20,62){\small GSS4}
\end{overpic}
\begin{overpic}[width = 0.23\textwidth,angle=-90,clip=true,trim={0cm 1.1cm 1.6cm 0cm}]{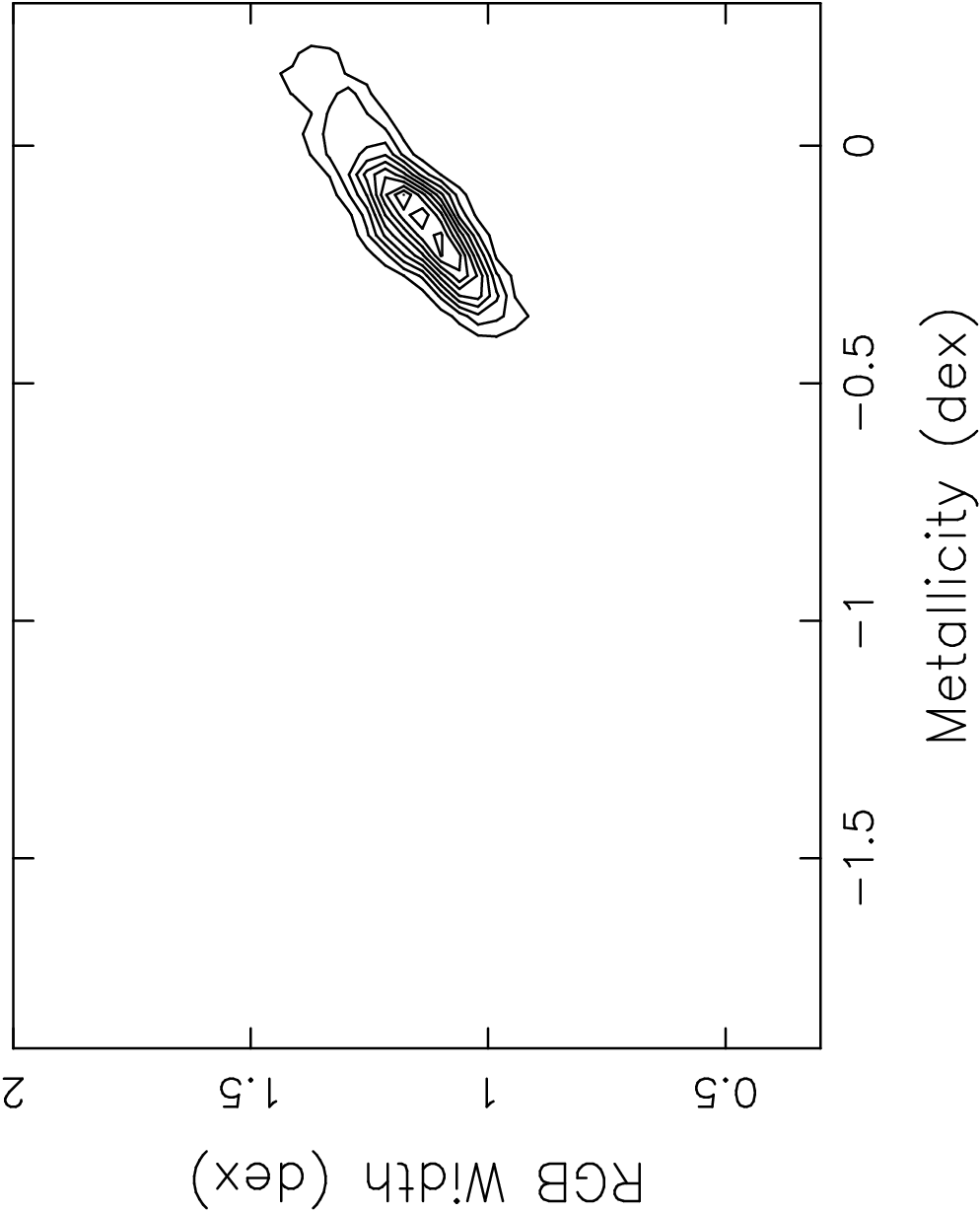}
\put(15,65){\small GSS4.5}
\end{overpic}
\begin{overpic}[width = 0.23\textwidth,angle=-90,clip=true,trim={0cm 1.1cm 1.6cm 0cm}]{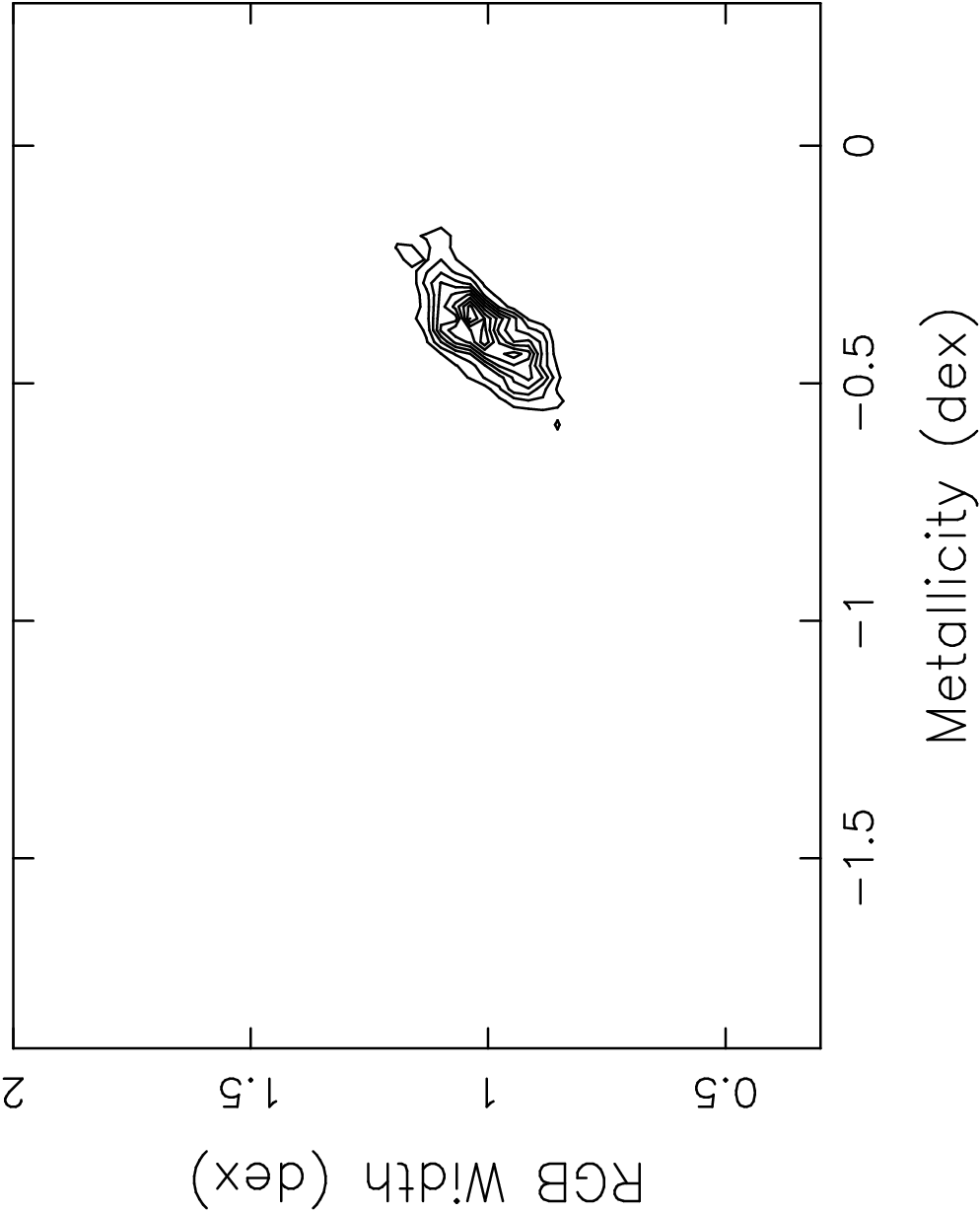}
\put(15,65){\small GSS5$\dagger$$^*$}
\end{overpic}
\\ \begin{overpic}[width = 0.23\textwidth,angle=-90,clip=true,trim={0cm 0cm 1.6cm 0cm}]{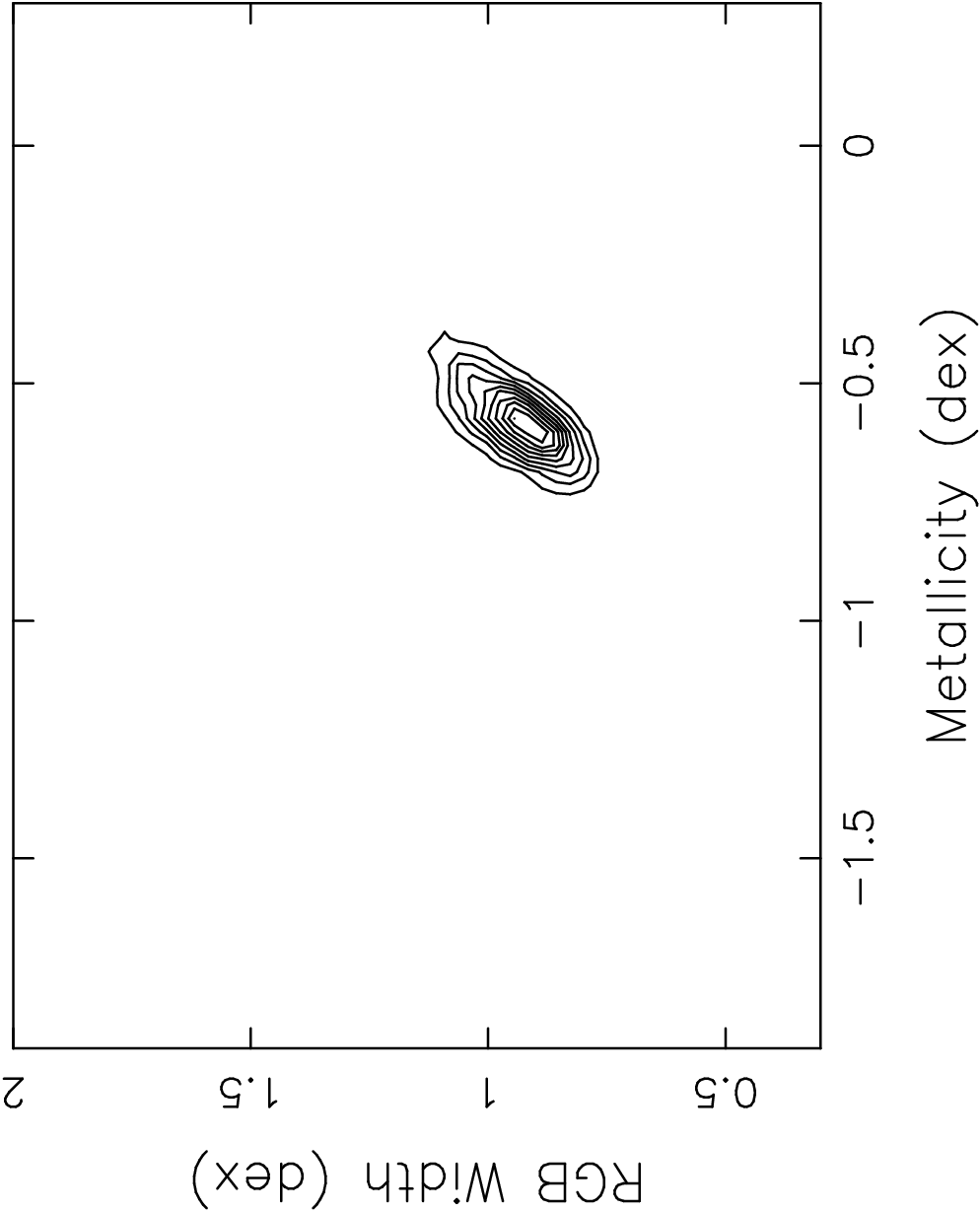}
\put(20,62){\small GSS5.5}
\end{overpic}
\begin{overpic}[width = 0.23\textwidth,angle=-90,clip=true,trim={0cm 1.1cm 1.6cm 0cm}]{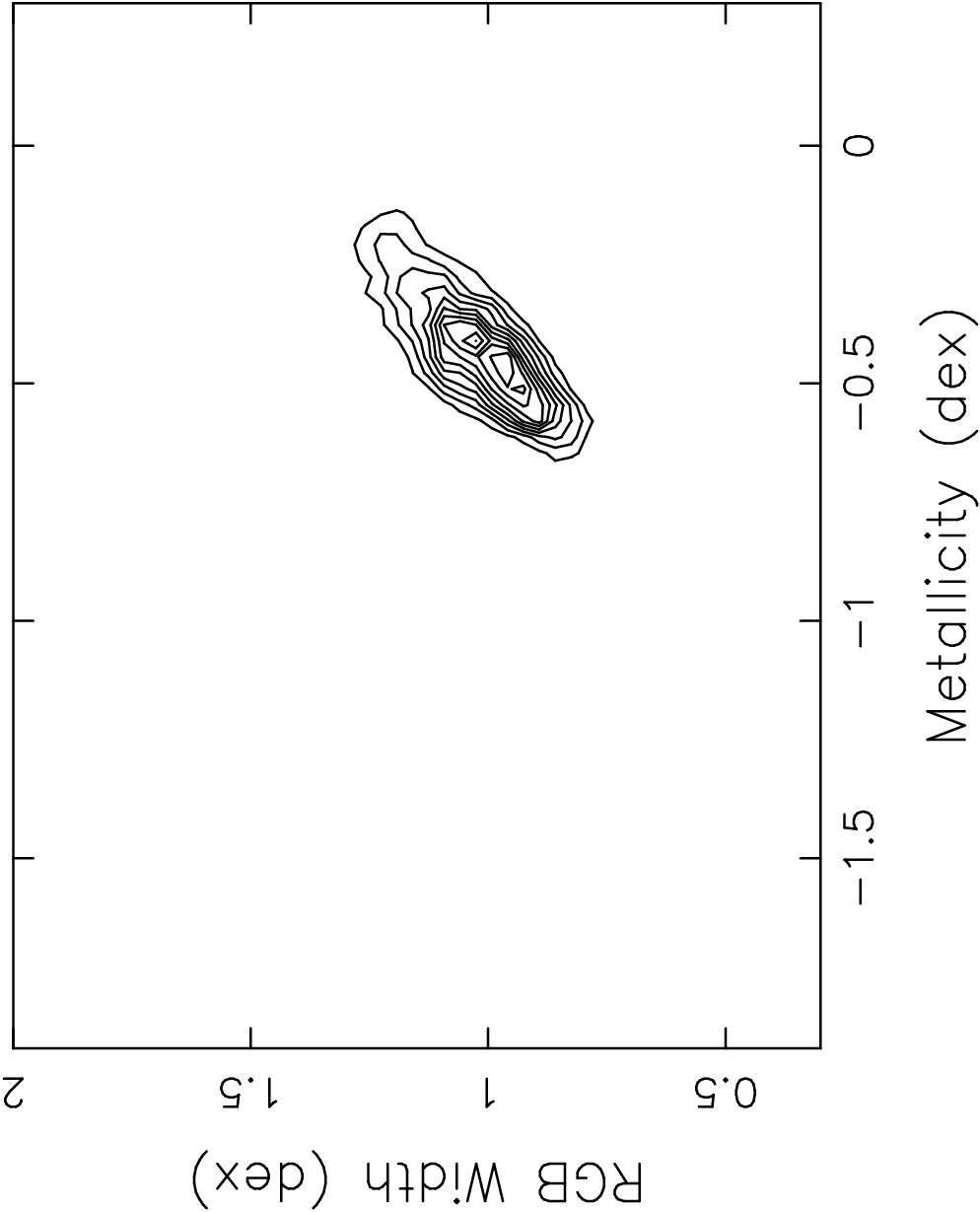}
\put(15,65){\small GSS6}
\end{overpic}
\begin{overpic}[width = 0.23\textwidth,angle=-90,clip=true,trim={0cm 1.1cm 1.6cm 0cm}]{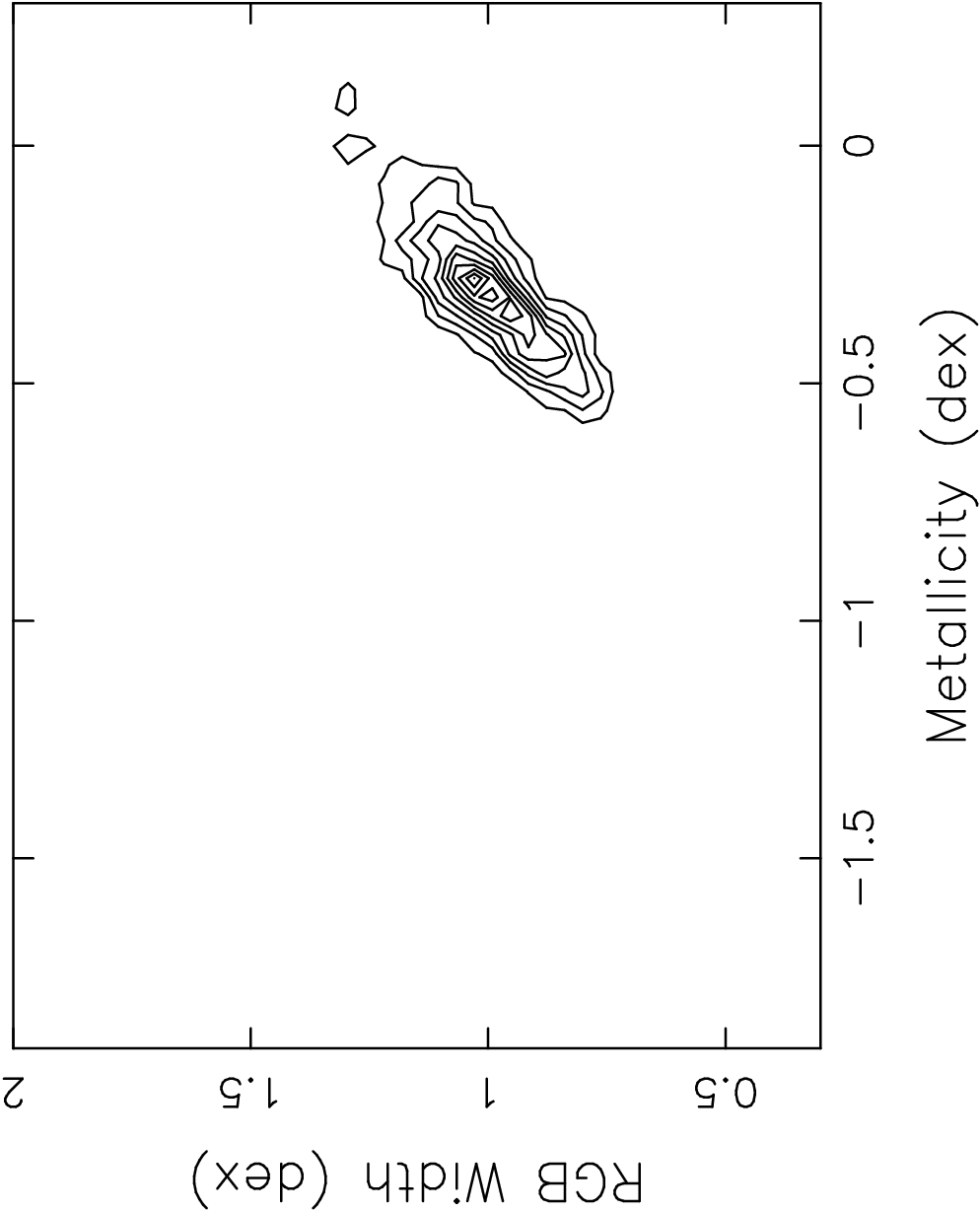}
\put(15,65){\small GSS6.5$\dagger$$^*$}
\end{overpic}
\\ \begin{overpic}[width = 0.258\textwidth,angle=-90,clip=true,trim={0cm 0cm 0cm 0cm}]{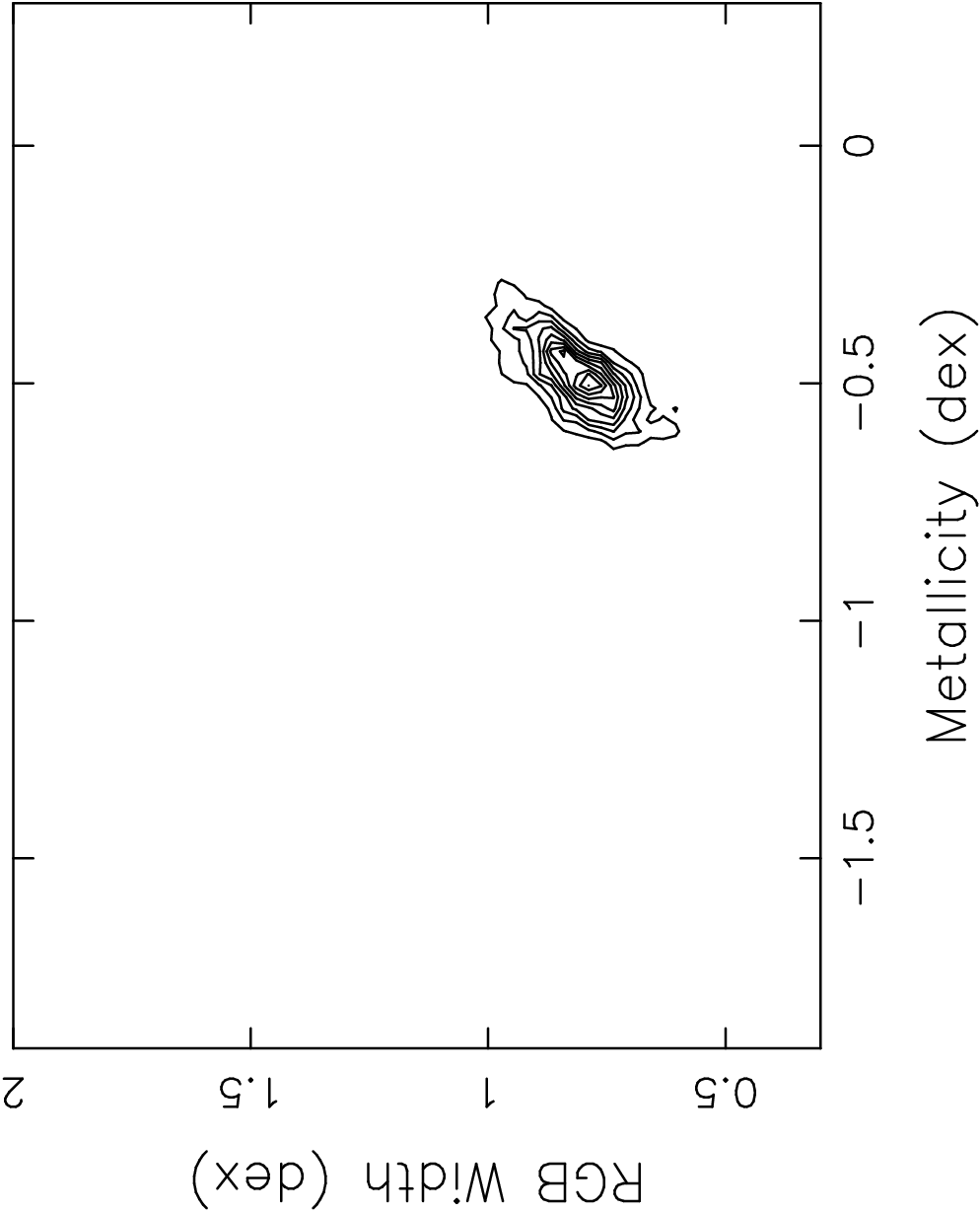}
\put(20,70){\small GSS7$\dagger$$^*$}
\end{overpic}
\begin{overpic}[width = 0.258\textwidth,angle=-90,clip=true,trim={0cm 1.1cm 0cm 0cm}]{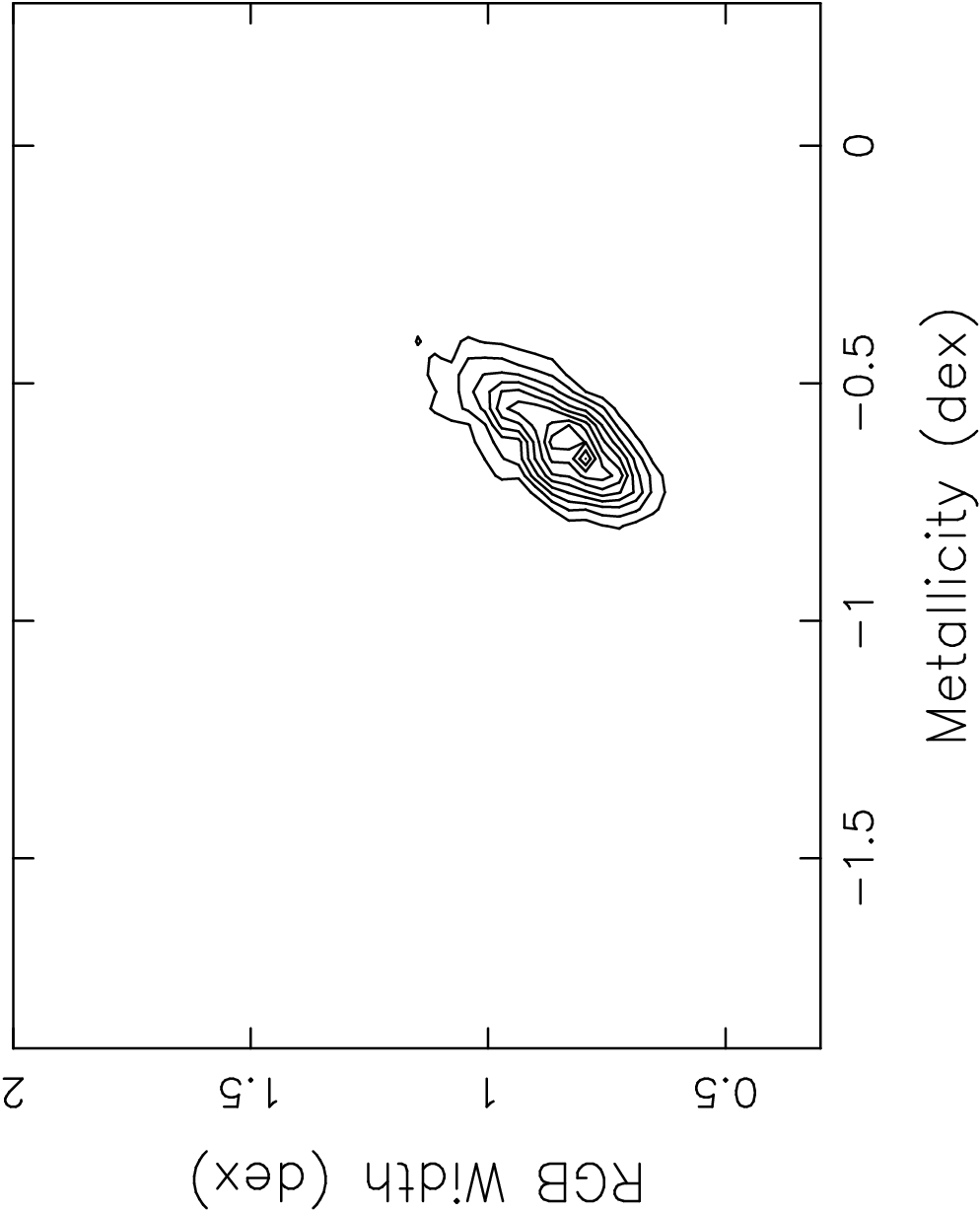}
\put(15,73){\small GSS7.5}
\end{overpic}
\begin{overpic}[width = 0.258\textwidth,angle=-90,clip=true,trim={0cm 1.1cm 0cm 0cm}]{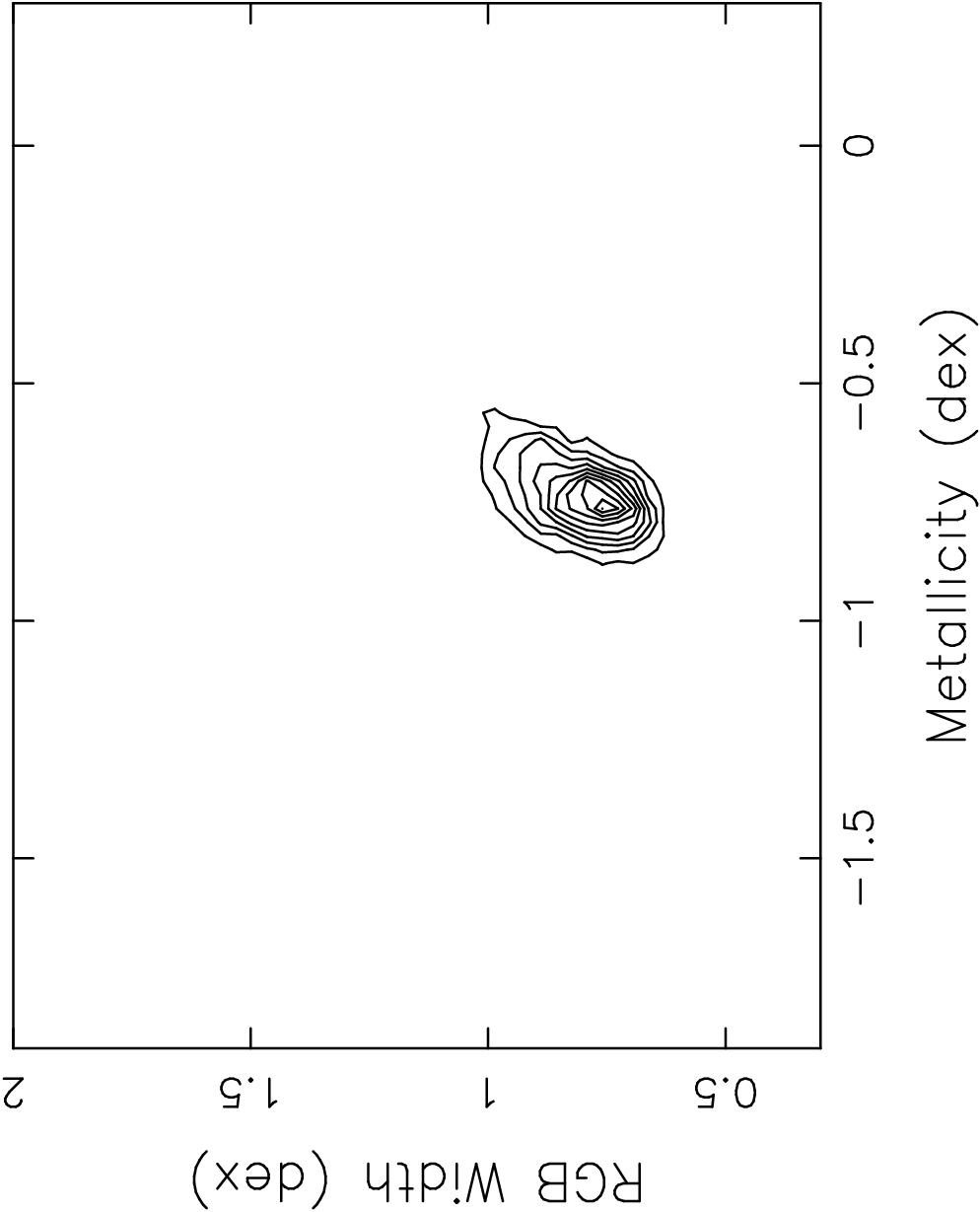}
\put(15,74){\small GSS8}
\end{overpic}

\end{array}$
\end{center}

\textbf{Appendix B Part I:} Contour plots illustrating the correlation between RGB width and metallicity for the fields listed in Table \ref{par_table1}. \\ Fields GSS1 through GSS8 are represented here. \\ Plots denoted $\dagger$$^*$ are generated by sampling only the parameter values consistent with a restricted TRGB range. The full, unrestricted versions denoted $\dagger$ are shown on the next page. The restricted ranges are: GSS5$\dagger$$^*$, $21.08 \le TRGB \le 21.18$; GSS6.5$\dagger$$^*$, $21.08 \le TRGB \le 21.30$; GSS7$\dagger$$^*$, $21.08 \le TRGB \le 21.30$. 

\label{AppBI}

\end{figure*}


\newpage


\begin{figure*}
\begin{center} 
$ \begin{array}{c}
\begin{overpic}[width = 0.23\textwidth,angle=-90,clip=true,trim={0cm 0cm 1.6cm 0cm}]{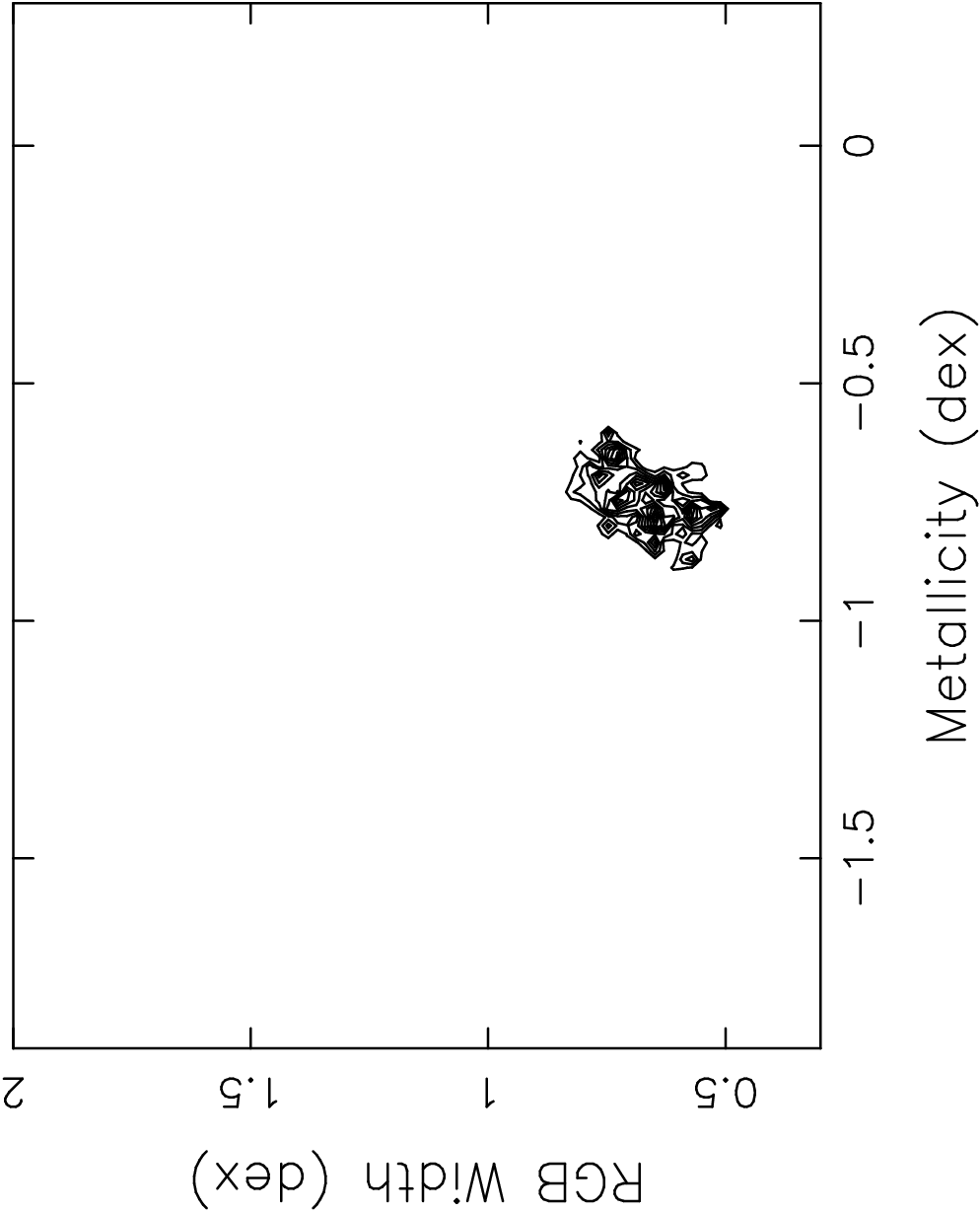}
\put(20,60){\small GSS8.5$\dagger$$^*$}
\end{overpic}
\begin{overpic}[width = 0.23\textwidth,angle=-90,clip=true,trim={0cm 1.1cm 1.6cm 0cm}]{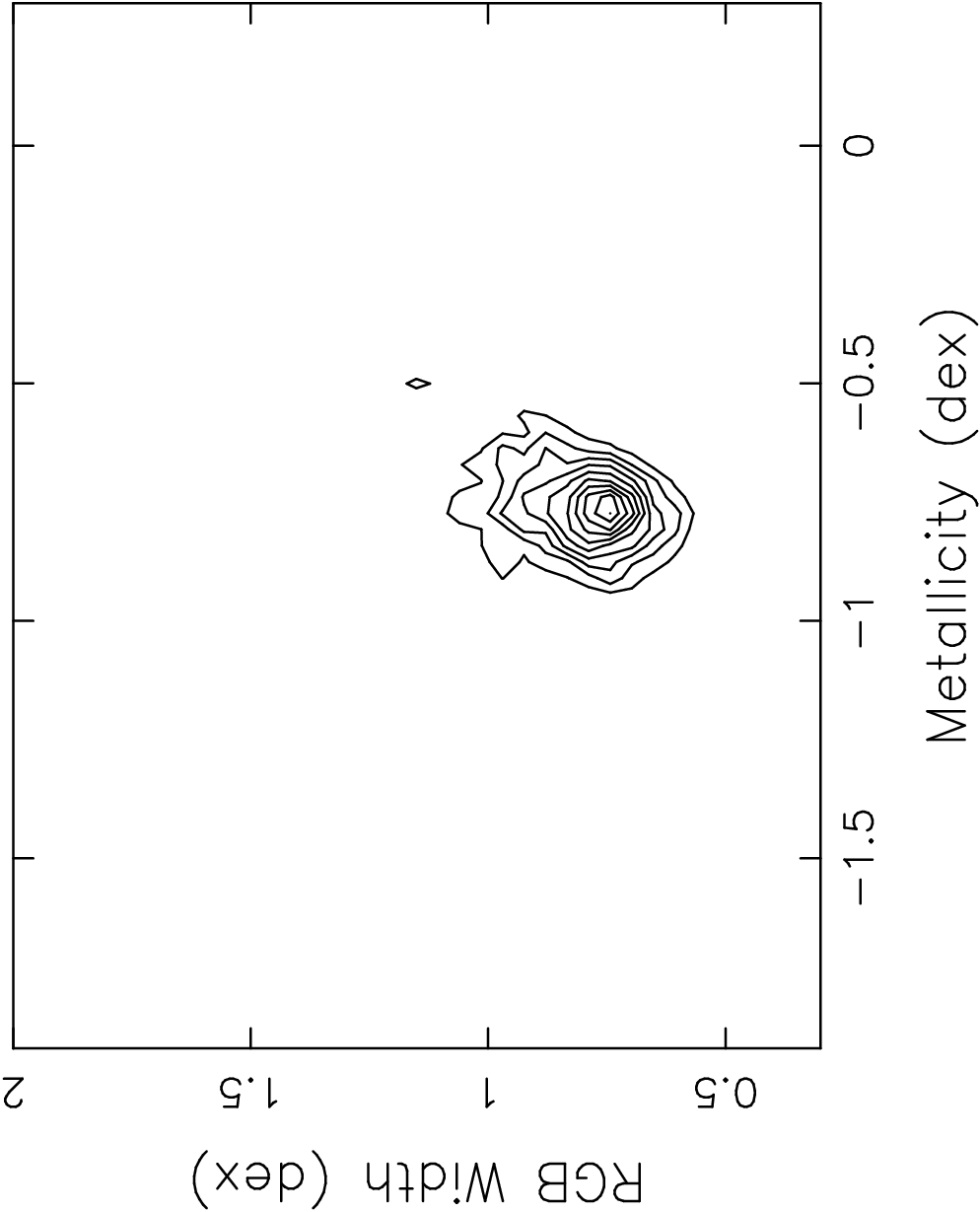}
\put(15,64){\small GSS9$\dagger$$^*$}
\end{overpic}
\begin{overpic}[width = 0.23\textwidth,angle=-90,clip=true,trim={0cm 1.1cm 1.6cm 0cm}]{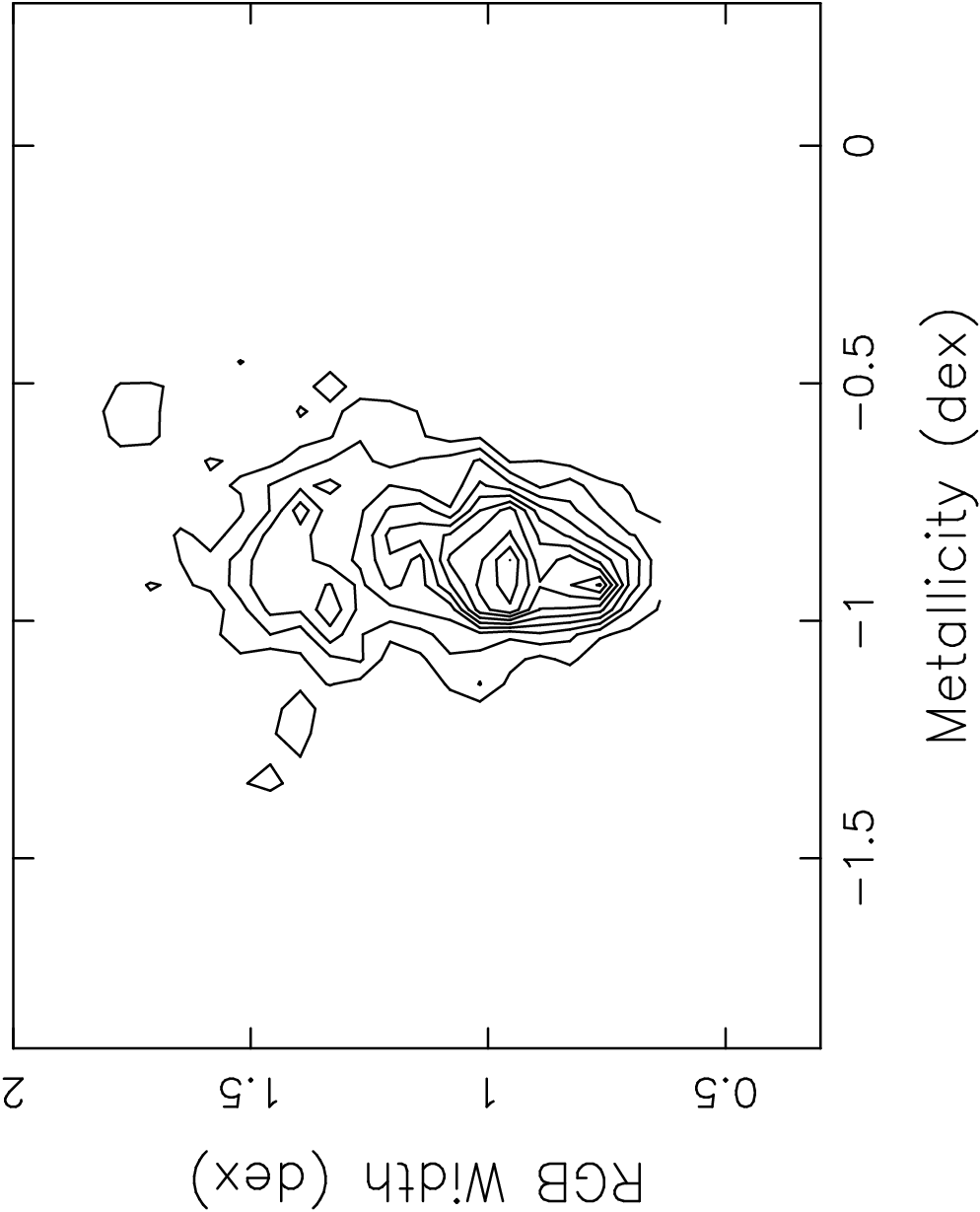}
\put(15,64){\small GSS9.5}
\end{overpic}
\\ \begin{overpic}[width = 0.23\textwidth,angle=-90,clip=true,trim={0cm 0cm 1.6cm 0cm}]{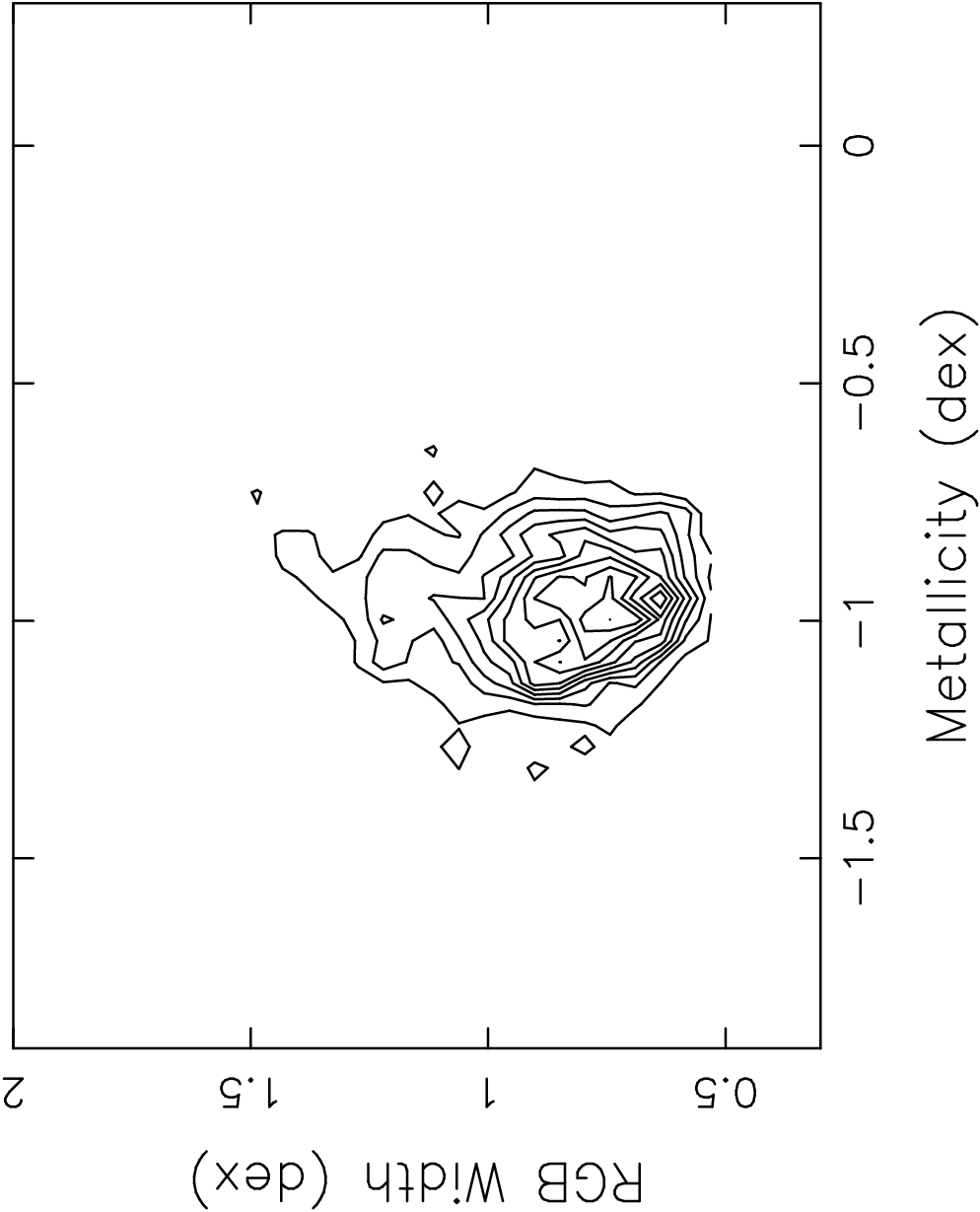}
\put(20,60){\small GSS10}
\end{overpic}
\begin{overpic}[width = 0.23\textwidth,angle=-90,clip=true,trim={0cm 1.1cm 1.6cm 0cm}]{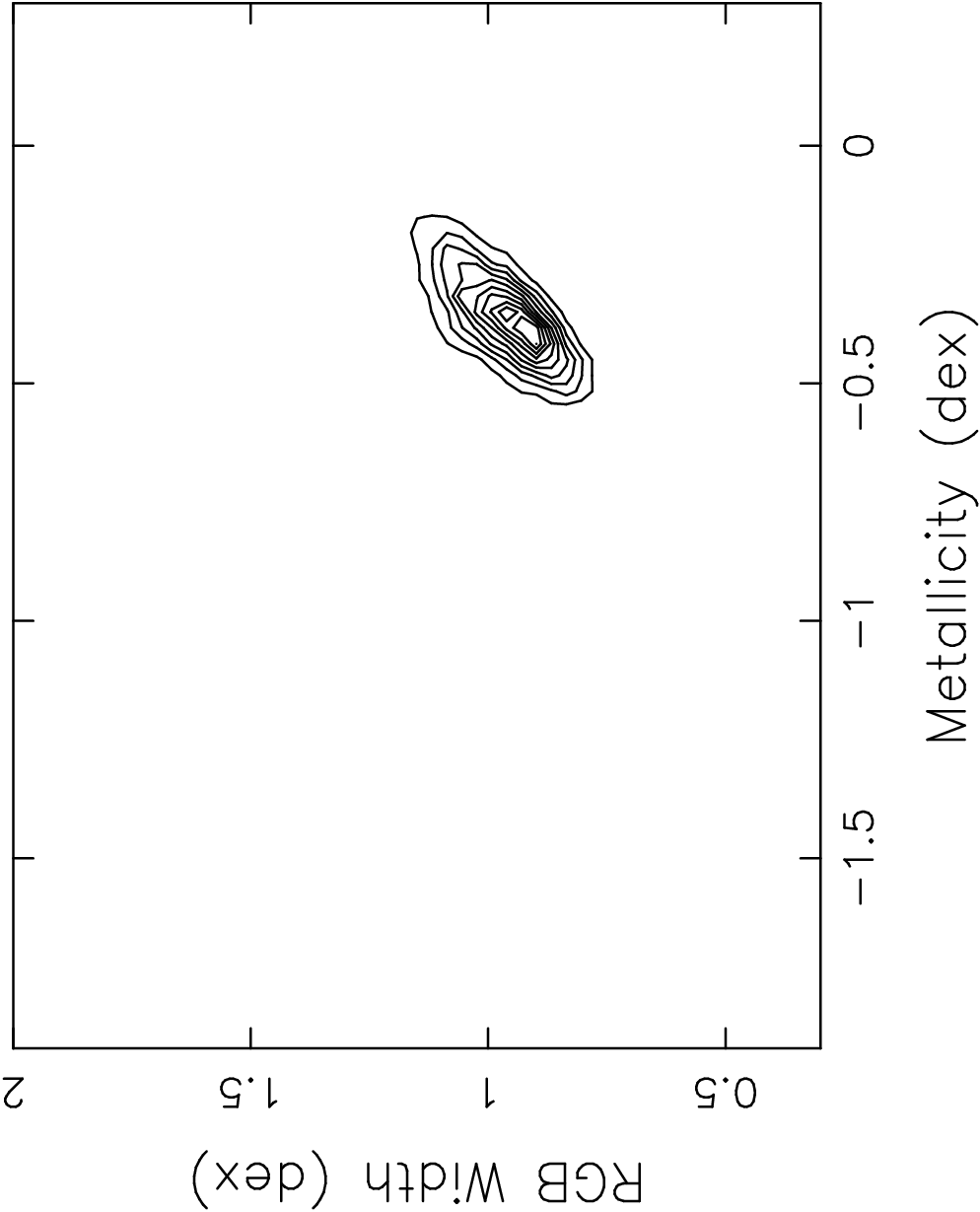}
\put(15,64){\small GSS5$\dagger$}
\end{overpic}
\begin{overpic}[width = 0.23\textwidth,angle=-90,clip=true,trim={0cm 1.1cm 1.6cm 0cm}]{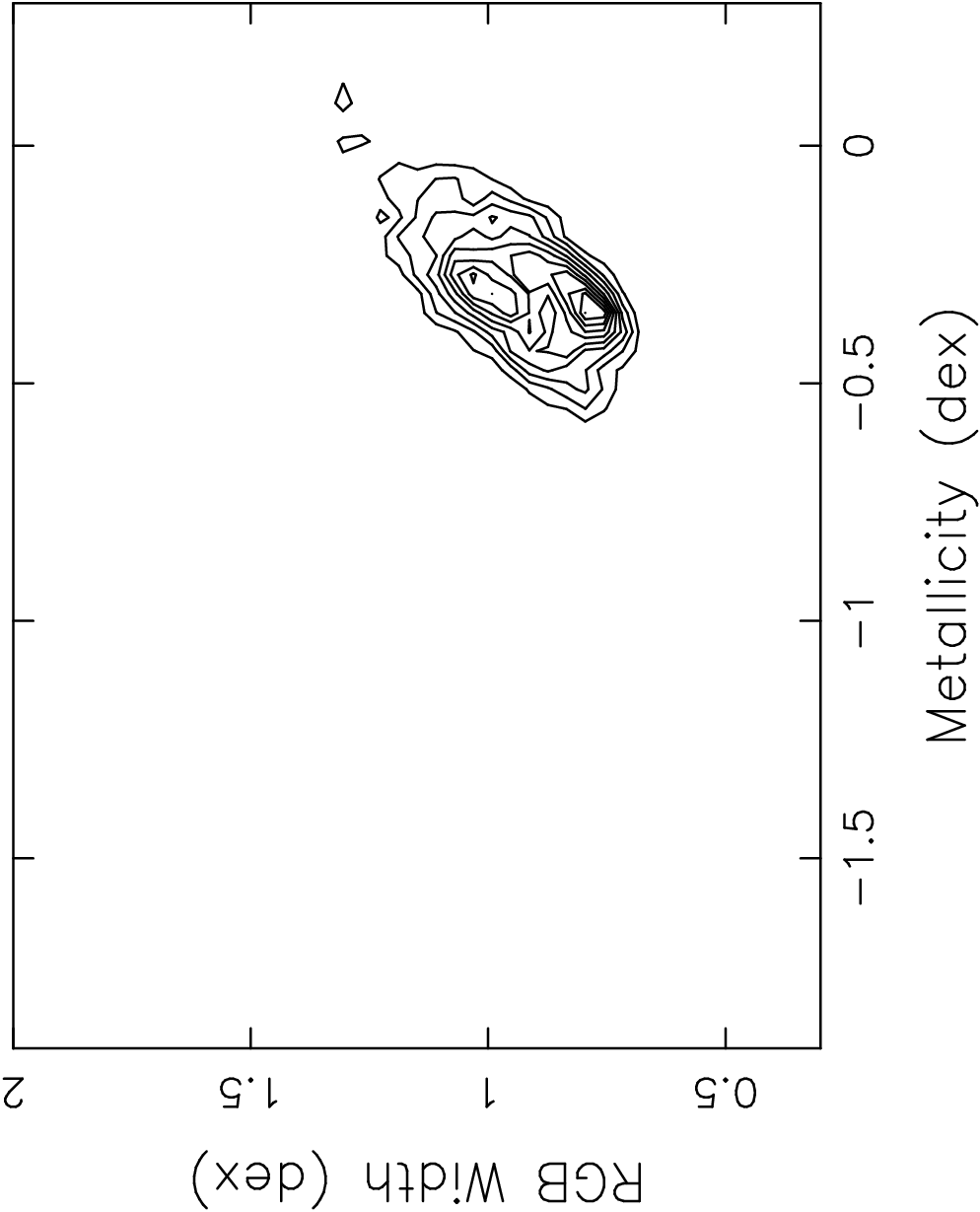}
\put(15,64){\small GSS6.5$\dagger$}
\end{overpic}
\\ \begin{overpic}[width = 0.258\textwidth,angle=-90,clip=true,trim={0cm 0cm 0cm 0cm}]{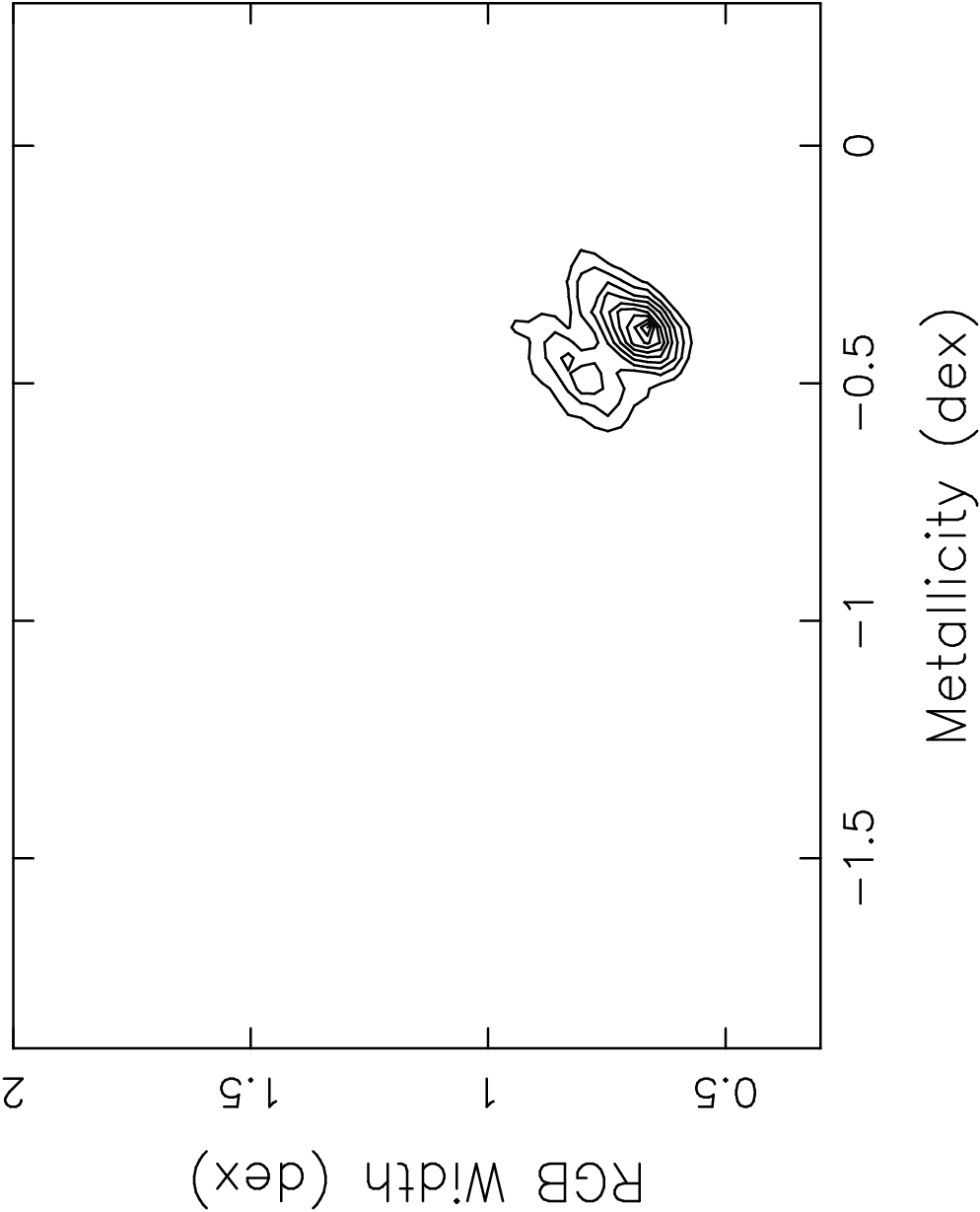}
\put(20,68){\small GSS7$\dagger$}
\end{overpic}
\begin{overpic}[width = 0.258\textwidth,angle=-90,clip=true,trim={0cm 1.1cm 0cm 0cm}]{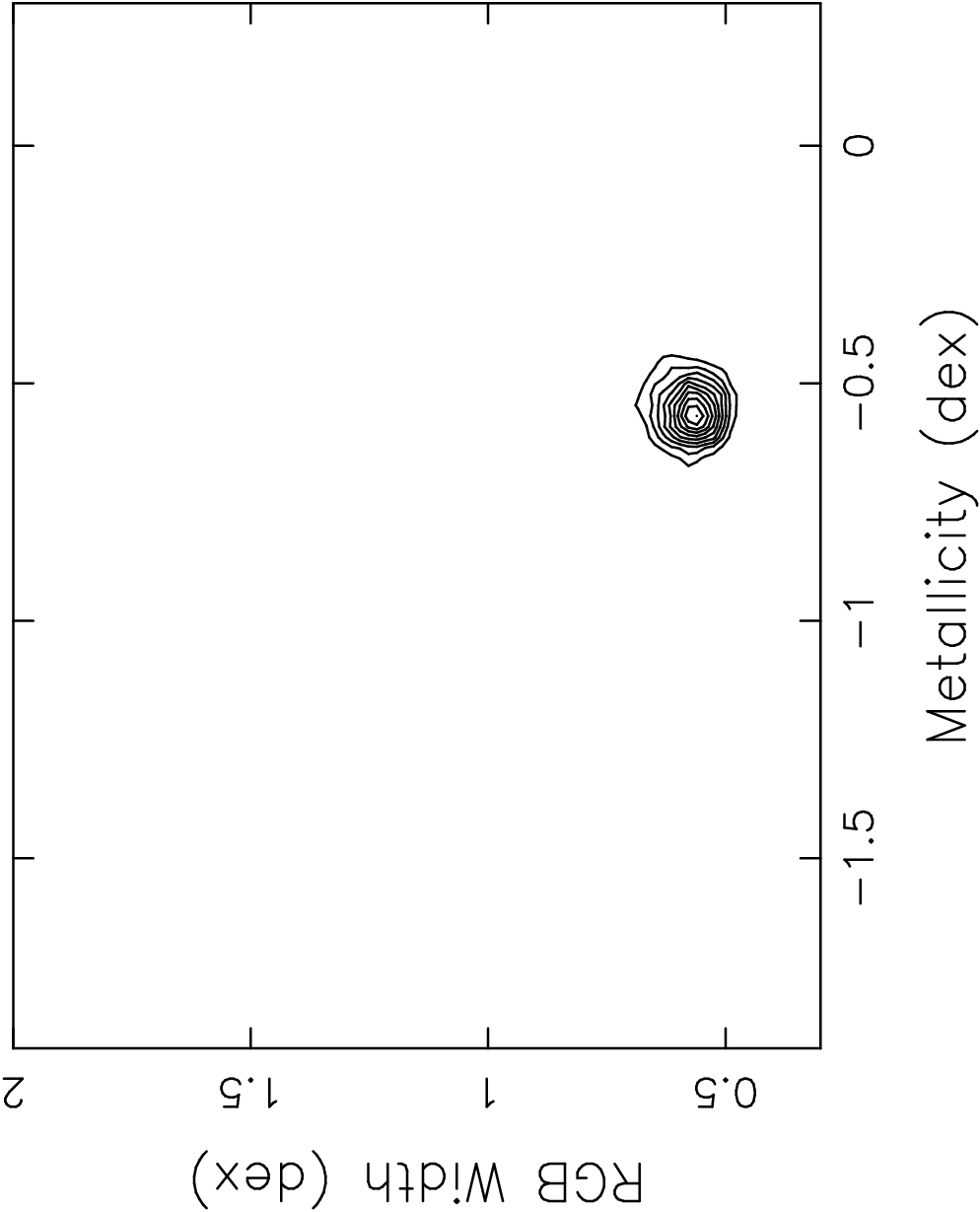}
\put(15,71){\small GSS8.5$\dagger$}
\end{overpic}
\begin{overpic}[width = 0.258\textwidth,angle=-90,clip=true,trim={0cm 1.1cm 0cm 0cm}]{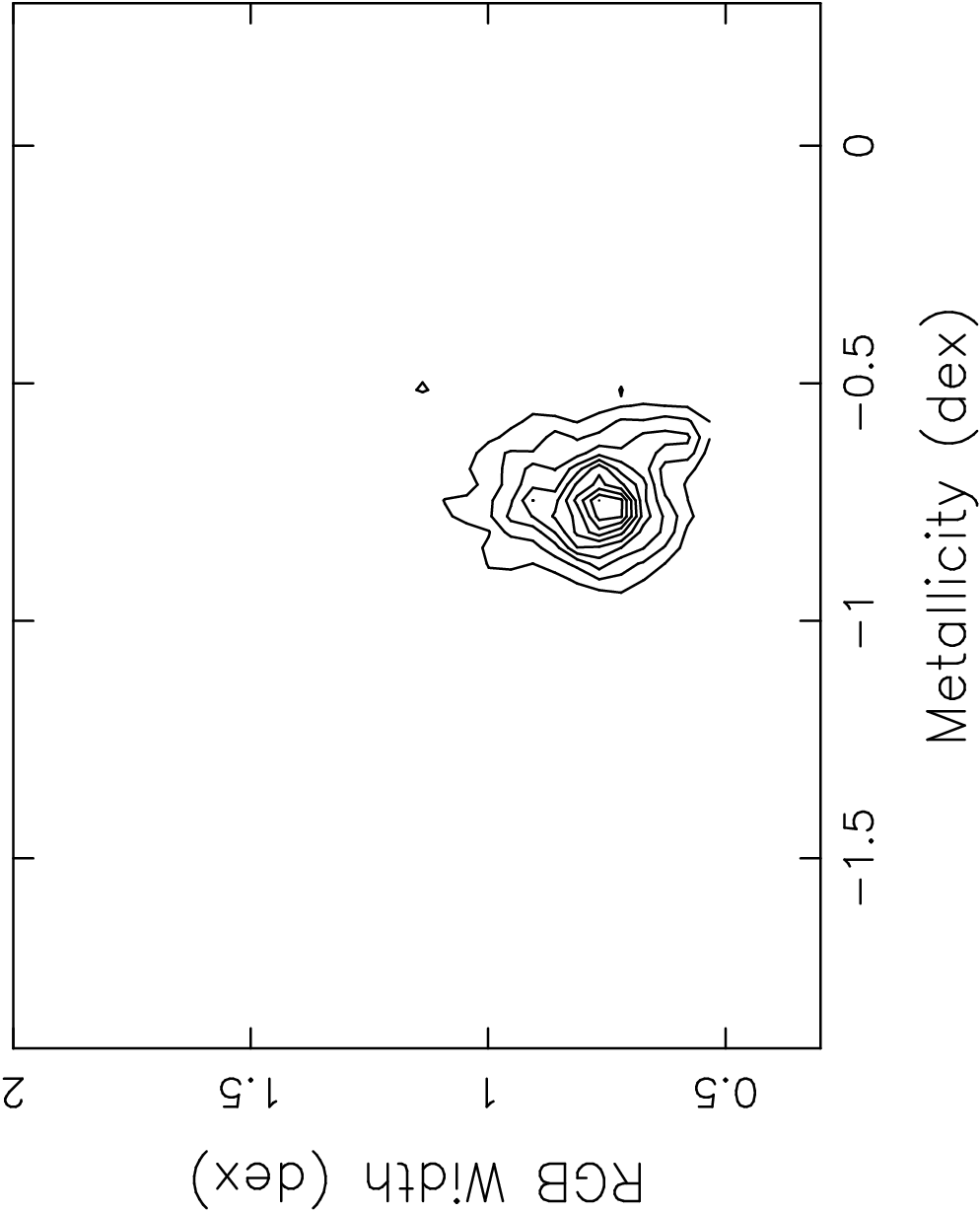}
\put(15,71){\small GSS9$\dagger$}
\end{overpic}

\end{array}$
\end{center}

\textbf{Appendix B Part II:} Contour plots illustrating the correlation between RGB width and metallicity for the fields listed in Table \ref{par_table1}. \\ Fields GSS8.5 through GSS10 are represented here. Plots denoted $\dagger$$^*$ are generated by sampling only the parameter values consistent with a restricted TRGB range. The full, unrestricted versions (for both Appendix B Parts I and II) are displayed here also and are denoted $\dagger$. \\ The restricted range plots are generated with the following limits: GSS8.5$\dagger$$^*$, $21.15 \le TRGB \le 21.30$; GSS9$\dagger$$^*$, $21.10 \le TRGB \le 21.30$. 

\label{AppBII}

\end{figure*}


\newpage


\begin{figure*}
\begin{center} 
$ \begin{array}{c}
\begin{overpic}[width = 0.23\textwidth,angle=-90,clip=true,trim={0cm 0cm 1.6cm 0cm}]{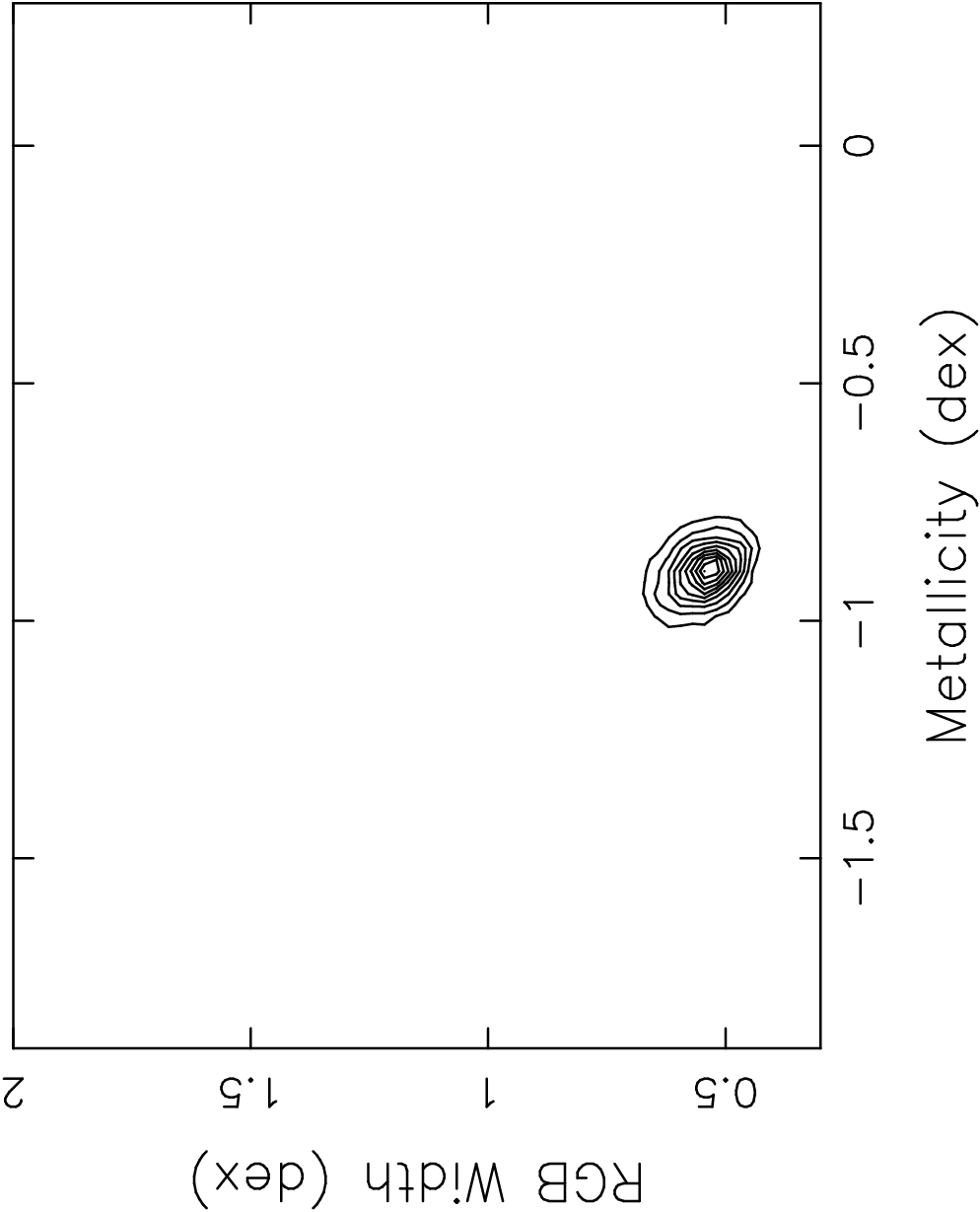}
\put(20,63){\small C1}
\end{overpic}
\begin{overpic}[width = 0.23\textwidth,angle=-90,clip=true,trim={0cm 1.1cm 1.6cm 0cm}]{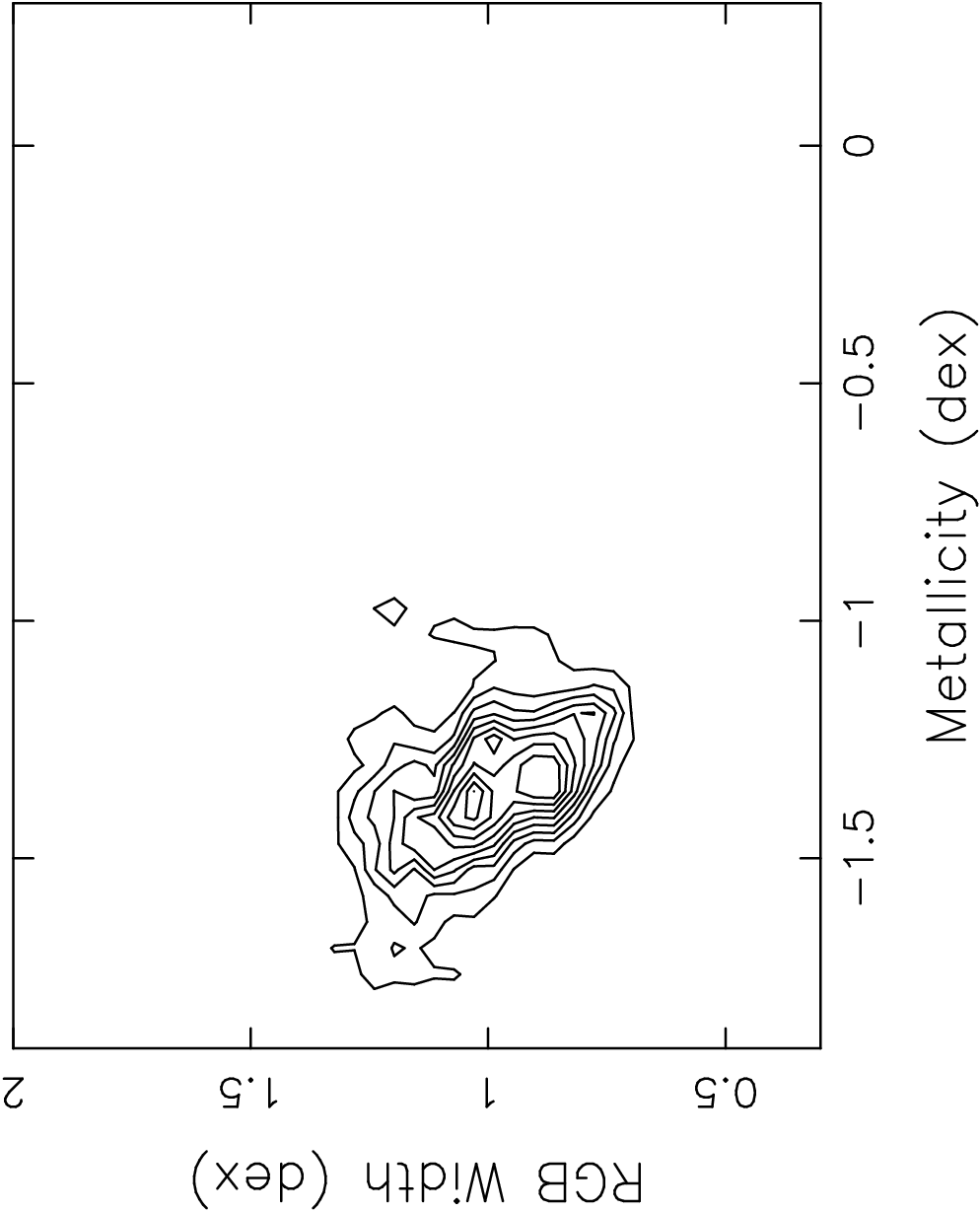}
\put(15,67){\small C2}
\end{overpic}
\\ \begin{overpic}[width = 0.23\textwidth,angle=-90,clip=true,trim={0cm 0cm 1.6cm 0cm}]{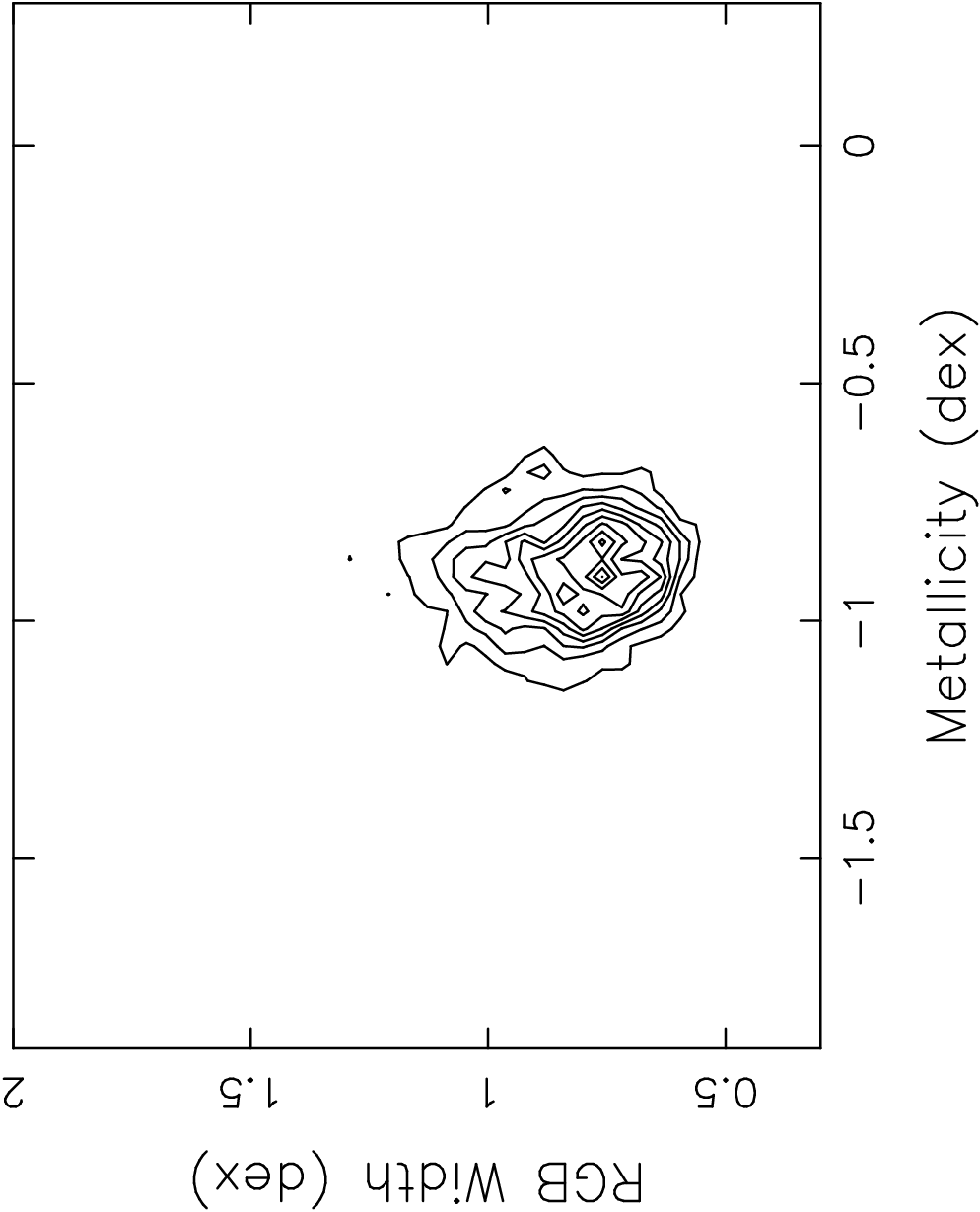}
\put(20,63){\small C3}
\end{overpic}
\begin{overpic}[width = 0.23\textwidth,angle=-90,clip=true,trim={0cm 1.1cm 1.6cm 0cm}]{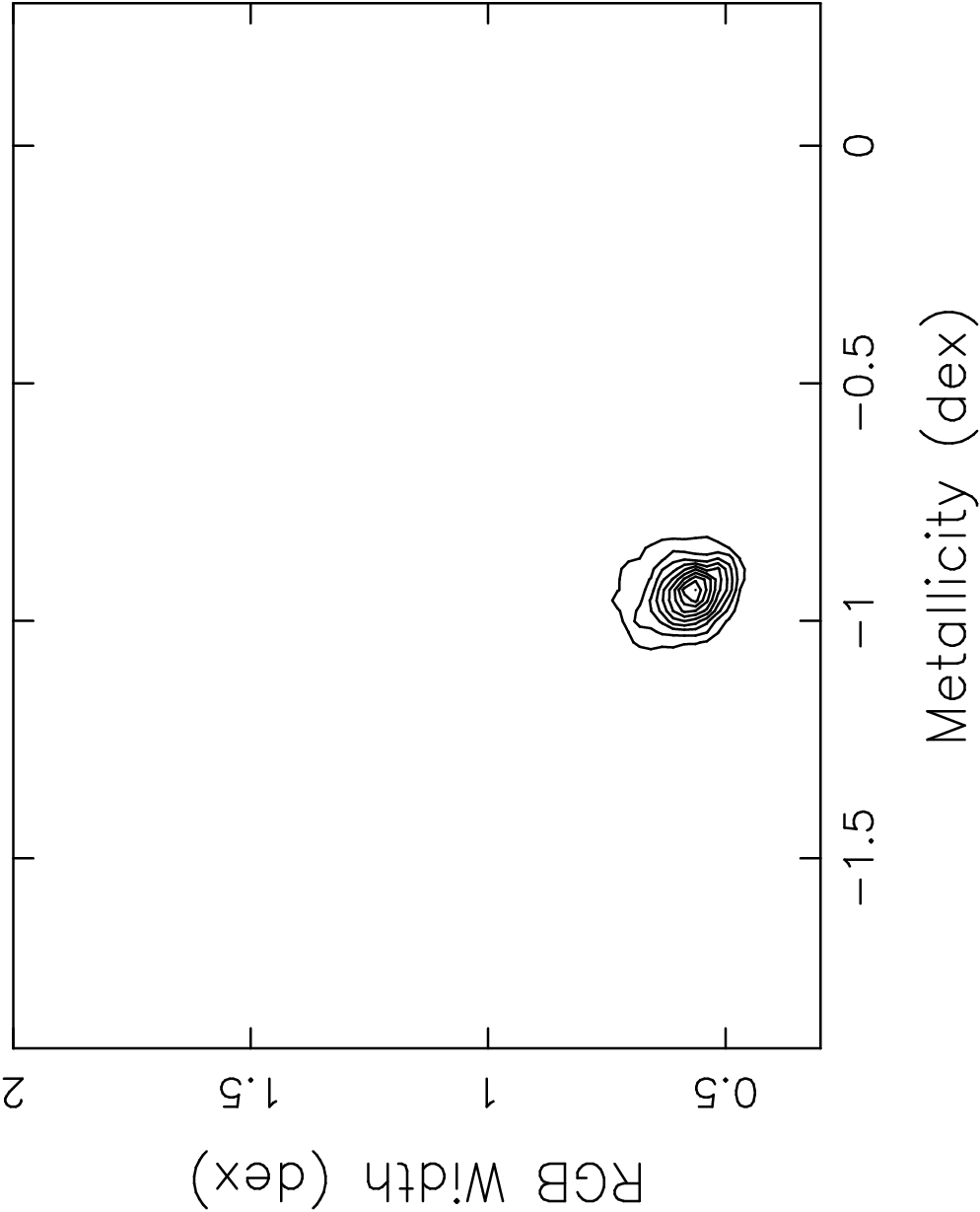}
\put(15,67){\small C4}
\end{overpic}
\\ \begin{overpic}[width = 0.23\textwidth,angle=-90,clip=true,trim={0cm 0cm 1.6cm 0cm}]{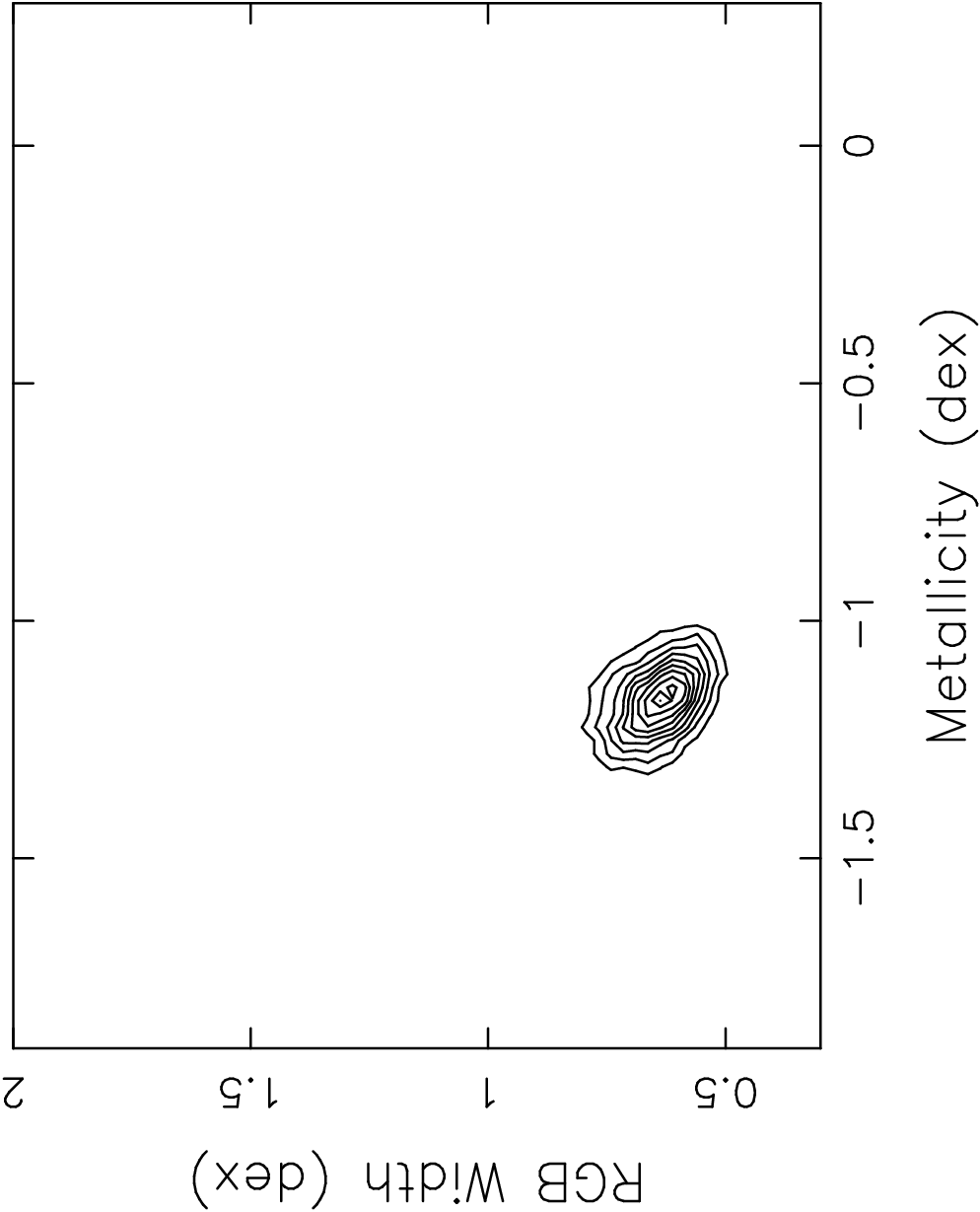}
\put(20,64){\small D1}
\end{overpic}
\begin{overpic}[width = 0.23\textwidth,angle=-90,clip=true,trim={0cm 1.1cm 1.6cm 0cm}]{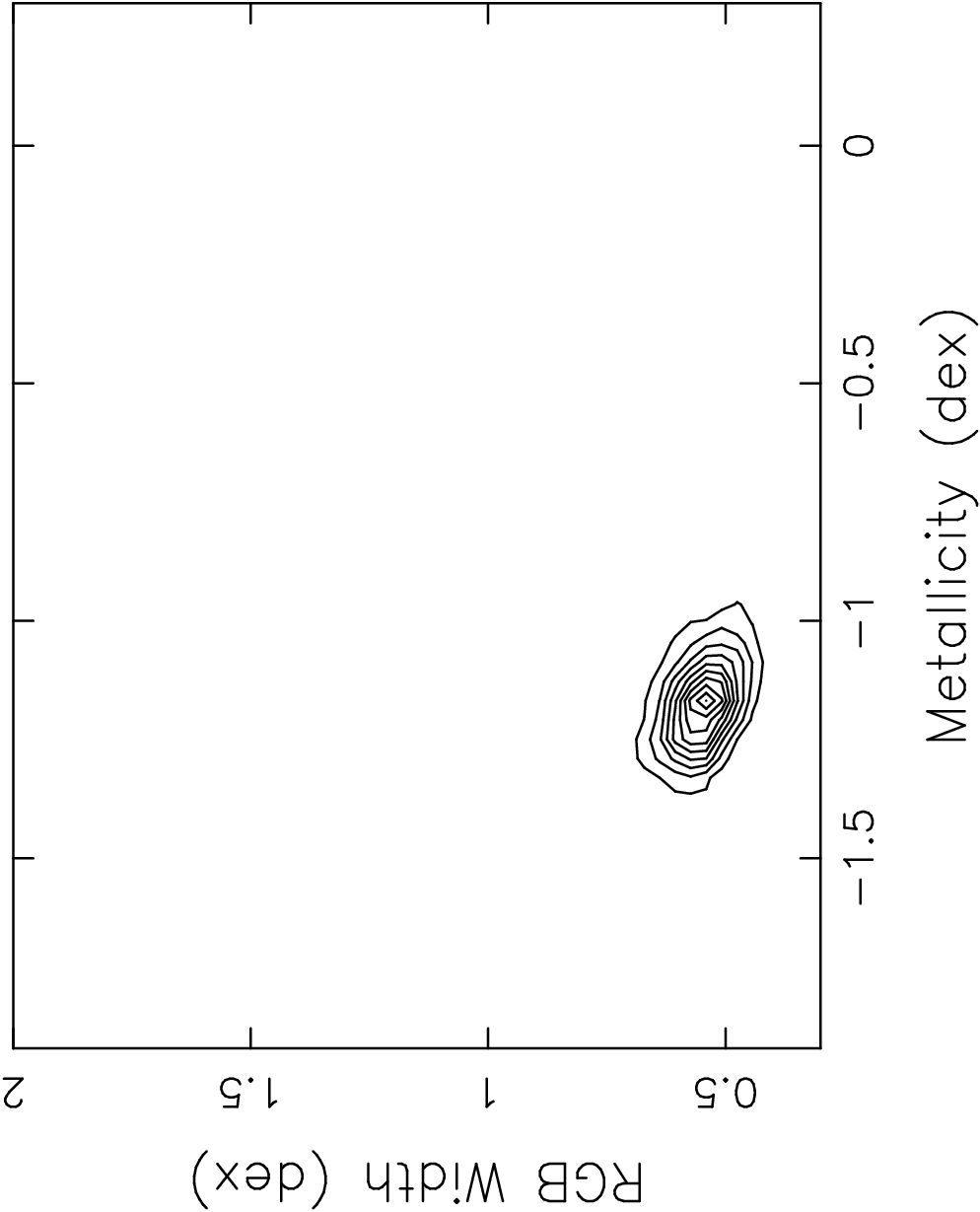}
\put(15,67){\small D2}
\end{overpic}
\begin{overpic}[width = 0.23\textwidth,angle=-90,clip=true,trim={0cm 1.1cm 1.6cm 0cm}]{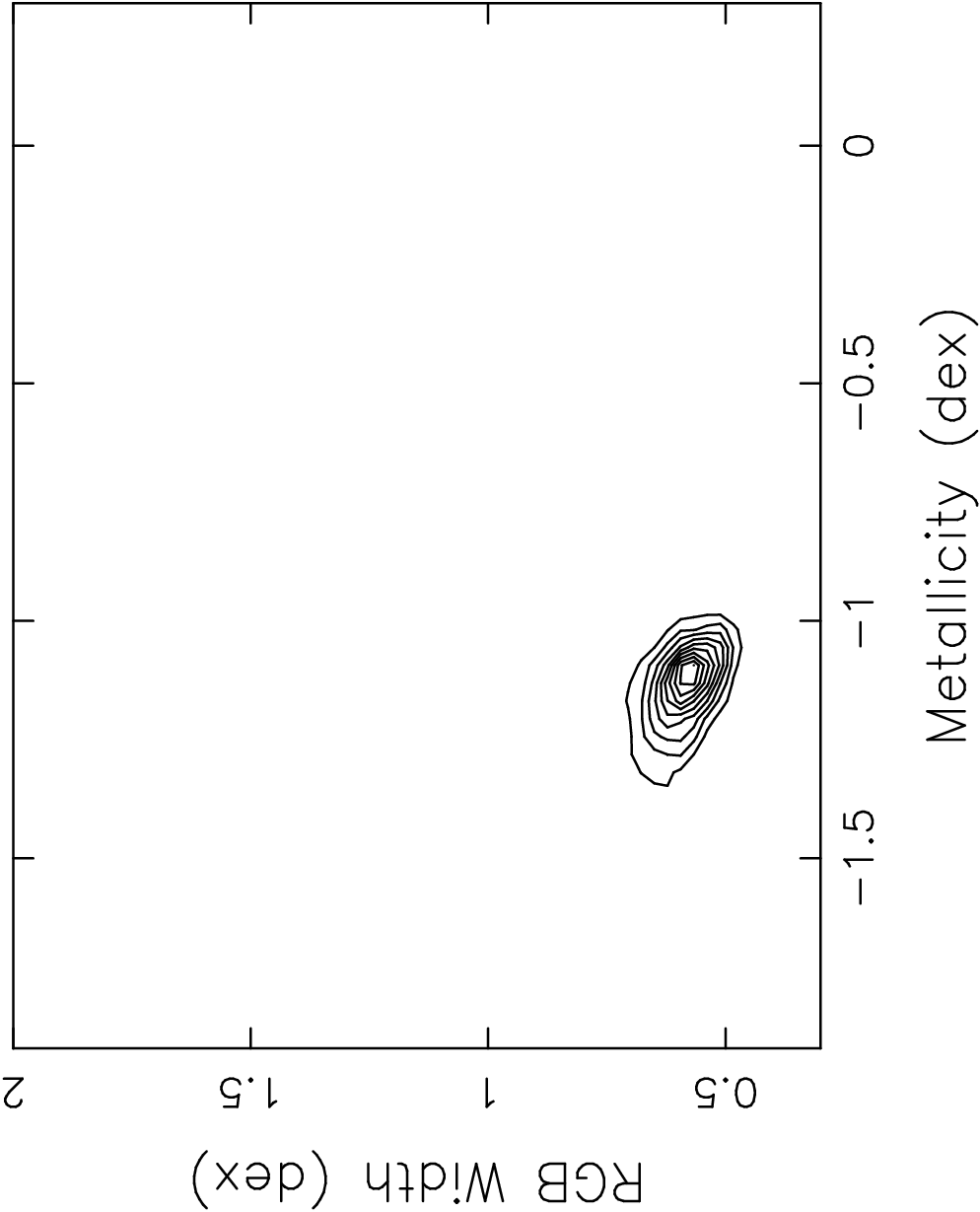}
\put(15,67){\small D3}
\end{overpic}
\\ \begin{overpic}[width = 0.252\textwidth,angle=-90,clip=true,trim={0cm 0cm 0cm 0cm}]{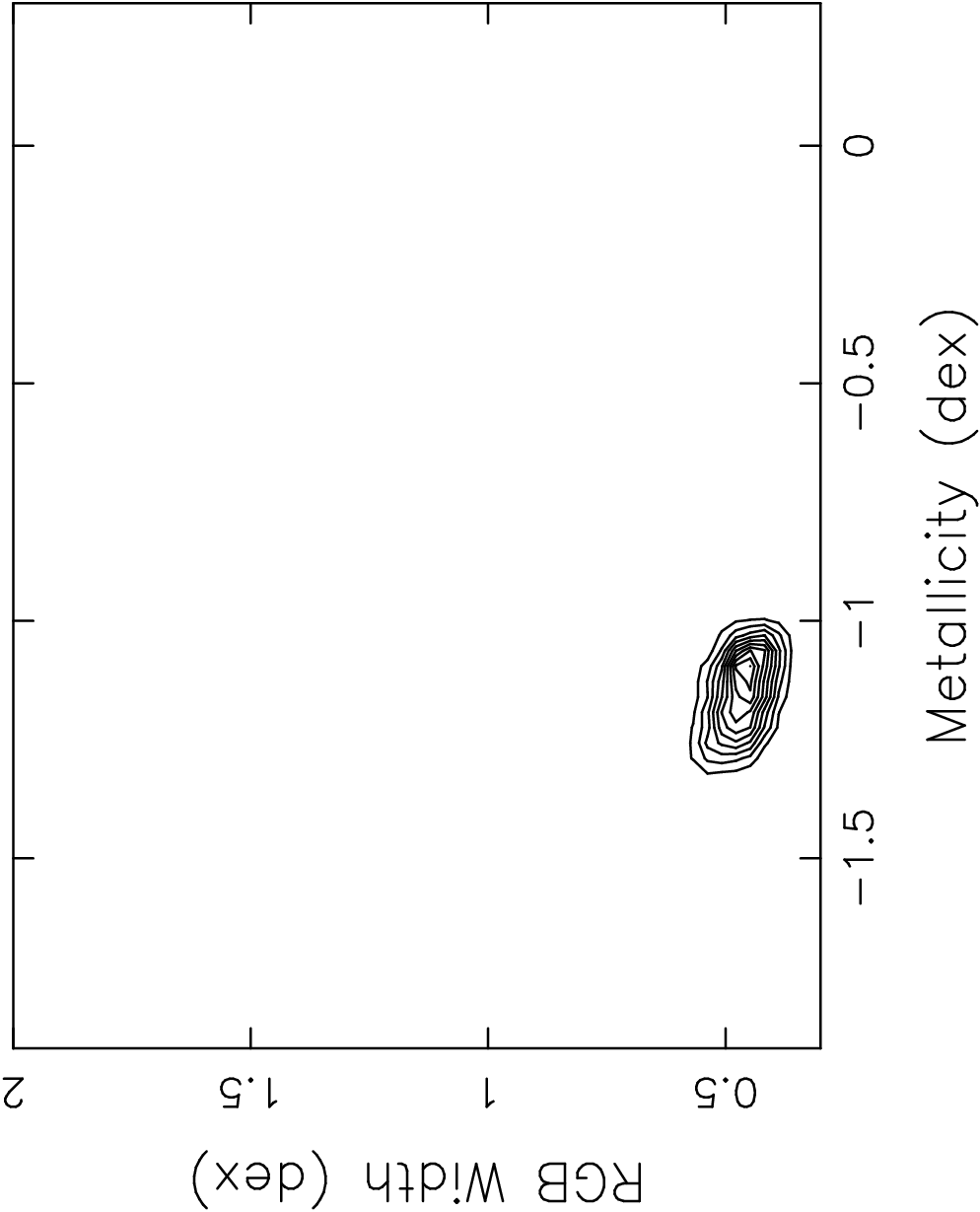}
\put(20,69){\small D4}
\end{overpic}
\begin{overpic}[width = 0.252\textwidth,angle=-90,clip=true,trim={0cm 1.1cm 0cm 0cm}]{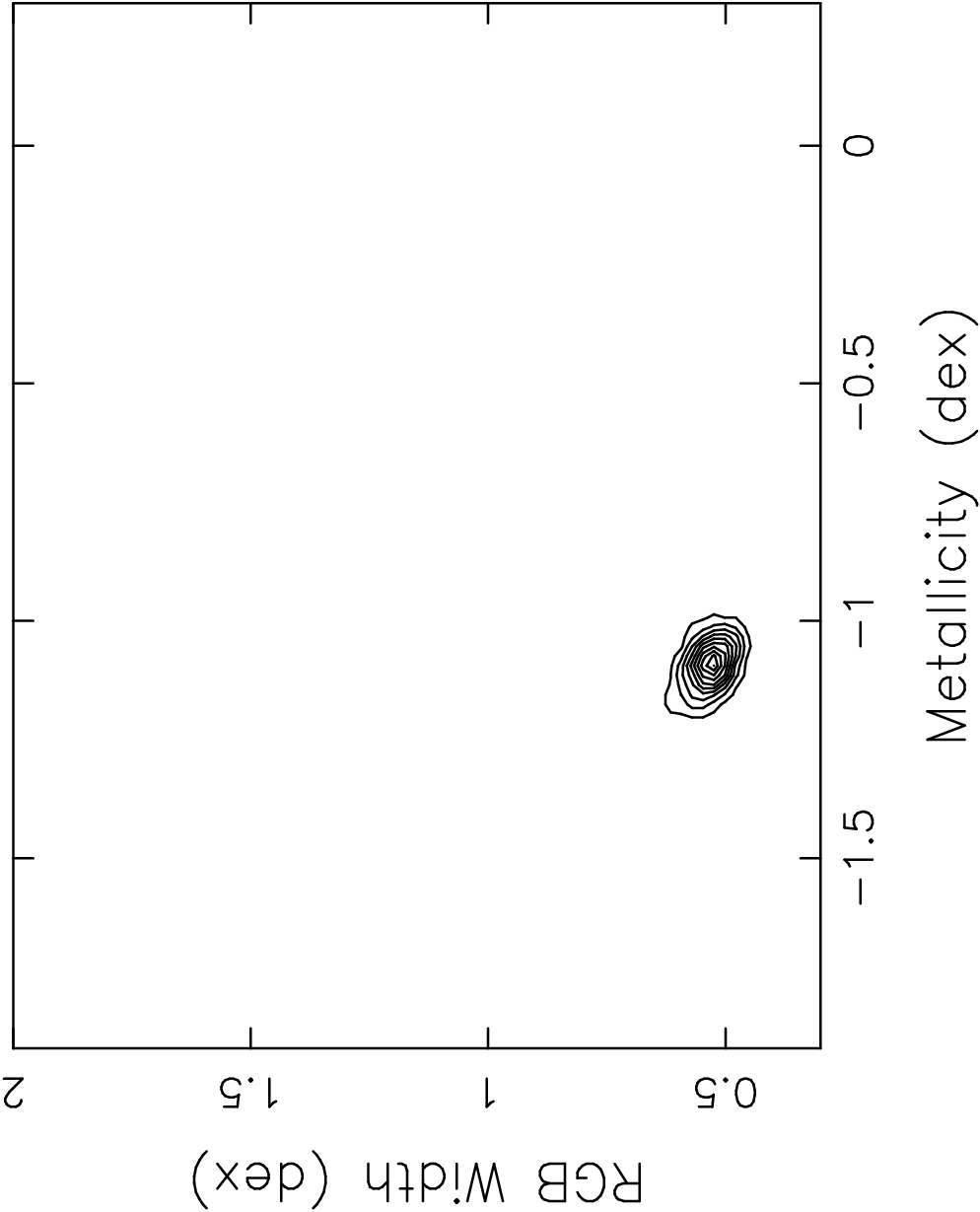}
\put(15,73){\small D5}
\end{overpic}

\end{array}$
\end{center}

\textbf{Appendix B Part III:} Contour plots illustrating the correlation between RGB width and metallicity for the fields listed in Table \ref{par_table2}. \\ Fields from Streams C \& D are represented here.

\label{AppBIII}

\end{figure*}

\newpage


\begin{figure*}
\begin{center} 
$ \begin{array}{c}
\begin{overpic}[width = 0.305\textwidth,angle=-90,clip=true,trim={0cm 0cm 1.6cm 0cm}]{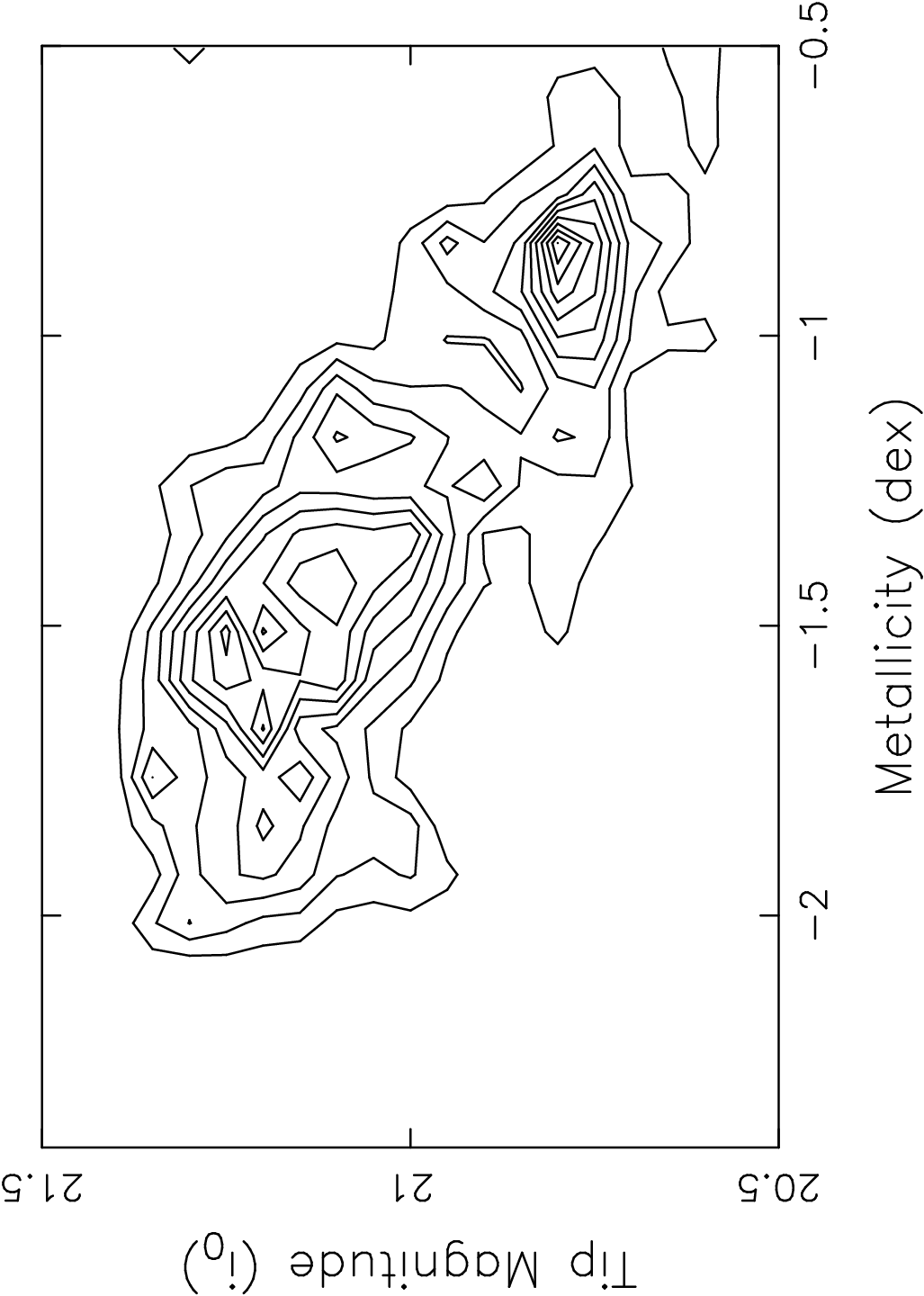}
\put(18,56){\small H1}
\end{overpic}
\begin{overpic}[width = 0.3\textwidth,angle=-90,clip=true,trim={0cm 0cm 1.6cm 0cm}]{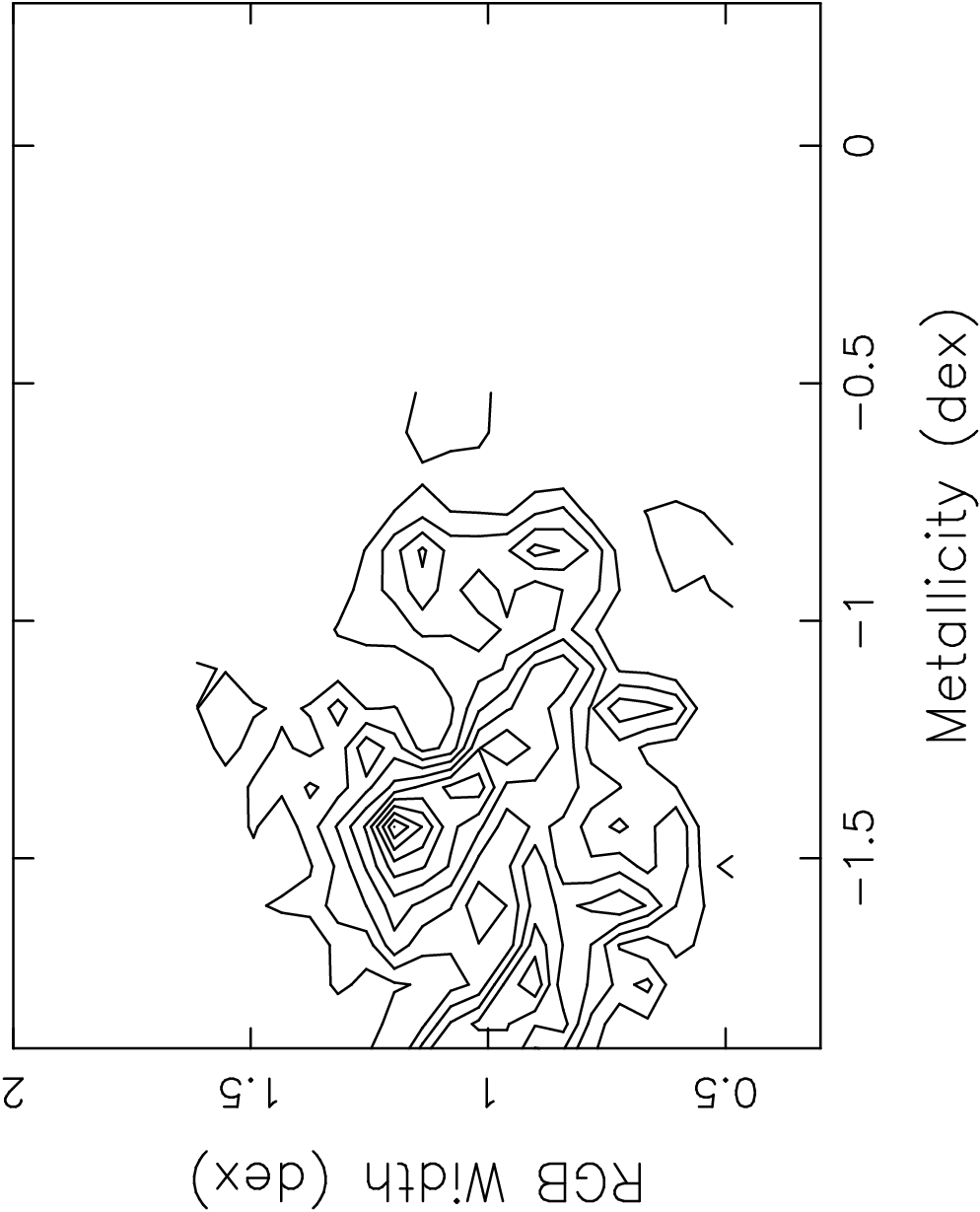}
\put(19,62){\small H1}
\end{overpic}
\\ \begin{overpic}[width = 0.305\textwidth,angle=-90,clip=true,trim={0cm 0cm 1.6cm 0cm}]{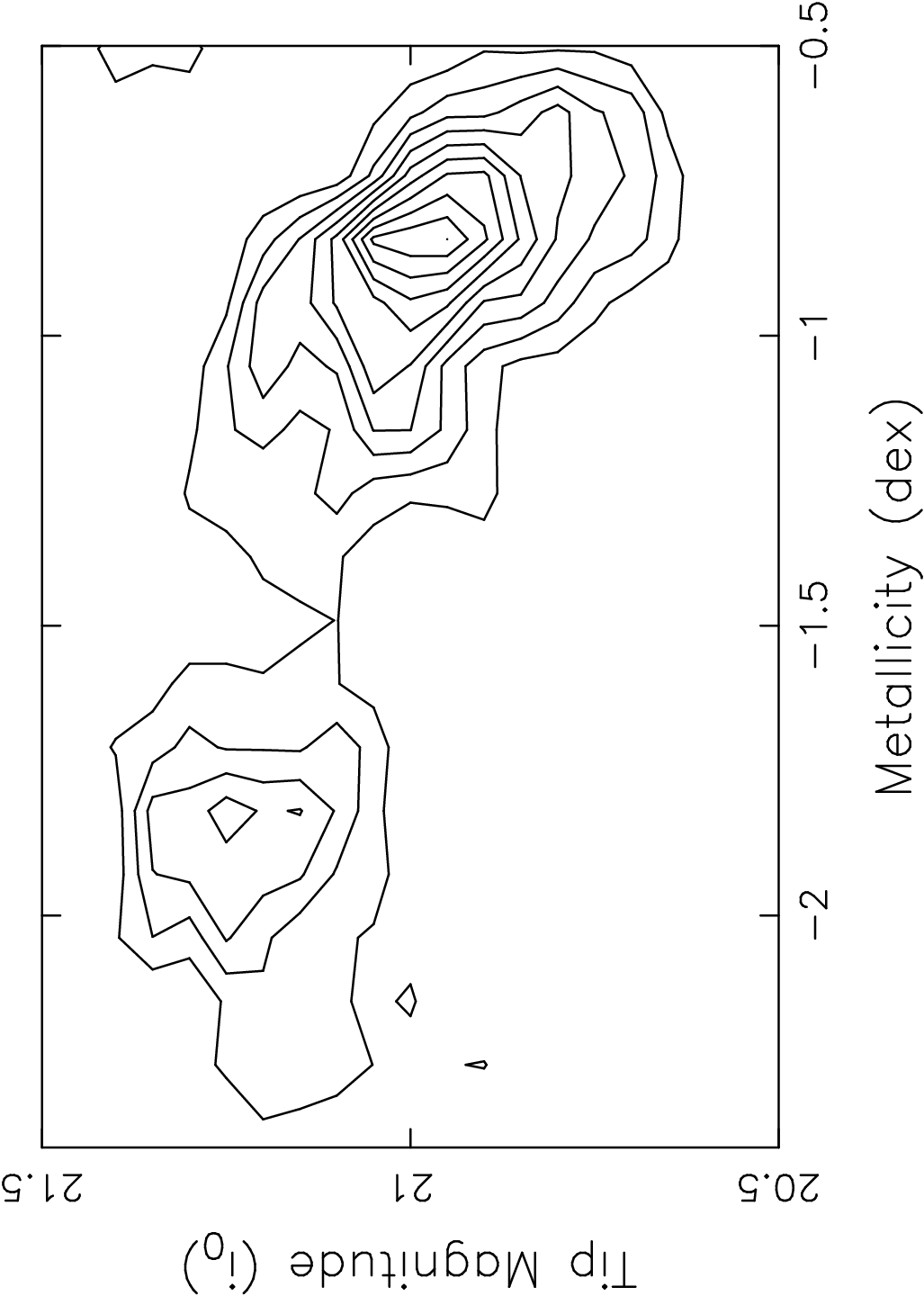}
\put(18,56){\small H2}
\end{overpic}
\begin{overpic}[width = 0.3\textwidth,angle=-90,clip=true,trim={0cm 0cm 1.6cm 0cm}]{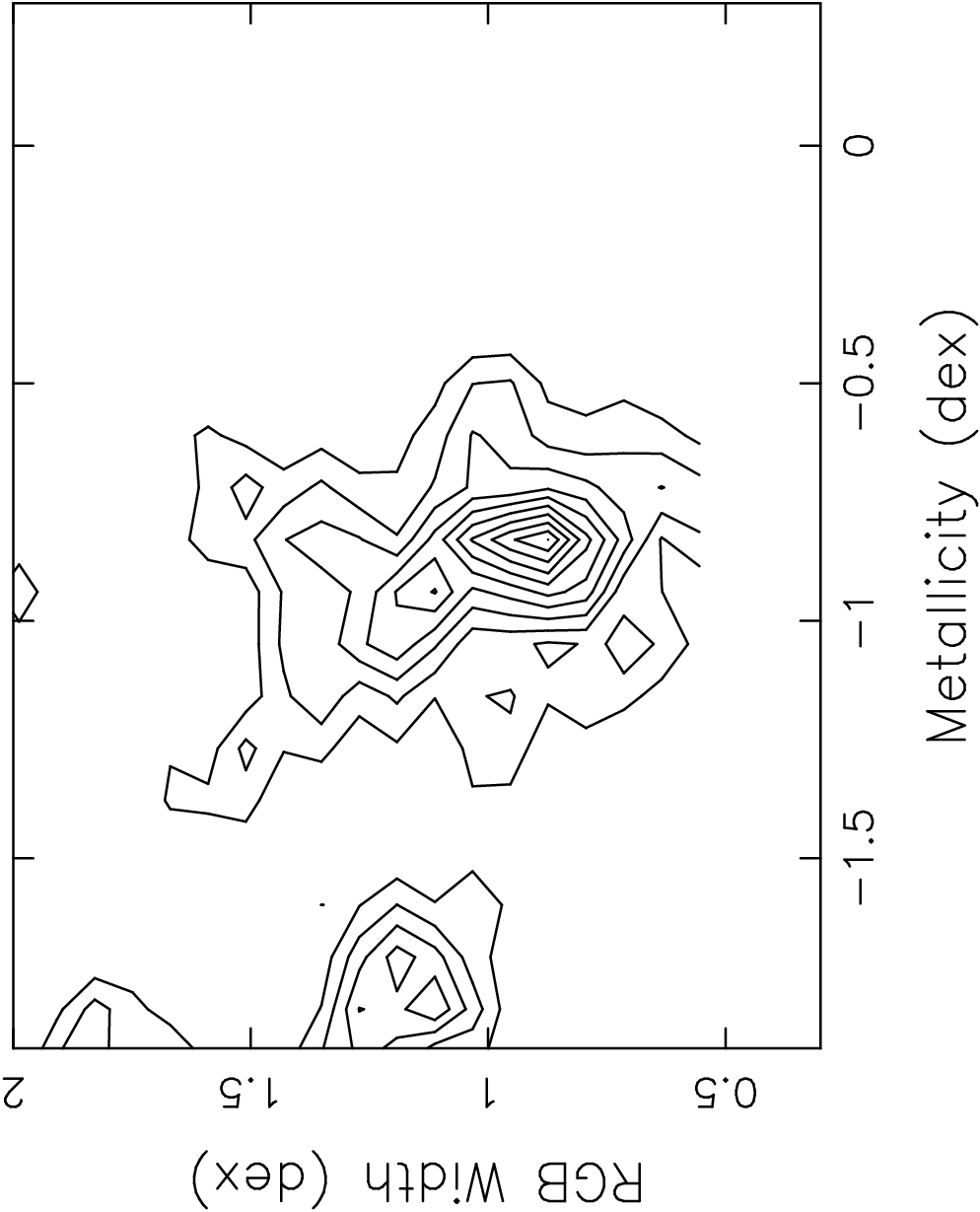}
\put(19,62){\small H2}
\end{overpic}
\\ \begin{overpic}[width = 0.335\textwidth,angle=-90,clip=true,trim={0cm 0cm 0cm 0cm}]{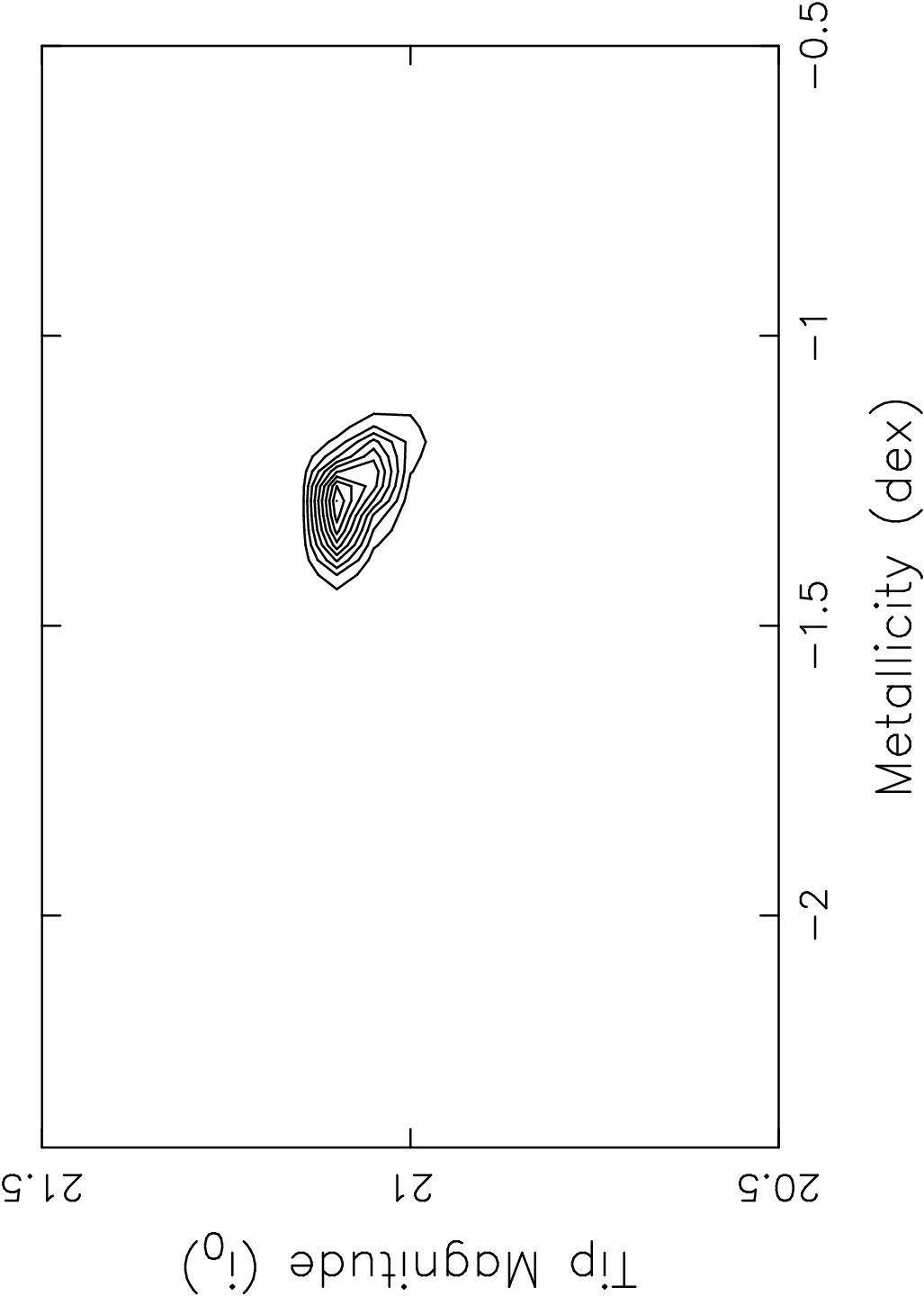}
\put(18,60){\small H3}
\end{overpic}
\begin{overpic}[width = 0.33\textwidth,angle=-90,clip=true,trim={0cm 0cm 0cm 0cm}]{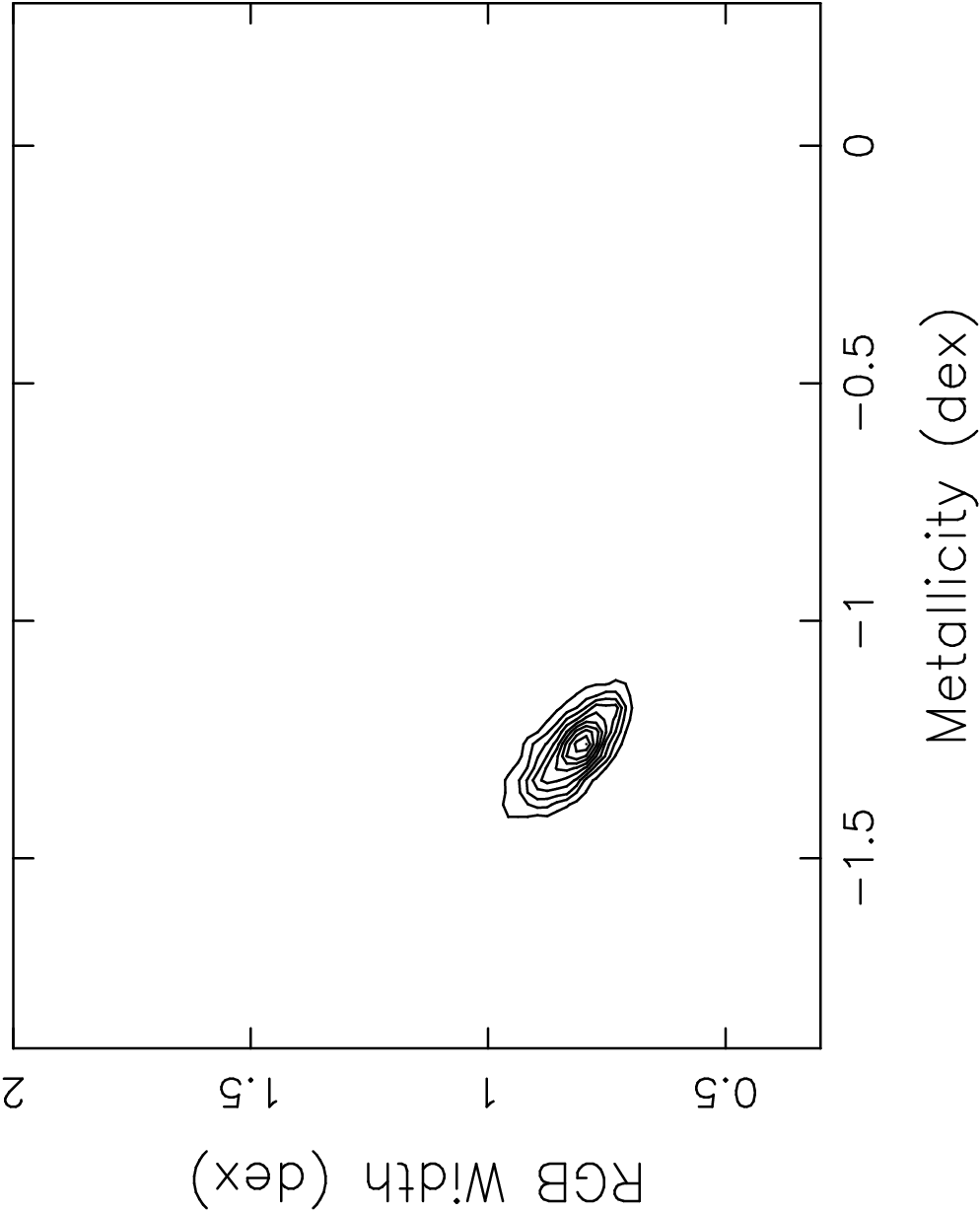}
\put(19,68){\small H3}
\end{overpic}

\end{array}$
\end{center}

\textbf{Appendix C:} Contour plots illustrating the correlation between tip magnitude and metallicity (left column) and between RGB width and metallicity (right column) for the 3 halo reference fields (see Table \ref{par_table2}). Note that the field H3 is much closer to the M31 disk than are H1 and H2 (see Fig. \ref{FieldMaps}), hence the markedly different distributions. 

\label{AppC}

\end{figure*}


\end{document}